\documentclass[a4paper,10pt]{article}
\pdfoutput=1

\usepackage{jcappub}

\usepackage{graphicx,natbib,amssymb,amsmath,stmaryrd,makecell}
\usepackage{txfonts}
\usepackage{xcolor}
\usepackage{hyperref}
\usepackage{xspace,dsfont}
\usepackage{caption, lipsum, comment}
\usepackage{textcomp}
\graphicspath{{Figures/}}
\usepackage{ulem}
\usepackage{cancel}
\newcommand{\beq}{\begin{equation}}   %

\newcommand{\eeq}{\end{equation}}   %

\newcommand{\beqa}{\begin{eqnarray}}   %

\newcommand{\eeqa}{\end{eqnarray}}   %

\newcommand{\bealf}[1]{\begin{align} #1 \end{align}}

\newcommand{\beal}{\begin{align}}
\newcommand{\enal}{\end{align}}

\newcommand{\bspl}{\begin{split}}

\newcommand{\espl}{\end{split}}

\newcommand{\bsub}{\begin{subequations}}

\newcommand{\esub}{\end{subequations}}

\newcommand{\bmulti}{\begin{multline}}   %

\newcommand{\beqm}{\begin{mathletters}}   %

\newcommand{\eeqm}{\end{mathletters}}   %

\newcommand{\me}{m_{\rm e}}

\newcommand{\recfastpp}{\ensuremath{\tt {RECFAST\!+\!+}}}
\newcommand{\fgspectra}{\ensuremath{\tt {fgspectra}}}
\newcommand{\camb}{\ensuremath{\tt {CAMB}}}

\usepackage{comment}

\title
{Shedding light on modified recombination scenarios with Rayleigh scattering of the CMB}

\author[1,2]{Debabrata Adak,}
\author[1,2]{Elena Vanetti,}
\author[1,2]{R.~T. Génova-Santos,}
\author[3]{Jens Chluba,}
\author[1,2]{Carlos Hern\'andez-Monteagudo}
\author[1,2]{J.~A. Rubi\~{n}o-Mart\'{\i}n}
\emailAdd{adak@iac.es}
\affiliation [1]{Instituto de Astrof\'{\i}sica de Canarias, E-38200 La Laguna, Tenerife, Spain}
\affiliation[2]{Departamento de Astrof\'{\i}sica, Universidad de La Laguna, E-38206 La Laguna, Tenerife, Spain}
\affiliation[3]{Jodrell Bank Centre for Astrophysics, The University of Manchester, Manchester M13 9PL}
\date{\today}
\abstract{
Rayleigh scattering (RS) of cosmic microwave background (CMB) photons by neutral hydrogen and helium, during and shortly after the last scattering surface, adds a frequency-dependent contribution to the CMB anisotropies that scales approximately as $\nu^4$ at leading order. The RS signal is most relevant at $200\,{\rm GHz}\lesssim\nu\lesssim800\,{\rm GHz}$ and could be detected using future CMB experiments. Because RS acts in a frequency-dependent manner which is sensitive to the ionisation history of the cosmological recombination process, it provides an additional probe of the recombination history. This makes it a natural test of modified recombination scenarios, such as variable electron mass $m_{\rm e}$ or fine-structure constant $\alpha$, which have been proposed to ease the Hubble tension. The correlation between the RS and primary CMB signal, both in temperature and polarisation, further enhances its usefulness for constraining cosmological parameters. Using a multi-frequency Fisher analysis with realistic noise and foregrounds, we forecast parameter constraints and biases for PICO and LiteBIRD + Planck. For PICO, adding RS improves the $\Lambda$CDM constraints on cosmological parameters by $\lesssim7\%$, but improves the constraints on some parameters ($H_0$, $m_{\rm e}$, $\alpha$, $\Omega_{\rm b}h^2$, $\Omega_{\rm c}h^2$) by $\simeq31\%$ for some modified scenarios in the recombination history. However, the improvement for LiteBIRD + Planck is limited to less than 2\% for most of the parameters. Neglecting RS biases the cosmological parameters by up to $\simeq1\sigma$ for PICO and below $1\sigma$ for LiteBIRD + Planck. Therefore RS must be modelled in the analyses of future CMB data.

}

\begin{document}
\maketitle
\flushbottom

\section{Introduction}
\label{sec:Intro}
In modern cosmology, our understanding of the physics of the cosmic microwave background (CMB) and the precise measurement of its temperature and polarisation anisotropies with many ground- and space-based experiments \citep{ SPTpol:2009, ACTPol:2010, Bennett:2013, Planck-I:2020} has led to the demonstration that the standard $\Lambda$CDM model of the Universe provides an extremely successful description of the current data. However, several cosmological tensions still persist, e.g., (i) mismatch between the Hubble constant $H_{0}$ from CMB data \citep{planck-VI:2020} and the same obtained from Cepheid-calibrated Type Ia Supernovae \citep{Riess:2022ApJ}; (ii) discrepancy between the amplitude of the growth of structure ($\sigma_8$) inferred from CMB data and the one from late-time large-scale structure observations \citep{Cosmology_Intertwined_III:2021APh, Pedreira:2024PhRvD, Shaikh_ACT:2023skz, Qu:2024sfu, Karim:2024luk}. Despite several efforts to attribute the Hubble tension to systematic uncertainties in the data \citep{Mortsell:2022ApJ,Mortsell:2022}, it has persisted. Several new-physics models have attempted to alleviate the Hubble tension \cite{Chiang:2018, Hart:2020Hubble,RoyChoudhury:2020dmd,Jedamzik:2020PhRvL,Ruchika:2023ugh,Erminia:2025,wang:2025EPJC}. However, no conclusive resolution has yet been reached. Since the CMB precisely measures the angular size of the sound horizon, $\theta_* = r_s/D_M(z_*)$, the value of $H_0$ inferred from it depends both on the comoving sound horizon at recombination, $r_s$, which is set by pre-recombination physics, and on the comoving distance to the last-scattering surface, $D_M(z_*)$, which depends on the post-recombination expansion history \citep{Knox2020}. Therefore, exploring modifications to the physics of the early or late Universe is particularly interesting to alleviate the Hubble tension.

Most attempts to alleviate the Hubble tension by modifying post-recombination physics rely on changes in the late-time expansion history, within the range allowed by BAO and Type Ia supernovae data \citep{DiValentino2021,Schoneberg:2021qvd}. These typically invoke an evolving effective dark energy component while leaving the comoving sound horizon at recombination unchanged, and include dynamical dark energy (DDE) \citep{Zhao:2017NatAs}, interacting dark sector \citep{DiValentino:2017}, and modified gravity models \citep{Pogosian:2021}. However, the freedom of these models to alleviate the $H_0$ tension is largely removed once BAO and supernovae data are included, since these models leave the sound horizon unchanged \citep{Knox2020,Schoneberg:2021qvd}. Late-time modifications have nevertheless received renewed attention after the DESI BAO results \citep{Adame:2025JCAP} which prefer DDE at $\sim2.8$--$4.2\sigma$ significance to reduce the $\Omega_{\rm m}$ tension, although this has been challenged by \citep{OColgain:2025MNRAS}, claiming such tension is due to unexplored systematics. 

Another compelling possibility is to change the sound horizon in the pre-recombination era. A well-studied class of models that achieves this relies on introducing an early dark energy (EDE) component that decays before the recombination epoch (see \cite{Kamionkowski:2023ARNPS} for a review).
Another promising avenue is to invoke early recombination, for instance via time-varying fundamental constants; in particular, varying electron mass, $m_{\rm e}$ or fine-structure constant $\alpha$ \citep{Hart:2017,Hart:2020PCA,Hart:2020Hubble, Sekiguchi2021PhRvD,Toda:2024PDU,Schoneberg:2025JCAP}. A larger electron mass during the early Universe makes the recombination process faster, which leads to a shorter sound horizon and therefore a larger value of $H_0$.
The varying electron mass model is well motivated in the literature and is consistent with other cosmological data sets that favour larger electron mass with greater significance than 2$\sigma$ \citep{Hoshiya2023,Seto:2023,TODA2025102035,SGhsosh:2025}.

In the literature, the impact of varying fundamental constants is studied mainly through their influence on Thomson scattering of CMB photons off of free electrons during the recombination era, at redshift $z\sim 1100$. In this work, we discuss the impact of varying fundamental constants on the scattering of CMB photons with newly formed hydrogen and helium atoms during the recombination era, namely Rayleigh scattering (RS, \citep{Yu:2001, Lewis2013Rayleigh}). The effect of RS is often neglected in the literature, assuming that CMB photons can travel freely after recombination. 
RS dampens CMB anisotropies on sub-horizon scales because photons are further scattered out of the line of sight, which leads to mixing between photons at hot and cold spots. 
In addition, the large-scale $E$-mode polarisation produced at recombination arises from the coherent scattering of the local temperature quadrupole into the line of sight. Since RS is increasingly important at higher frequencies and relevant only in the later stages of recombination, when the photons have free-streamed further, and the quadrupole has grown, the additional RS increases the polarisation amplitude on large scales at high frequencies. Thus RS introduces an asymptotic part in the CMB  intensity and polarisation anisotropies which scales as $\nu^4$ at leading order, making observations at frequencies 200 GHz $\lesssim \nu \lesssim$ 800 GHz particularly important for future CMB experiments. In this work, we model the effect of RS on CMB anisotropies and study its cosmological importance for future CMB experiments, e.g., LiteBIRD \citep{LiteBIRD:2022cnt} and PICO \citep{PICO:2019}.

The paper is organised as follows. In Section~\ref{sec:Rayleigh_MODREC}, we describe the effect of RS on the CMB temperature and polarisation anisotropies, and show how a variation of $m_{\rm e}$ or $\alpha$ modifies the RS opacity, the Thomson and Rayleigh visibility functions, and the resulting frequency-dependent power spectra. In Section~\ref{sec:methodology} we describe
 the multi-frequency Fisher formalism (Section~\ref{sec:fromalism}) used to forecast parameter constraints and to estimate the biases arising from an incorrect model, and the foreground model (Section~\ref{sec:fg_model}) included in the analysis. Section~\ref{sec:results} presents our results:
the forecasts cover constraints with and without RS for $\Lambda$CDM and the modified recombination scenarios (Section~\ref{sec:res_constraints}), and the parameter biases that arise when RS or the variation of the fundamental constants is not modelled consistently (Section~\ref{sec:res_bias}). We summarise our conclusions in
Section~\ref{sec:conclusion}. The derivation of the Fisher matrix and of the parameter bias is given in Appendix~\ref{sec:derivations}, and the corresponding results for LiteBIRD + Planck are presented in Appendix~\ref{app:B}.

\section{Rayleigh scattering in $\Lambda$CDM and  modified recombination scenarios}
\label{sec:Rayleigh_MODREC}
In this section, we illustrate the mechanism of RS and its impact on  CMB power spectra at different frequencies, both in the standard $\Lambda$CDM and modified recombination scenarios. The fiducial values of the six $\Lambda$CDM cosmological parameters are set to the best-fit values of the Planck 2018 results \citep{planck-VI:2020} throughout this paper.

\subsection{Rayleigh Scattering effects on CMB temperature and polarisation anisotropies}\label{subsec:rs_physics}
The elastic scattering of long-wavelength photons by neutral particles is known as \textit{Rayleigh scattering} \citep{Rayleigh01081881}. By the time recombination occurs, the vast majority of CMB photons have energies well below the hydrogen ionisation energy and therefore wavelengths well above the atomic sizes of the hydrogen and helium atoms that are produced during recombination.

The dipole induced in neutral hydrogen and helium scatters these long-wavelength photons with a cross-section that is proportional to the fourth power ($\propto\nu^{4}$) of their frequency at leading order. The approximate Rayleigh scattering cross-section for photons at frequency $\nu$ scattered by a hydrogen atom in the ground state is given by the Kramers–Heisenberg formula \citep{Lee:2005},

\begin{equation}\label{eq:rs_crosssection}
    \sigma_{\rm R}(\nu) = \sigma_{\rm T}\,\left(\frac{\nu}{\nu_{\rm eff}}\right)^4\left [  1 +  2.626 \left(\frac{\nu}{\nu_{\rm eff}}\right)^2 +  5.502 \left (\frac{\nu}{\nu_{\rm eff}} \right)^4+\ldots\right ],
\end{equation}
where~$\sigma_{\rm T}$ is the Thomson scattering cross-section and~$\nu_{\rm eff} = \sqrt{8/9}cR_{\infty}$ (where~$R_{\infty}$ is the Rydberg constant).
Therefore, the total comoving opacity of the photons at frequency~$\nu$ can be approximated as
\begin{equation}\label{eq:rs_opacity}
    \dot{\tau} (\nu) = \dot{\tau}_{\rm T} + \dot{\tau}_{\rm R}(\nu) = a\,n_{\rm e}\sigma_{\rm T} + a\,\sigma_{\rm R}(\nu) \Big[n_{\rm H} + R_{\rm He}n_{\rm He}\Big],
\end{equation}
where $\dot{\tau}_{\rm T}$ and $\dot{\tau}_{\rm R}(\nu)$ are the comoving opacities for Thomson and Rayleigh scattering respectively, $a$ is the scale factor, $n_{\rm e}$, $n_{\rm H}$ and $n_{\rm He}$ are the electron, neutral hydrogen and neutral helium number densities, respectively, and $R_{\rm He} \simeq 0.1$ is the relative strength of Rayleigh scattering of helium with respect to hydrogen \citep{Tarafdar:1967}.  Rayleigh scattering becomes increasingly important relative to Thomson scattering as $n_{\rm e}$  drops and $n_{\rm H}$ and $n_{\rm He}$ increase during recombination. Furthermore, once recombination is complete, the number density of neutral hydrogen evolves with redshift as $(1+z)^3$ and the CMB photon frequency as $(1+z)$, so that the Rayleigh scattering rate changes as $(1+z)^7$ at leading order. This makes the process efficient near the last scattering surface only. This additional scattering reduces the photon mean free paths, thus increasing the diffusion damping acting on small scales \citep{Yu:2001}.

\begin{figure}
   \includegraphics[width=0.5\textwidth]{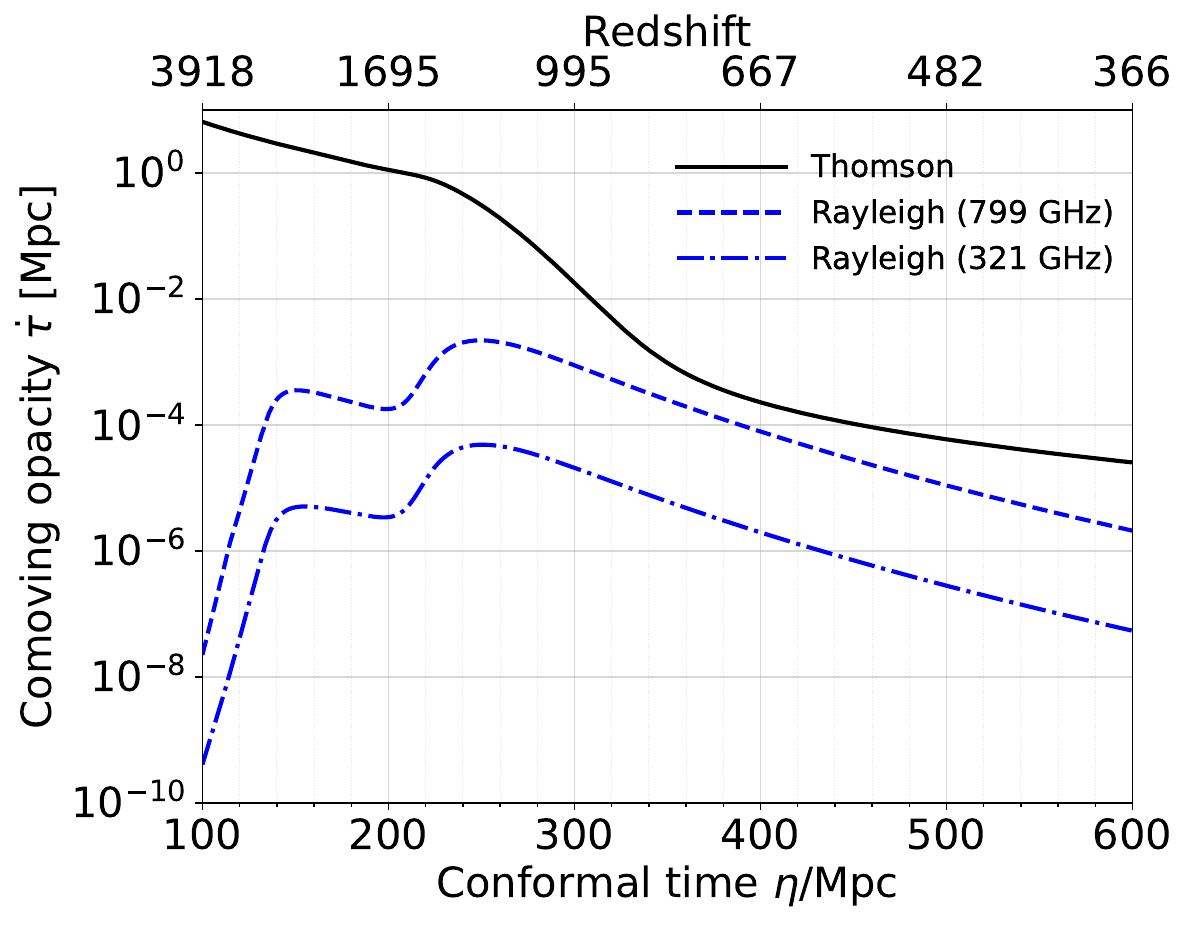}
    \includegraphics[width=0.5\textwidth]{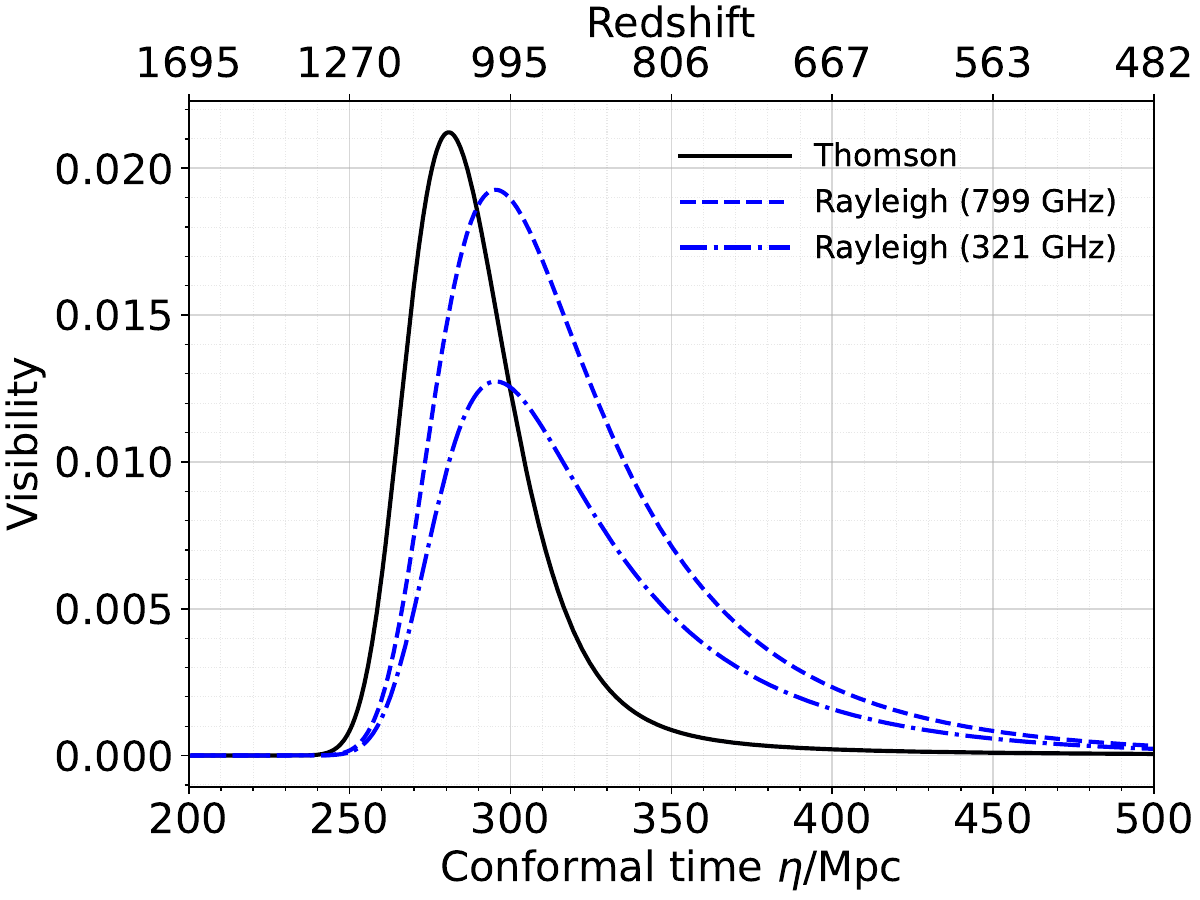}
    \caption{Left panel: Comoving opacity of Thomson scattering, $\dot{\tau}_{\rm T}$ (solid line), and of Rayleigh scattering, $\dot{\tau}_{\rm R}$, at 321\,GHz (dash-dotted line) and 799\,GHz (dashed line). At low frequencies, the RS contribution to the total opacity is negligible compared with Thomson scattering, whereas at high frequencies it becomes relevant towards the end of recombination, before decreasing rapidly as~$(1+z)^7$ once recombination is complete. Right panel: The corresponding normalised visibility functions. For clarity, the visibility function at 799\, GHz has been multiplied by a factor of 1.5.}
    \label{fig:visibility}
\end{figure}

The left panel of Figure~\ref{fig:visibility} shows the comoving Thomson scattering opacity (solid line), together with the frequency-dependent RS contribution to $\dot{\tau}$ in Eq.~\eqref{eq:rs_opacity} at two representative frequencies: 321\,GHz (dash-dotted line) and 799\,GHz (dashed line), for the fiducial $\Lambda$CDM cosmology. We have chosen these two frequencies since they correspond to PICO channels \citep{PICO:2019} and lie in the range where the RS signal is strongest. At low frequencies the RS comoving opacity is highly subdominant compared with Thomson scattering, but its steep $\nu^4$ dependence makes it increasingly important at higher frequencies. The right panel shows the corresponding visibility functions, $g_{\rm T}(\eta) = \dot{\tau}_{\rm T}(\eta)\,e^{-\tau_{\rm T}(\eta)}$, for Thomson scattering alone (solid line) and $g_{\rm R}(\nu,\eta) = \dot{\tau}_{\rm R}(\nu,\eta)\,e^{-\tau_{\rm R}(\nu,\eta)}$ for RS at 321\,GHz (dash-dotted line) and 799\,GHz (dashed line). Since RS acts mainly in the later stages of recombination, it shifts the peak of the visibility function towards lower redshift, an effect present at all frequencies but most pronounced at high frequencies. Different observing frequencies therefore probe slightly different stages of the evolution of the anisotropies during recombination. At these later times, the local quadrupole is larger, which boosts the $E$-mode polarisation on scales $10 \lesssim \ell \lesssim 300$ \citep{Takahara:1991, Yu:2001}, while the sound horizon is slightly larger, resulting in a shift of the acoustic peaks to lower multipoles in the higher-frequency channels.
This shift is potentially dependent on the ionisation history and the expansion rate of the Universe. By adding frequency-dependent information on the positions of the acoustic peaks, the RS signal can therefore help break parameter degeneracies, improving CMB constraints on the Hubble constant while also tightening constraints on the ionisation history.
With the inclusion of the RS effect, CMB harmonic coefficients at frequency $\nu$ can be perturbatively approximated as
\begin{equation}
    a_{\ell m}^X(\nu) \simeq a^X_{\ell m} + \left(\frac{\nu}{\nu_0}\right)^4 \Delta a_{\ell m}^{X_4} + \left(\frac{\nu}{\nu_0}\right)^6 \Delta a_{\ell m}^{X_6} + \left(\frac{\nu}{\nu_0}\right)^8 \Delta a_{\ell m}^{X_8},
\end{equation}
where $\nu_{0}$ is some reference frequency of RS and X$\in$ \{T, E\},  $a_{\ell m}$ are the harmonic coefficients of the primary CMB due to Thomson scattering and $\Delta a_{\ell m}^{X_n}$ are the harmonic coefficients of the RS contribution at the reference frequency $\nu_{0}$. Therefore, the cross-power spectra of the CMB computed between two channels can be written as
\begin{equation}\label{eq:cmb_power}
\begin{split}
C_{\ell}^{XY}(\nu_i,\nu_j) \simeq{}& C_{\ell}^{XY}
  + \sum_{n=4,6,8} \left(\frac{1}{\nu_0}\right)^n\left[\nu_j^{n} C_{\ell}^{X\Delta Y_n}
  + \nu_i^{n} C_{\ell}^{\Delta X_n Y}\right] 
+ \left(\frac{\nu_i \nu_j}{\nu_{0}^{2}}\right)^{4} C_{\ell}^{\Delta X_4 \Delta Y_4} + ...
\end{split}
\end{equation}
The first, second, and third terms represent the primary CMB power spectrum, the cross-spectrum between the primary CMB\footnote{Throughout this paper, we use `primary CMB' to refer to the CMB anisotropies (including lensing) generated by Thomson scattering alone, i.e.\ excluding the frequency-dependent contribution from Rayleigh scattering.} and the RS signal, and the cross-spectrum of the RS signal between two observing frequencies, respectively.

\subsection{Dependencies of Rayleigh scattering signal on varying fundamental constants}
\label{sec:Var_FC}

\begin{figure}
   
    \includegraphics[width=0.5\textwidth]{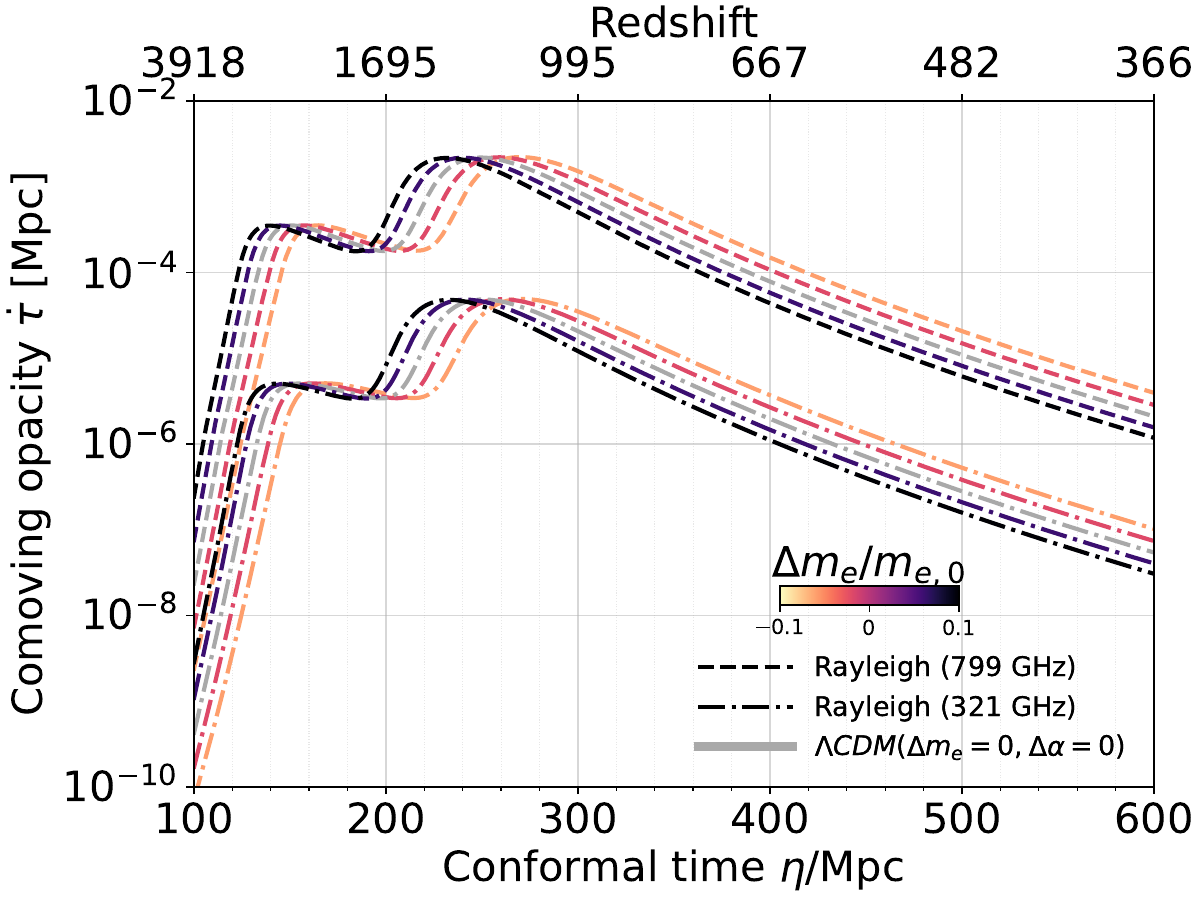}
     \includegraphics[width=0.5\textwidth]{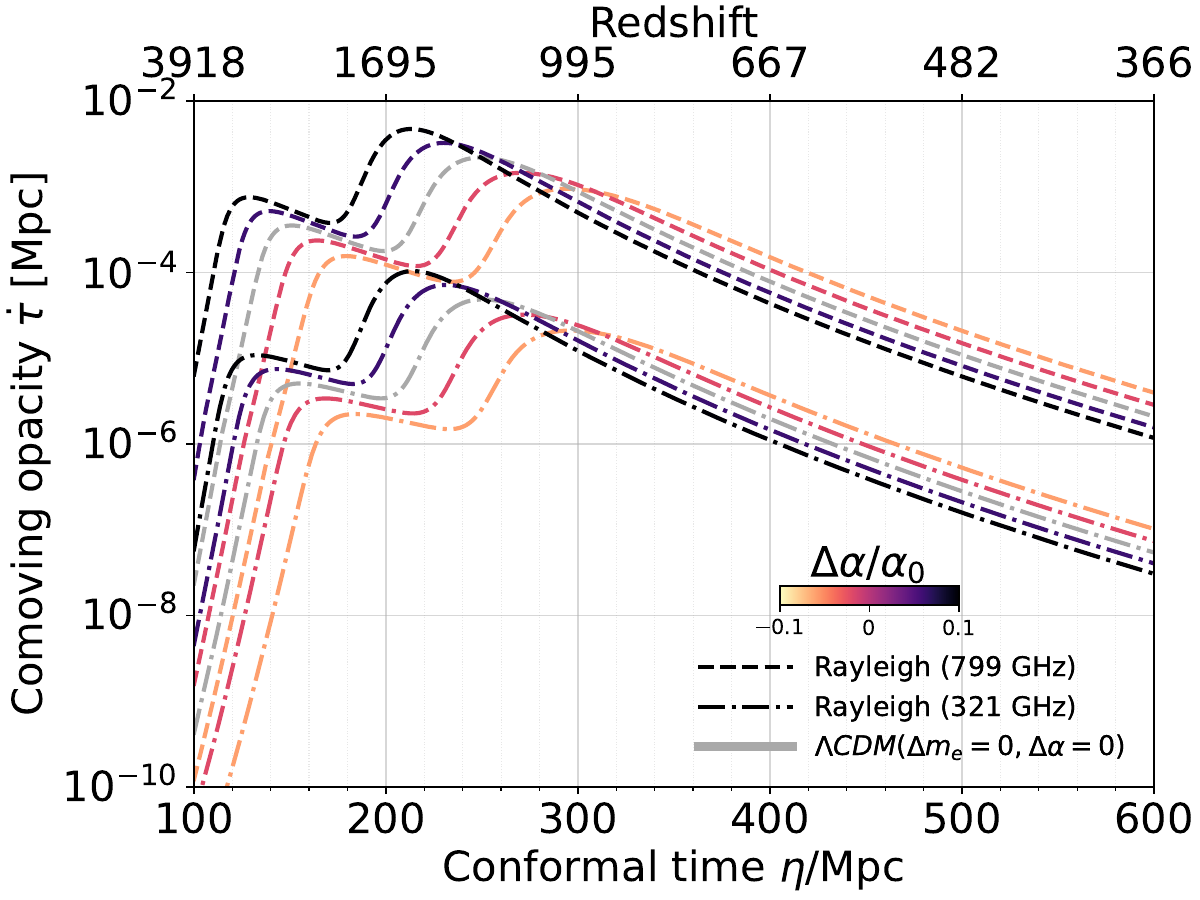}\\
    \caption{
    Effect of variations in $m_{\rm e}$ and $\alpha$ on the Rayleigh comoving opacity, $\dot{\tau}_\mathrm{R}(\nu)$, at two representative observing frequencies, 321\,GHz (dash-dotted lines) and 799\,GHz (dashed lines). Left panel: For different values of $\Delta m_{\rm e}/m_{\mathrm{e},0}$. Right panel: For different values of $\Delta\alpha/\alpha_0$. The corresponding curves for the $\Lambda$CDM scenario are highlighted in grey.}
    \label{fig:opacity_change_for_me_alpha}
\end{figure}
In this section, we illustrate the effect of varying fundamental constants focusing on the electron mass, $m_{\rm e}$ and the fine-structure constant, $\alpha$, on the recombination history and  CMB power spectra, with particular emphasis on the contribution from RS. We used the modified version of \recfastpp\ \citep{Chluba2010b,Chluba2010} to model the recombination history and propagate the resulting changes in the free-electron, $n_{\rm e}$, neutral hydrogen, $n_{\rm H}$ and neutral helium $n_{\rm He}$ number densities (cf. Eq.~\ref{eq:rs_opacity}) to a modified version of \camb\ \citep{CAMB} to compute the CMB power spectra contributed by Thomson and Rayleigh scattering across frequencies. 

Variations in $m_{\rm e}$ or $\alpha$ modify the recombination process through several distinct physical mechanisms. These include changes in atomic binding energies, atomic transition rates, photoionisation/recombination rates, and the radiative transfer of photons through modified Thomson, Compton, and resonance scattering \citep{Scoccola2009,Seager2000,Chluba2016CosmoSpec}. The resulting modifications in the recombination history in an effective three-level atom approach \citep{Zeldovich68,SeagerRecfast1999} and their impact on the primary CMB power spectra are discussed in detail by \cite{Hart:2017}. We consider deviations of $m_{\rm e}$ and $\alpha$ from their current values ($m_{\mathrm{e},0}=511$ keV and $\alpha\simeq 1/137$) during the recombination epoch, with both reaching their present values after recombination has completed. We parameterise these variations during the recombination era as, $m_{\rm e} = m_{\mathrm{e},0} (1 + \frac{\Delta m_{\rm e}}{m_{\mathrm{e},0}})$ and $\alpha = \alpha_{0} (1 + \frac{\Delta \alpha}{\alpha_{0}})$. 
The most important effect of these variations is the increase in the atomic binding energies of hydrogen and helium, $E_{i} \propto m_{\rm e}\alpha^2$, for $\frac{\Delta m_{\rm e}}{m_{\mathrm{e},0}}>0$ or $\frac{\Delta \alpha}{\alpha_{0}}>0$, which makes photoionisation less efficient at a given temperature. Since CMB temperature drops with the expansion of the Universe as $T_{\gamma} \propto (1+z)$, CMB temperature falls below the ionisation threshold at a higher redshift for $\frac{\Delta m_{\rm e}}{m_{\mathrm{e},0}}>0$ or $\frac{\Delta \alpha}{\alpha_{0}}>0$. Consequently, recombination occurs earlier, leading to a larger abundance of neutral hydrogen and helium at higher redshifts compared to the standard recombination history. 

The earlier production of neutral atoms initially enhances the Rayleigh scattering opacity at high redshifts. In addition, variations in $m_{\rm e}$ and $\alpha$ also modify the RS opacity through changes in the Rayleigh scattering cross-section, $\sigma_{\rm R}(\nu)$. The Rayleigh scattering cross-section, to leading order in its frequency dependence, is given by 
\bealf{
\label{eq:test}
\sigma_{\rm R}(\nu)
&\simeq \sigma_{\rm T} \,\left(\frac{\nu}{\nu_{\rm eff}}\right)^4\propto \nu^4\alpha^{-6} \me^{-6},}
where $\sigma_{\rm T}\propto \alpha^2m_{\rm e}^{-2}$ is the Thomson scattering cross-section and the effective Rayleigh resonance frequency scales as $\nu_{\rm eff}\propto \alpha^2m_{\rm e}$. Consequently, increasing $m_{\rm e}$ or $\alpha$ decreases the Rayleigh scattering cross-section, which leads to a reduction in the Rayleigh scattering opacity with respect to the $\Lambda$CDM case once recombination is essentially completed. Figure~\ref{fig:opacity_change_for_me_alpha} shows the evolution of the Rayleigh comoving opacity, $\dot{\tau}_{\rm R}(\nu)$ (cf. Eq.~\ref{eq:rs_opacity}) for different values of $\Delta m_{\rm e}/m_{\mathrm{e},0} \in \{-0.1,0.1\}$ (left panel) and $\Delta \alpha/\alpha_{0} \in \{-0.1,0.1\}$ (right panel) at two representative frequencies, 799 GHz (dashed curves) and 321 GHz (dashed-dotted curves). The corresponding curve for the $\Lambda$CDM model is shown in grey. Increasing $m_{\rm e}$ or $\alpha$ shifts the onset of Rayleigh scattering opacity to higher redshifts compared to the $\Lambda$CDM scenario, due to earlier recombination. The effect is exactly the opposite for decreasing values of $m_{\rm e}$ and $\alpha$. The suppression of $\sigma_{\rm R} \propto\alpha^{-6}m_{\rm e}^{-6}$, which becomes apparent only after recombination is complete, produces a steeper decline in comoving opacity than in $\Lambda$CDM. 
As expected, the effect is slightly greater for changes in $\alpha$ than for $m_{\rm e}$, due to their different impact on the enhancement of energy transition levels and other intermediate processes of recombination as described in \citep{Scoccola2009,Hart:2017}. At lower frequencies, $\dot{\tau}_{\rm R}$ responds in the same way to variations of $m_{\rm e}$ and $\alpha$, but its amplitude is substantially smaller since it decreases according to the scaling of the Rayleigh scattering cross-section $\sigma_{\rm R}(\nu) \propto \nu^4$.

\subsection{Effect of varying fundamental constants on visibility function and CMB power spectra}
\label{sec:cmb_spectra_for_me_alpha}

\begin{figure}
     \includegraphics[width=0.5\textwidth]{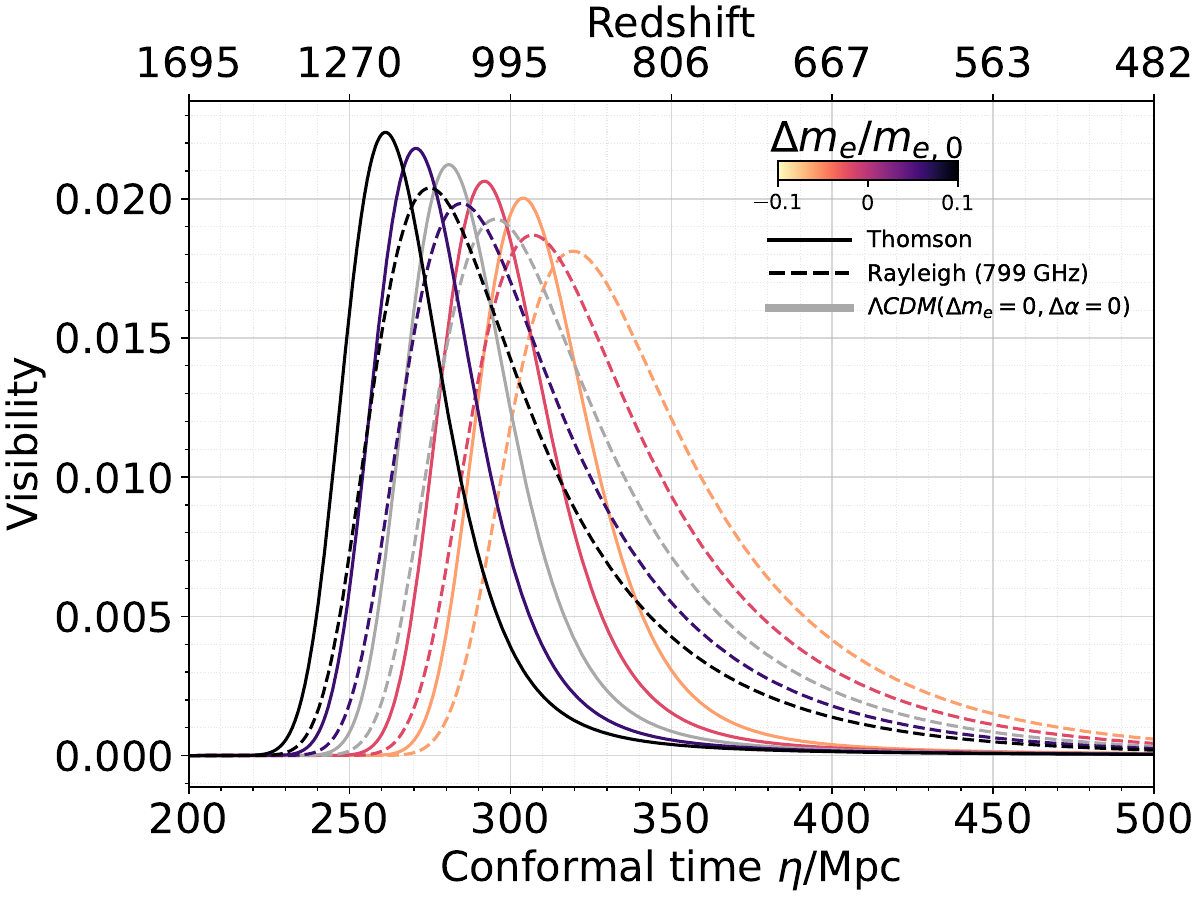}
     \includegraphics[width=0.5\textwidth]{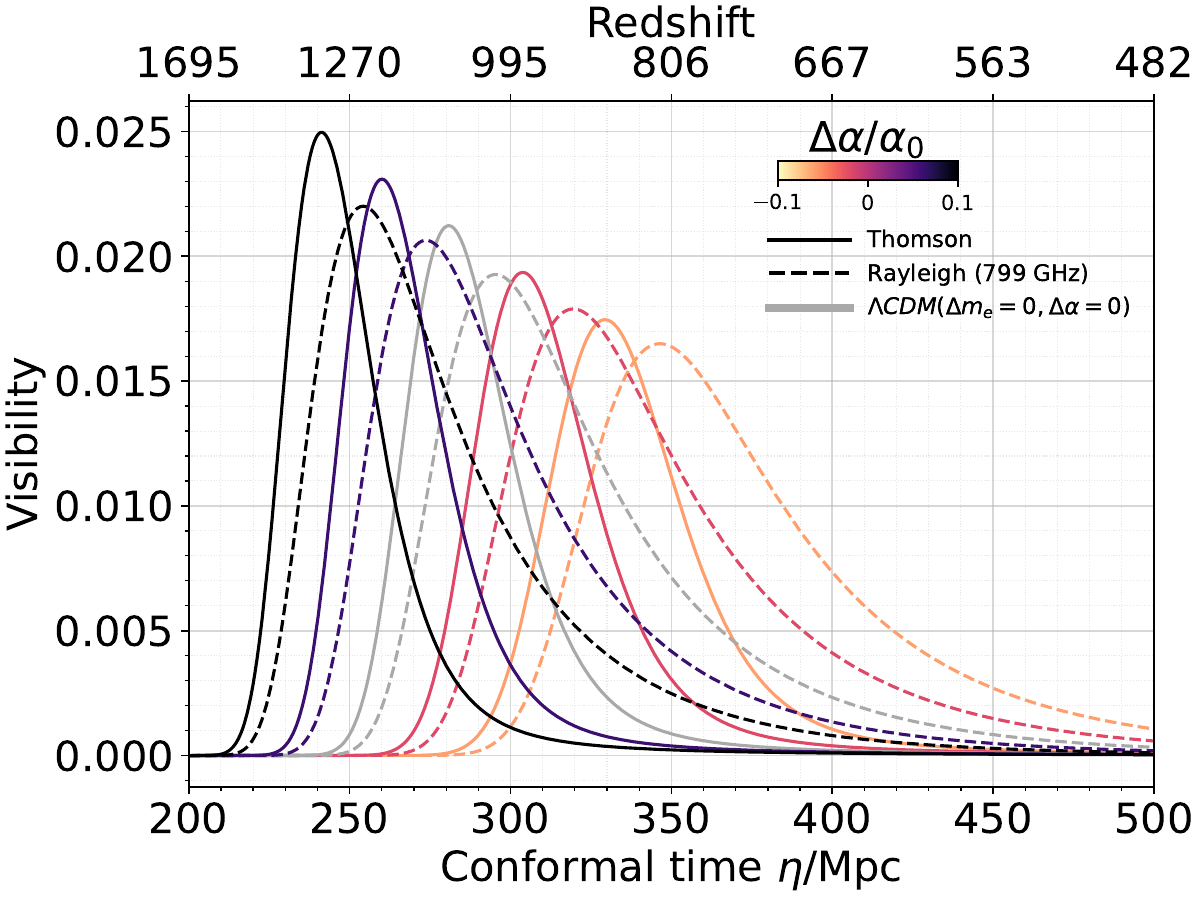}
    \caption{
    Effect of variations in $m_{\rm e}$ (left panel) and $\alpha$ (right panel), on the Thomson (solid lines) and Rayleigh (dashed lines) visibilities (normalised).  For clarity,  Rayleigh visibilities at 799 GHz are multiplied by a factor of 1.5. The corresponding curves for the $\Lambda$CDM scenario are highlighted in grey.
    }
    \label{fig:visibility_change_for_me_alpha}
\end{figure}
We investigate the changes in the RS visibility function,  $g_{\rm R}(\eta,\nu) = \dot{\tau}_{\rm R}(\eta,\nu)e^{-\tau_{\rm R}(\eta,\nu)}$ and the Thomson visibility function, $g_{\rm T}(\eta) = \dot{\tau}_{\rm T}(\eta)e^{-\tau_{\rm T}(\eta)}$ for different recombination histories, as these are governed by the dynamics of the recombination process and ultimately determine the Thomson and Rayleigh contributions to the CMB temperature and polarisation power spectra. We computed visibilities using a modified version of \camb\ and in  Figure~\ref{fig:visibility_change_for_me_alpha} we present Thomson (solid lines) and Rayleigh (dashed lines) visibilities (normalised) at the observing frequency of 799 GHz for different values of $\Delta m_{\rm e}/m_{\mathrm{e},0} \in \{-0.1,0.1\}$ (left panel) and $\Delta\alpha/\alpha_0 \in \{-0.1,0.1\}$ (right panel). The visibilities corresponding to the $\Lambda$CDM model are coloured grey. For $\Delta m_{\rm e}/m_{\mathrm{e},0}>0$ and $\Delta \alpha/\alpha_{0}>0$, the visibility peak shifts towards higher redshifts compared to the $\Lambda$CDM case, for Thomson and Rayleigh scattering. The peak shifts in the opposite direction for $\Delta m_{\rm e}/m_{\mathrm{e},0}<0$ and $\Delta \alpha/\alpha_{0}<0$. Moreover, lowering $m_{\rm e}$ or $\alpha$ broadens the Rayleigh visibility function more than the Thomson one, since the Rayleigh opacity is more sensitive to these variations.

\begin{figure}
    \centering
    \includegraphics[width=\textwidth]{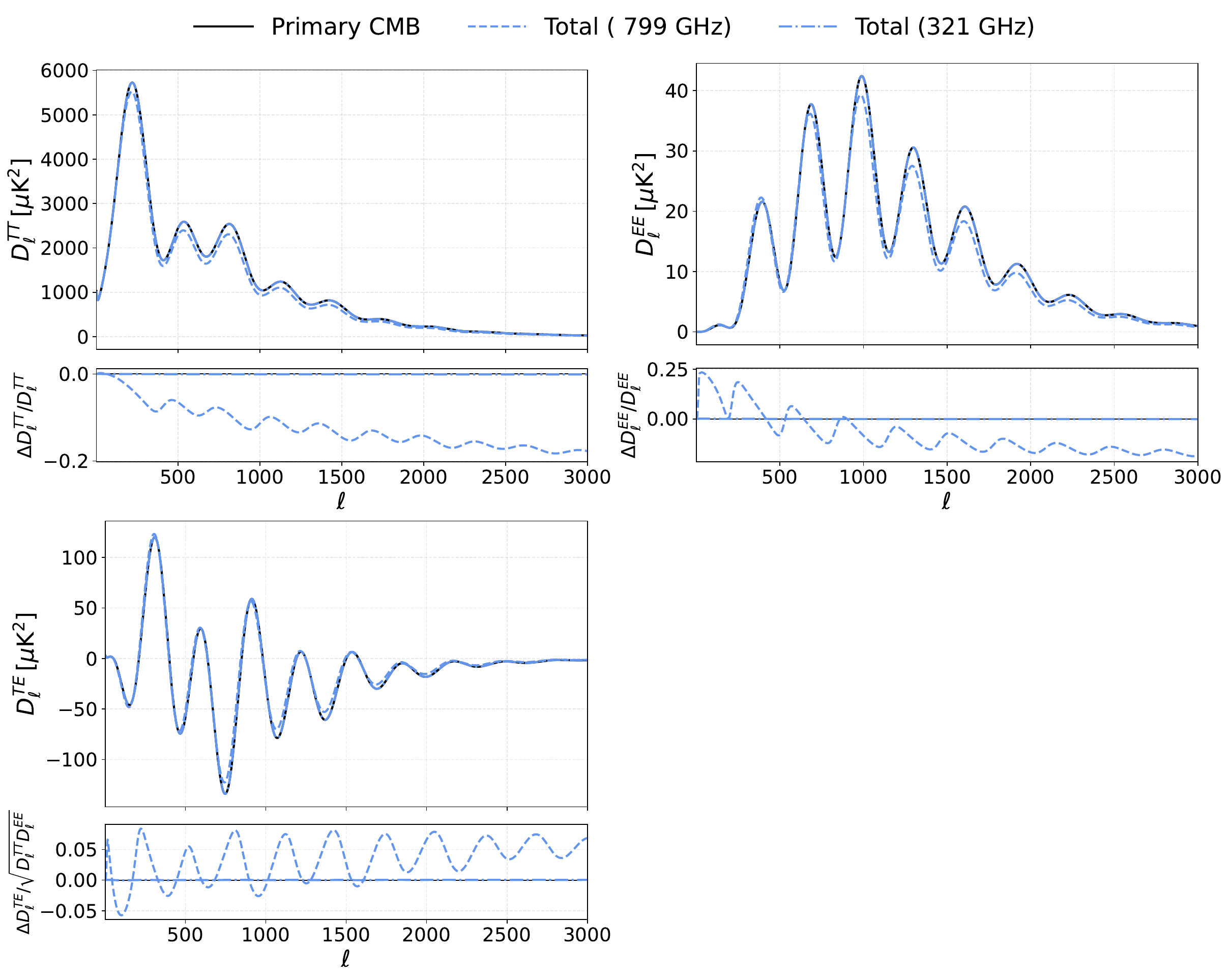}
   
    \caption{Total power spectra including both Thomson and RS effects at two observing frequencies: 799 GHz (dashed line) and 321 GHz (dash-dotted line). The grey solid lines correspond to primary CMB power spectra due to Thomson scattering alone. The $TT$, $EE$ and $TE$ spectra are shown in the upper left, upper right and lower left panels, respectively. All cases shown here are for the $\Lambda$CDM model. The lower panel of each plot shows the relative difference between the total power spectrum at each frequency and the primary CMB-only power spectrum. }
    \label{fig:diff_power_spectra_lcdm}
\end{figure}

We then propagate the effects of varying fundamental constants through a modified version of \texttt{CAMB} and calculate the auto- and cross-frequency CMB power spectra across the observing frequencies. We first investigate the power spectra for the $\Lambda$CDM scenario. In Figure~\ref{fig:diff_power_spectra_lcdm}, we illustrate the total $TT$, $EE$, and $TE$ power spectra contributed by both Thomson and Rayleigh scattering at two representative frequencies, 321~GHz (dashed-dotted lines) and 799~GHz (dashed lines) and compare them with the primary CMB power spectra arising from Thomson scattering alone (solid grey lines). The lower panels show the relative difference between the total power spectra at each frequency and the primary CMB power spectra. At the lower frequency of 321~GHz, the total power spectra remain close to the primary CMB power spectra, since Rayleigh scattering is relatively inefficient at low frequencies. However, at higher frequencies, such as 799~GHz, the effects of Rayleigh scattering become more significant. In particular, the acoustic peaks of the total power spectra are shifted towards lower values of $\ell$ (larger scales), reflecting the shift of the visibility peak towards lower redshifts, which causes an increase in the sound horizon and a decrease in the angular diameter distance to recombination at this frequency \citep{Beringue:2021}. In addition, the amplitude of the anisotropies is suppressed since increased Rayleigh scattering leads to further mixing and redistribution of photons. 
We further investigate the frequency dependence of this effect by examining the angular power spectra at other observing frequencies and find that the most pronounced effect of Rayleigh scattering occurs in the $EE$ power spectrum. Nevertheless, for observing frequencies below $500$~GHz, the relative modification remains small, with $\left|\frac{\Delta D_\ell}{D_\ell}\right| \lesssim 1\%$.

\begin{figure}
    \centering
    \includegraphics[width=\textwidth]{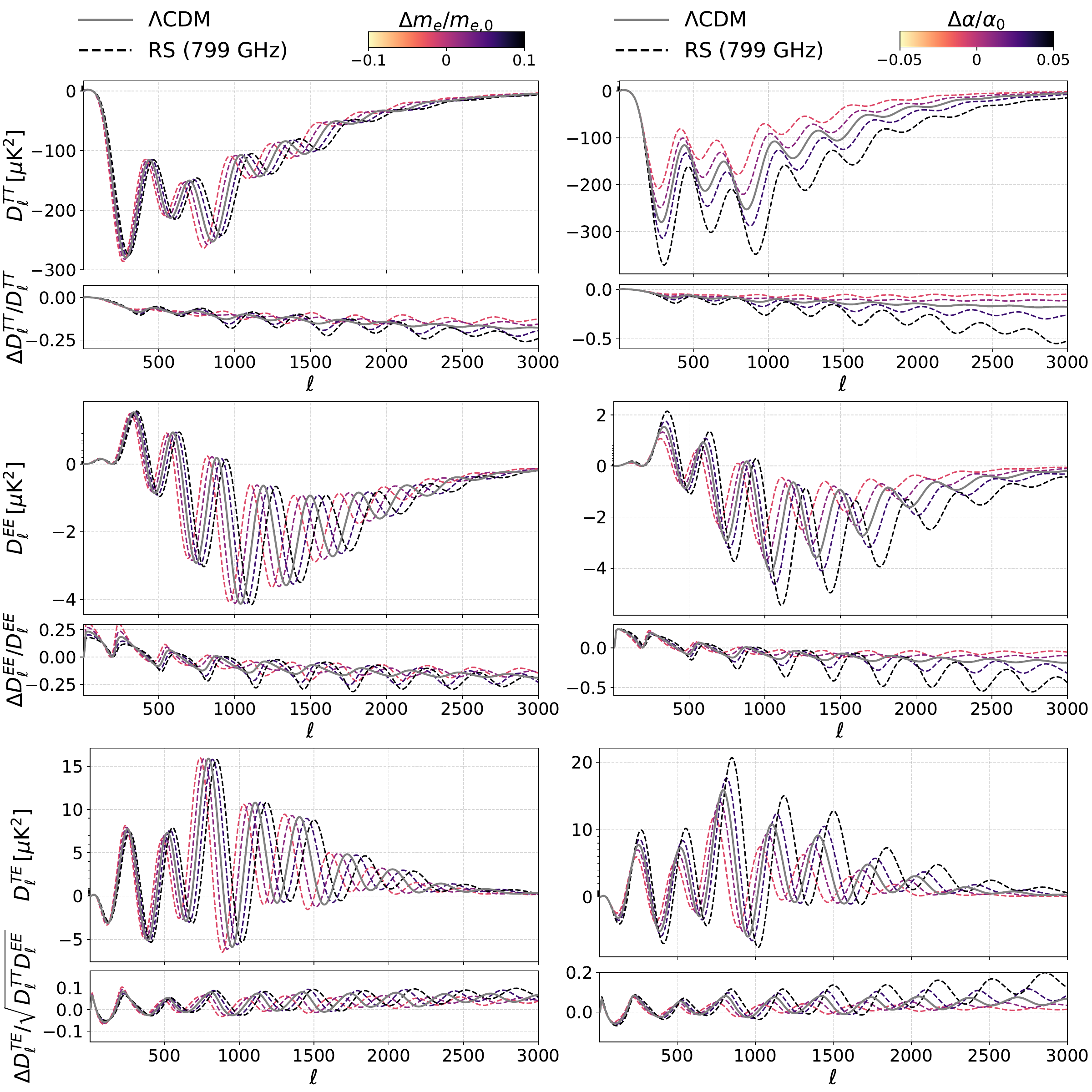}
   
    \caption{ Effect of varying the electron mass (left column) and $\alpha$ (right column) on the frequency-dependent part of the CMB power spectra (cf. Eq~\ref{eq:cmb_power}) i.e., the combined Rayleigh-primary cross-correlation and Rayleigh-Rayleigh auto-correlation contributions at observing frequency, 799 GHz. The upper, middle, and lower rows represent the $TT$, $EE$, and $TE$ power spectra, respectively. In each panel, the grey curves denote the power spectra for the $\Lambda$CDM scenario. The lower panels in each plot present the Rayleigh scattering contribution with respect to the corresponding primary CMB power spectra for the $\Lambda$CDM case as a reference, denoted by $\Delta D_{\ell}/D_{\ell}$ for different values of $\Delta m_{\rm e}/m_{\mathrm{e},0}\in \{-0.1,0.1\}$ and $\Delta \alpha/\alpha_{0}\in \{-0.05,0.05\}$. For the $TE$ power spectrum, the relative change is normalised as $\Delta D_{\ell}^{TE}/\sqrt{D_{\ell}^{TT}D_{\ell}^{EE}}.$ }
    \label{fig:diff_power_spectra_varying_me_and_alpha}
\end{figure}

We then investigated the effects of varying fundamental constants on the Rayleigh-scattering contribution to the total CMB power spectra. In Figure~\ref{fig:diff_power_spectra_varying_me_and_alpha}, we illustrate the frequency-dependent contributions to the $TT$, $EE$, and $TE$ power spectra, as defined in Eq.~\ref{eq:cmb_power}, for $\Delta m_{\rm e}/m_{\mathrm{e},0} \in \{-0.1,\,0.1\}$ (left column) and $\Delta\alpha/\alpha_0 \in \{-0.05,\,0.05\}$ (right column) at a representative frequency channel of 799 GHz. These spectra are compared with the corresponding $\Lambda$CDM case, shown by the solid grey curves. The lower panel of each subplot shows the relative Rayleigh-scattering contributions with respect to the primary CMB power spectra in the fiducial $\Lambda$CDM scenario for the different cases. We find that increasing $m_{\rm e}$ or $\alpha$ shifts the acoustic peaks towards higher $\ell$ values (smaller scales). This behaviour reflects the shift of the effective last-scattering surface towards higher redshifts \citep{Kaplinghat1999,Battye2001}. We also find that the shift of the acoustic peaks due to changes in $m_{\rm e}$ or $\alpha$ remains largely independent of the observing frequency. The amplitudes of the acoustic peaks are also boosted when the fundamental constants are increased. For example, for $\Delta \alpha/\alpha_0 = 0.02$, we find that the relative change in the amplitude of acoustic peaks with respect to the $\Lambda$CDM scenario is $\lesssim$ 12\%.




\section{Methodology}\label{sec:methodology}
\subsection{Fisher Formalism}\label{sec:fromalism}
We are interested in quantifying the improvement in the constraints on the parameters of the fiducial $\Lambda$CDM cosmological model and several of its extensions due to the inclusion of the RS signal in addition to the primary CMB for future CMB experiments. To do so, we use a standard Fisher matrix approach \citep{Tegmark:1997ApJ, Carron.J:2013}. For a set of zero-mean Gaussian fields (in our case CMB spherical harmonic coefficients, $a_{\ell m}^{X}$ for $X=T,E$) having parameter-dependent covariance matrix, $\textbf{C} =\rm  diag\{\textbf{C}_{\ell}({\theta})\}$, the Fisher matrix is defined as the expectation value of the second derivative of the log-likelihood, $\rm  \mathcal{L} \propto \frac{1}{\sqrt{\textbf{C}}}exp{(-\frac{1}{2}{\textbf{a}^{\dag}\textbf{C}^{-1}\textbf{a}})} $ with respect to the model parameters of interest (denoted here as $\theta_i$), evaluated at their fiducial values $\theta_{i,\rm fid}$,
\begin{equation}\label{eq:fish}
     \mathrm{F}_{ij} = -\Biggl \langle \frac{\partial^2 \ln \mathcal{L}}{\partial\theta_i\partial\theta_j} \biggr  \rangle\Bigg|_{\theta_{\rm fid}}= \sum_{\ell}\frac{(2\ell + 1)}{2}f_{\rm sky}\mathrm{ Tr}\Bigg[\textbf{C}_{\ell}^{-1}\frac{\partial \textbf{C}_{\ell}}{\partial \theta_i}\textbf{C}_{\ell}^{-1}\frac{\partial \textbf{C}_{\ell}}{\partial \theta_j}\Bigg].
\end{equation}
The $\textbf{C}_{\ell}$ matrix for CMB data can be written as,

\begin{equation}\label{eq:cl_matrix}
    \textbf{C}_{\ell} =
\begin{pmatrix}
\textbf{C}_{\ell}^{TT} & \textbf{C}_{\ell}^{TE} \\
\textbf{C}_{\ell}^{TE} & \textbf{C}_{\ell}^{EE}
\end{pmatrix},
\end{equation}
where $\textbf{C}_{\ell}^{XY}$ are $N_{\nu} \times N_{\nu}$ matrices for $N_{\nu}$ frequency channels and each element of $\textbf{C}_{\ell}^{XY}$ incorporates cross-frequency angular power spectra of CMB, foregrounds and noise. The noise power spectrum at frequency $\nu$ is modelled as 
\begin{equation}
    N_\ell^{XY,\nu\nu^{'}} = \Delta_X^2 \exp\left[\ell(\ell+1)\Bigg(\frac{\theta_{\mathrm{FWHM}}}{\sqrt{8\ln 2}}\Bigg)^2\right] \delta_{\nu,\nu^{'}}\delta_{X,Y},
\end{equation}
where $\Delta_X$ (in $\mu\mathrm{K}\cdot\mathrm{rad}$) is the noise sensitivity for X$\in \{T,E\}$ and $\theta_{\mathrm{FWHM}}$ (in radians) is the beam FWHM at the respective frequency. In this analysis, we adopt the instrument configuration for PICO (see Table 3.2 of \citep{PICO:2019}), LiteBIRD and Planck (see Table 13 of \citep{LiteBIRD:2022cnt} and Table 4 of \citep{Planck-I:2020}). We use only the frequency channels at $\nu \gtrsim 90$\,GHz, since the RS signal is
negligible at lower frequencies and since this choice avoids the need to model some of the low-frequency foregrounds, such as 
free-free and anomalous microwave emission, which dominate frequencies below 90 GHz. The modelling of foreground power spectra is discussed in Section~\ref{sec:fg_model}.
Once the Fisher matrix is computed, we obtain the lower limit of the parameter covariance matrix,
\begin{equation}
    \mathrm{M}_{ij} = (\mathrm{F}^{-1})_{ij},
\end{equation}
and the marginalised 1$\sigma$ uncertainties of the parameters are $\sigma_{\theta_{i}} = \sqrt{\mathrm{M}_{ii}}$.

Moreover, we aim to quantify the bias in the parameter estimation if we incorrectly model the CMB power spectra, for instance ignore modelling the RS contribution in the covariance matrix $\textbf{C}_{\ell}$. In that case, the spherical harmonic coefficients of the data are drawn from $\textbf{a}_{\ell m} \sim \mathcal{N} (\textbf{0}, \textbf{C}_{\ell}^{\rm true})$ where $\textbf{C}_{\ell}^{\rm true}$ is the true data covariance matrix.  However, the log-likelihood, $\rm ln \mathcal{L} $ assumes the model covariance $\textbf{C}_{\ell}$, which differs from  $\textbf{C}_{\ell}^{\rm true}$.  Therefore, $-\rm ln \mathcal{L} $ is minimised with respect to the drifted parameter values, $\hat{\theta}_{i} = \theta_{i,\rm fid} + \delta \theta_{i}$ with associated bias $\delta \theta_{i}$ for respective parameters.  Minimising the first derivative of the log-likelihood, we can obtain the bias in each parameter as
\begin{equation}\label{eq:bias}
    \delta \theta_{i} = \sum_{j}  (\mathrm{F}^{-1})_{ij}\sum_{\ell}\frac{(2\ell + 1)}{2}f_{\rm sky}\mathrm{ Tr}\Bigg[\textbf{C}_{\ell}^{-1}\frac{\partial \textbf{C}_{\ell}}{\partial \theta_{j}}\textbf{C}_{\ell}^{-1} \Delta \textbf{C}_{\ell}\Bigg],
\end{equation}
where $\Delta\textbf{C}_{\ell} =   \textbf{C}_{\ell}^{\rm true} - \textbf{C}_{\ell} (\theta_{\rm fid})$ is the difference in the covariance matrix due to the mismatch between the model and the data. 
A detailed derivation of Eq.~\ref{eq:fish} and Eq.~\ref{eq:bias} is shown in Appendix~\ref{sec:derivations}.

\subsection{Foreground modelling}\label{sec:fg_model}
We modelled the foreground contamination following the baseline model of the ACT DR6 analysis~\cite{Louis:2025JCAP,Beringue:2025JCAP}. In temperature, we include the kinematic and thermal Sunyaev--Zel'dovich (kSZ, tSZ) effects, the Poisson and clustered components of the cosmic infrared background (CIB), the tSZ$\times$CIB correlation, unresolved radio sources, Galactic dust and Galactic synchrotron emission. However, since we use only channels $\gtrsim 90$ GHz in our analysis, we do not include a very detailed modelling of synchrotron (only for $TE$), free-free and spinning dust emissions. The foreground power spectra are computed with the \fgspectra\footnote{\url{https://github.com/simonsobs/fgspectra/tree/v1.3.0}}  code, assuming delta-function passbands at the channel frequencies. All components are normalised at $\ell_0=3000$ and $\nu_0=150\,$GHz except synchrotron. The synchrotron is normalised at $\ell_0=80$ and $\nu_0=30\,$GHz. The functional forms and the values of the fiducial parameters are summarised in Table~\ref{table:fg_modelling}. The foreground spectra are added to the CMB and RS spectra in the covariance matrix of Eq.~\ref{eq:cl_matrix}.

\begin{table*}[h]
\centering
\footnotesize
\renewcommand{\arraystretch}{1.6}
\begin{tabular}{l c p{5.4cm} p{3.3cm} l}
\hline\hline
Component & Spectra & Model, $D_\ell^{\nu_i\nu_j}$ [$\mu{\rm K}^2$] & Fiducial parameters & References \\
\hline
kSZ & $TT$ &
$a_{\rm kSZ}\,D^{\rm kSZ}_{\ell,\ell_0}$ &
$a_{\rm kSZ}=1.48$ &
\cite{Battaglia_2013,Louis:2025JCAP} \\
tSZ & $TT$ &
$a_{\rm tSZ}\,D^{\rm tSZ}_{\ell,\ell_0}\left(\frac{\ell}{\ell_0}\right)^{\alpha_{\rm tSZ}}\frac{f(\nu_i)f(\nu_j)}{f^2(\nu_0)}$ &
$a_{\rm tSZ}=3.3$, $\alpha_{\rm tSZ}=-0.53$ &
\cite{Battaglia2012,Louis:2025JCAP} \\
CIB, Poisson & $TT$ &
$a_{\rm p}\,\frac{\ell(\ell+1)}{\ell_0(\ell_0+1)}\,\frac{\mu(\nu_i;\beta_{\rm p},T_{\rm d})\,\mu(\nu_j;\beta_{\rm p},T_{\rm d})}{\mu^2(\nu_0;\beta_{\rm p},T_{\rm d})}$ &
$a_{\rm p}=7.65$, $\beta_{\rm p}=1.87$, $T_{\rm d}=9.6\,$K &
\cite{Louis:2025JCAP, Beringue:2025JCAP} \\
CIB, clustered & $TT$ &
$a_{\rm c}\,D^{{\rm CIB}\mbox{-}{\rm c}}_{\ell,\ell_0}\,\frac{\mu(\nu_i;\beta_{\rm c},T_{\rm d})\,\mu(\nu_j;\beta_{\rm c},T_{\rm d})}{\mu^2(\nu_0;\beta_{\rm c},T_{\rm d})}$ &
$a_{\rm c}=3.69$, $\beta_{\rm c}=1.87$ &
\cite{Planck-XXX:2014,Choi:2020JCAP} \\
tSZ$\times$CIB & $TT$ &
$-\xi\sqrt{a_{\rm tSZ}a_{\rm c}}\,D^{\rm tSZ\times CIB}_{\ell,\ell_0}\,\frac{f(\nu_i)\mu(\nu_j;\beta_{\rm c},T_{\rm d})+f(\nu_j)\mu(\nu_i;\beta_{\rm c},T_{\rm d})}{f(\nu_0)\,\mu(\nu_0;\beta_{\rm c},T_{\rm d})}$ &
$\xi=0.09$ &
\cite{Addison:2012MNRAS,Louis:2025JCAP} \\
Radio sources & $TT$ &
$a_{\rm s}\,\frac{\ell(\ell+1)}{\ell_0(\ell_0+1)}\left(\frac{\nu_i\nu_j}{\nu_0^2}\right)^{\beta_{\rm s}}\frac{g(\nu_i)g(\nu_j)}{g^2(\nu_0)}$ &
$a_{\rm s}=2.86$, $\beta_{\rm s}=-2.5$ &
\cite{Dunkley:2011ApJ,Louis:2025JCAP,Beringue:2025JCAP} \\
Galactic dust & $TT$, $TE$, $EE$ &
$a_{\rm g}^{XY}\left(\frac{\ell}{500}\right)^{\alpha_{\rm d}^{XY}}\frac{\mu(\nu_i;\beta_{\rm d},T^{\rm eff}_{\rm d})\,\mu(\nu_j;\beta_{\rm d},T^{\rm eff}_{\rm d})}{\mu^2(\nu_0;\beta_{\rm d},T^{\rm eff}_{\rm d})}$ &
$a_{\rm g}^{TT}=7.9$, $a_{\rm g}^{TE}=0.42$, $a_{\rm g}^{EE}=0.167$, $\alpha_{\rm d}^{TT}=-0.6$, $\alpha_{\rm d}^{TE,EE}=-0.4$, $\beta_{\rm d}=1.5$, $T^{\rm eff}_{\rm d}=19.6\,$K &
\cite{Planck-xi:2020,Louis:2025JCAP} \\
Synchrotron & $TT$, $EE$ &
$a_{\rm sync}^{XX}\left(\frac{\ell}{80}\right)^{\alpha_{\rm sync}}\left(\frac{\nu_i\nu_j}{\nu_{\rm 0}^2}\right)^{\beta_{\rm sync}}\frac{g(\nu_i)g(\nu_j)}{g^2(\nu_{\rm 0})}$, $\nu_{\rm 0}=30\,$GHz &
$a_{\rm sync}^{TT}=9.8$, $a_{\rm sync}^{EE}=2.37$, $\alpha_{\rm sync}=-0.81$, $\beta_{\rm sync}=-3.1$ &
\cite{Planck-IV:2020} 
\\
\hline\hline
\end{tabular}
\caption{Foreground components included in the model, written as $D^{\nu_i,\nu_j}_\ell=\ell(\ell+1)C^{\nu_i,\nu_j}_\ell/2\pi$ in CMB thermodynamic units for channels $i,j$. The pivot scale and pivot frequency are $\ell_0=3000$ and $\nu_0=150\,$GHz for all components except Synchrotron, which is normalised at $\ell_0=80$ and $\nu_0=30\,$GHz. Where $D^{\rm X}_{\ell,\ell_0}$ denotes the template power spectrum of component X, normalised at the pivot multipole $\ell_0$. Here $g(\nu)$ is the conversion factor from Rayleigh--Jeans units to CMB thermodynamic temperature unit, $\mu(\nu;\beta,T)\equiv\nu^{\beta+1}g(\nu)/(e^{h\nu/k_{\rm B}T}-1)$ is a modified blackbody in CMB thermodynamic units, and $f(\nu)=x\coth(x/2)-4$ with $x=h\nu/k_{\rm B}T_{\rm CMB}$ is the tSZ spectral function. Fiducial values of the extragalactic and Galactic-dust parameters are taken from the given references. }
\label{table:fg_modelling}
\end{table*}

\section{Results}\label{sec:results}
In this section we present the forecasted constraints on the cosmological parameters obtained with and without the inclusion of the RS signal, and quantify the parameter biases that arise when RS is neglected or when the recombination history is mis-modelled. We present the results for PICO \citep{PICO:2019} noise levels in the main text. The corresponding results for LiteBIRD + Planck \citep{LiteBIRD:2022cnt, Planck-I:2020} are presented in Appendix~\ref{app:B}. The results for both LiteBIRD+Planck and PICO noise levels are obtained with $f_{\rm sky}=0.7$, and with the foreground model of Section~\ref{sec:fg_model}, using the $TT$, $TE$ and $EE$ spectra of all frequency channels and their cross-spectra. For PICO we set $\ell_{\rm max} = 3000$. We use  $\ell_{\rm max} = 200$ for LiteBIRD and for Planck we use $\ell_{\rm max} = 2500$ and $\ell_{\rm max} = 2000$ for temperature and polarisation, respectively.

We consider three cosmological models. Our baseline model is $\Lambda$CDM with six parameters, $\{\Omega_{\rm b}h^2, \Omega_{\rm c}h^2, $ $\ln(10^{10}A_{\rm s}), H_0, n_{\rm s}, \tau\}$ and fiducial values are set to the best-fit values of the Planck 2018 results \citep{planck-VI:2020}. Besides the $\Lambda$CDM model, we study two extensions of the $\Lambda$CDM model where we vary either $m_{\rm e}/m_{\mathrm{e},0}$ or $\alpha/\alpha_0$ in two ways. In the model labelled $\Lambda{\rm CDM}+x_{\rm e}$, the variation of the fundamental constant enters only through the modified recombination history $x_{\rm e}(z)$ computed with \recfastpp\ \citep{Chluba2010, Chluba2015PMF}, while the Thomson and Rayleigh cross-sections are held at their standard values. In the model labelled $\Lambda{\rm CDM}+x_{\rm e}+\sigma_{\rm T}+\sigma_{\rm R}$, the variation is propagated consistently to the scattering cross-sections as well, i.e.\ $\sigma_{\rm T}\propto\alpha^2 m_{\rm e}^{-2}$ and $\sigma_{\rm R}\propto \nu^4\alpha^{-6}m_{\rm e}^{-6}$ (cf. Eq.~\ref{eq:test}). The fiducial values are $ m_{\rm e}/m_{ \mathrm{e},0}=\alpha/\alpha_0=1$ for the forecasts of parameter constraints in Section~\ref{sec:res_constraints}, while the bias analysis of Section~\ref{sec:res_bias} assumes $m_{\rm e}/m_{\mathrm{e},0}=1.02$ (or $\alpha/\alpha_0=1.01$) in the simulated data.

\subsection{Forecast constraints with and without Rayleigh scattering}
\label{sec:res_constraints}
\begin{figure}
    \includegraphics[width=.5\linewidth]{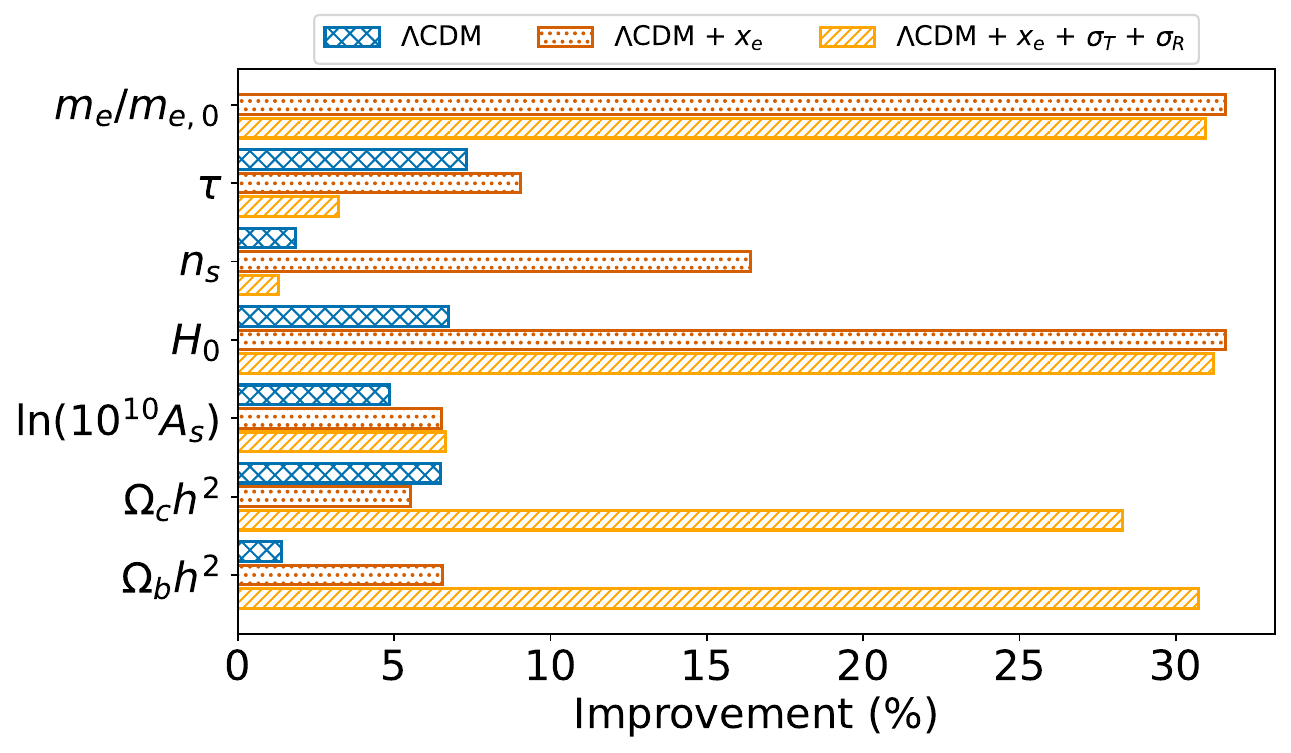}
    \includegraphics[width=.5\linewidth]{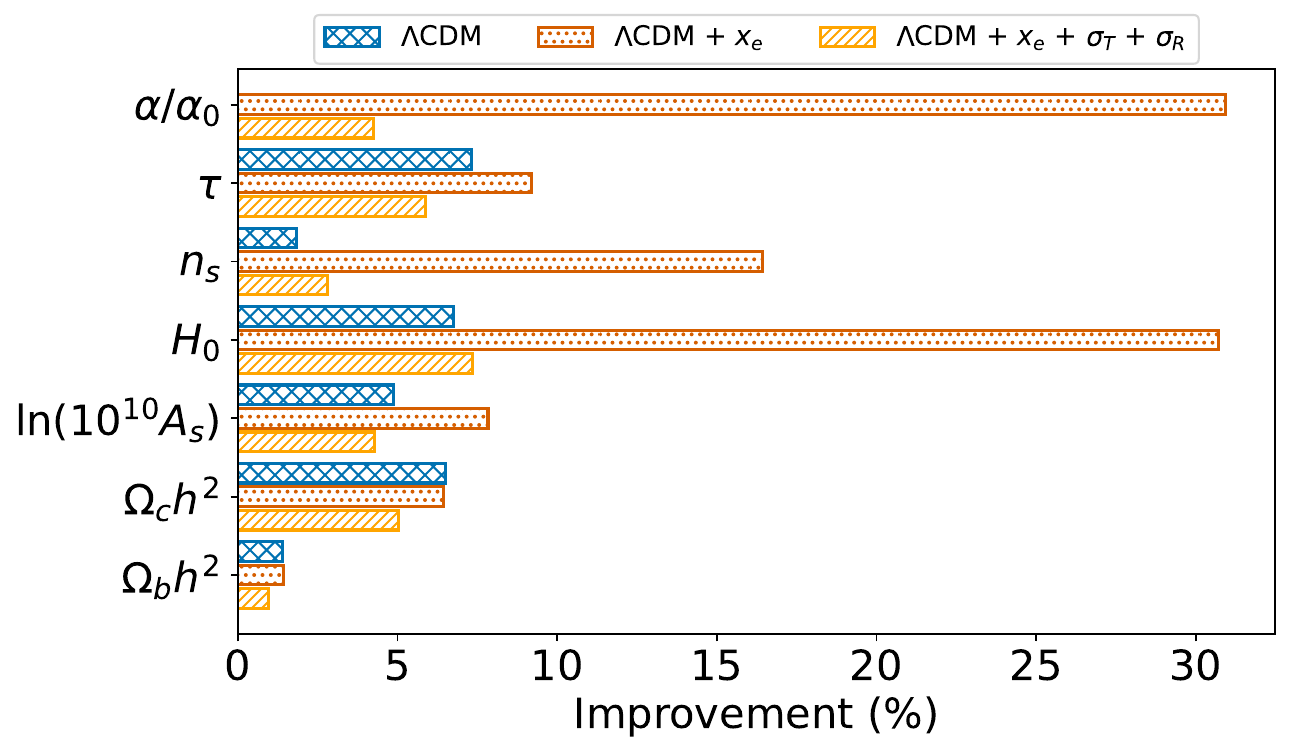}
    
    \caption{Improvement in the error bars $1-\sigma_\theta^{\rm with\,RS}/\sigma_\theta^{\rm without\,RS}$ of the cosmological parameters and electron mass (left) or fine-structure constant (right) when we include the RS signal alongside the primary CMB signal for PICO. The blue, red and orange bars correspond to the $\Lambda$CDM and two modified recombination cases, respectively.}
    \label{fig:pico_errorbar_improve}
\end{figure}

\paragraph{{\textit{For PICO:}}} Table~\ref{table:standard_dev_params} quotes the marginalised $68\%$ uncertainties of the related parameters for all models, and Figure~\ref{fig:pico_errorbar_improve} shows the corresponding fractional improvement, $1-\sigma_{\theta}^{\rm with\,RS}/\sigma_{\theta}^{\rm without\,RS}$, obtained by adding the RS signal to the primary CMB for PICO. The improvement factors are quoted in Table~\ref{tab:improvement}.

\begin{table*}[h]
\centering
\renewcommand{\arraystretch}{1.15}
\begin{tabular}{l c c c c c c c c}
\hline\hline
\multicolumn{1}{c}{Model} & $\Omega_\mathrm{b}h^2$ & $\Omega_\mathrm{c}h^2$ & $\ln(10^{10}A_\mathrm{s})$ &
$H_0$ & $n_\mathrm{s}$ & $\tau$ & $m_{\rm e}/m_{\mathrm{e},0}$ & $\alpha/\alpha_0$ \\
 & [$\times10^{-5}$] & [$\times10^{-4}$] & [$\times10^{-3}$] & & [$\times10^{-3}$] & [$\times10^{-3}$] &
[$\times10^{-3}$] & [$\times10^{-4}$] \\
\hline\hline
\textbf{$\Lambda$CDM} & & & & & & & & \\
\quad Without RS & 2.7 & 3.9 & 3.2 & 0.15 & 1.4 & 1.8 & -- & -- \\
\quad With RS    & 2.6 & 3.6 & 3.0 & 0.14 & 1.4 & 1.6 & -- & -- \\
\hline\hline
\textbf{$\Lambda$CDM + $x_{\rm e}$} & & & & & & & & \\
\quad Without RS & 2.8 & 4.0 & 3.2 & 0.33 & 2.1 & 1.8 & 1.6 & -- \\
\quad With RS    & 2.6 & 3.7 & 3.0 & 0.23 & 1.8 & 1.6 & 1.1 & -- \\
\hline
\quad Without RS & 2.7 & 3.8 & 3.2 & 0.31 & 2.1 & 1.7 & -- & 5.6 \\
\quad With RS    & 2.7 & 3.6 & 3.0 & 0.21 & 1.7 & 1.6 & -- & 3.8 \\
\hline\hline
\textbf{$\Lambda$CDM + $x_{\rm e}$ + $\sigma_\mathrm{T}$ + $\sigma_\mathrm{R}$} & & & & & & & & \\
\quad Without RS & 31  & 18  & 3.3 & 2.8  & 1.4 & 1.7 & 14  & -- \\
\quad With RS    & 21  & 13  & 3.1 & 1.9  & 1.4 & 1.7 & 9.5 & -- \\
\hline
\quad Without RS & 2.7 & 3.9 & 3.3 & 0.24 & 2.1 & 1.8 & -- & 4.1 \\
\quad With RS    & 2.7 & 3.6 & 3.2 & 0.23 & 2.1 & 1.7 & -- & 3.9 \\
\hline\hline
\end{tabular}
\caption{68\% confidence intervals for the six $\Lambda$CDM parameters and the additional parameter $m_{\rm e}/m_{\mathrm{e}, 0}$ (and $\alpha/\alpha_0$) obtained for the $\Lambda$CDM model and its possible extensions, comparing analyses with and without including the RS signal alongside the Primary CMB for PICO.}\label{table:standard_dev_params}
\end{table*}

\begin{table*}[h]
\centering
\renewcommand{\arraystretch}{1.1}
\begin{tabular}{l c c c c c c c c}
\hline\hline
Model & $\Omega_\mathrm{b}h^2$ & $\Omega_\mathrm{c}h^2$ & $\ln(10^{10}A_{\rm s})$ & $H_0$ & $n_\mathrm{s}$ & $\tau$ & $m_{\rm e}/m_{\mathrm{e},0}$ & $\alpha/\alpha_0$\\
\hline
$\Lambda$CDM                                              & 1.4  & 6.5  & 4.9 & 6.7  & 1.8  & 7.3 & --   & --   \\
$\Lambda$CDM + $x_{\rm e}$ (for $m_{\rm e}$)                              & 6.5  & 5.5  & 6.5 & 31.6 & 16.4 & 9.0 & 31.6 & --   \\
$\Lambda$CDM + $x_{\rm e}$ (for $\alpha$)                           & 1.4  & 6.5  & 7.8 & 30.7 & 16.4 & 9.2 & --   & 30.9 \\
$\Lambda$CDM + $x_{\rm e}$ + $\sigma_{\rm T}$ + $\sigma_{\rm R}$ (for $m_{\rm e}$)    & 30.7 & 28.3 & 6.6 & 31.2 & 1.3  & 3.2 & 30.9 & --   \\
$\Lambda$CDM + $x_{\rm e}$ + $\sigma_{\rm T}$ + $\sigma_{\rm R}$ (for $\alpha$) & 0.9  & 5.0  & 4.3 & 7.4  & 2.8  & 5.9 & --   & 4.2  \\
\hline\hline
\end{tabular}
\caption{Fractional improvement (in \%), $1-\sigma_\theta^{\rm with\,RS}/\sigma_\theta^{\rm without\,RS}$, in the marginalised $1\sigma$ uncertainties quoted in Table~\ref{table:standard_dev_params} when the RS signal is included. }
\label{tab:improvement}
\end{table*}

\newlength{\imgh}
\begin{figure}[p]
    \centering
    \setlength{\imgh}{0.325\textheight}   
    \includegraphics[width=0.42\textwidth]{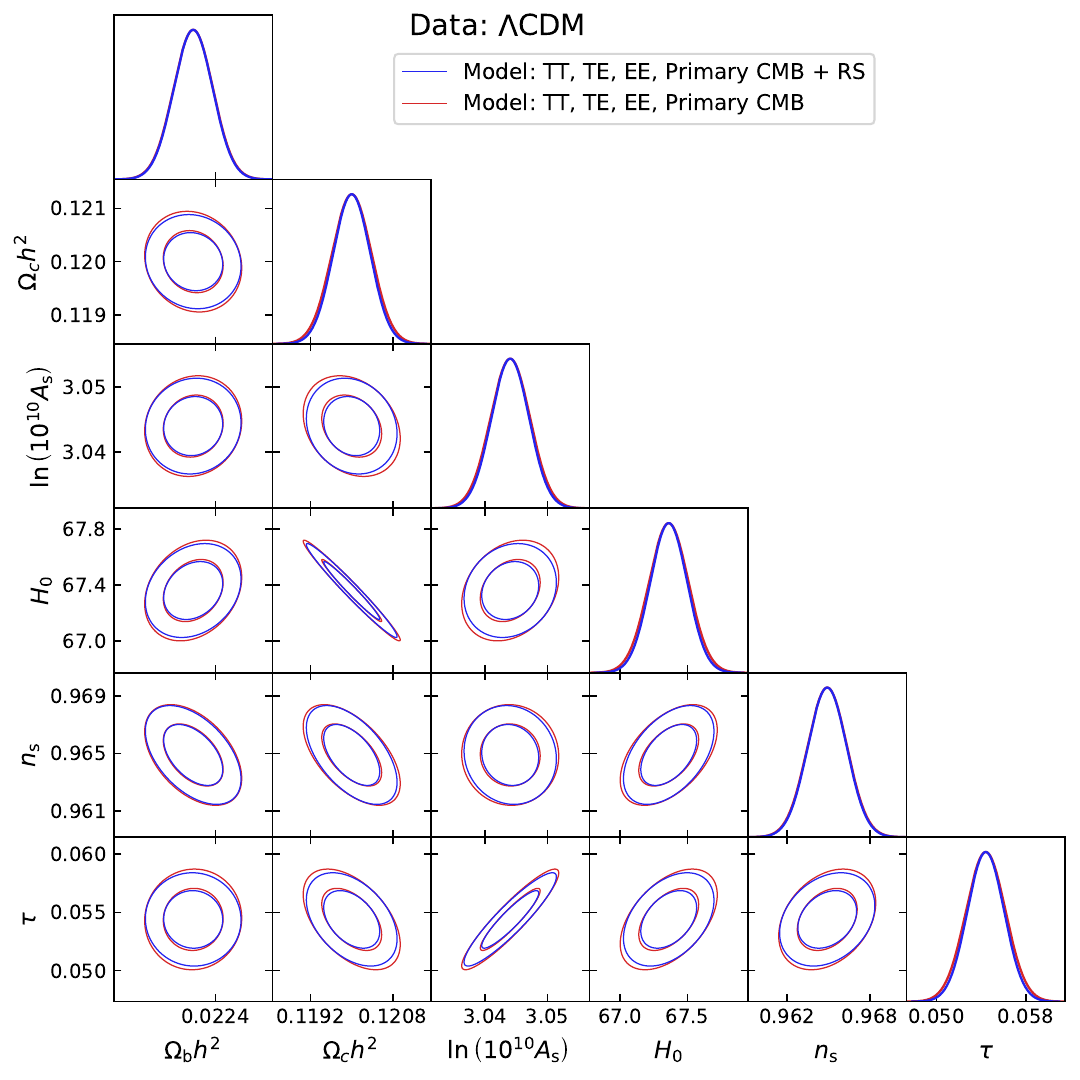} 
    \makebox[0.565\textwidth]{}\\
    \includegraphics[height=\imgh]{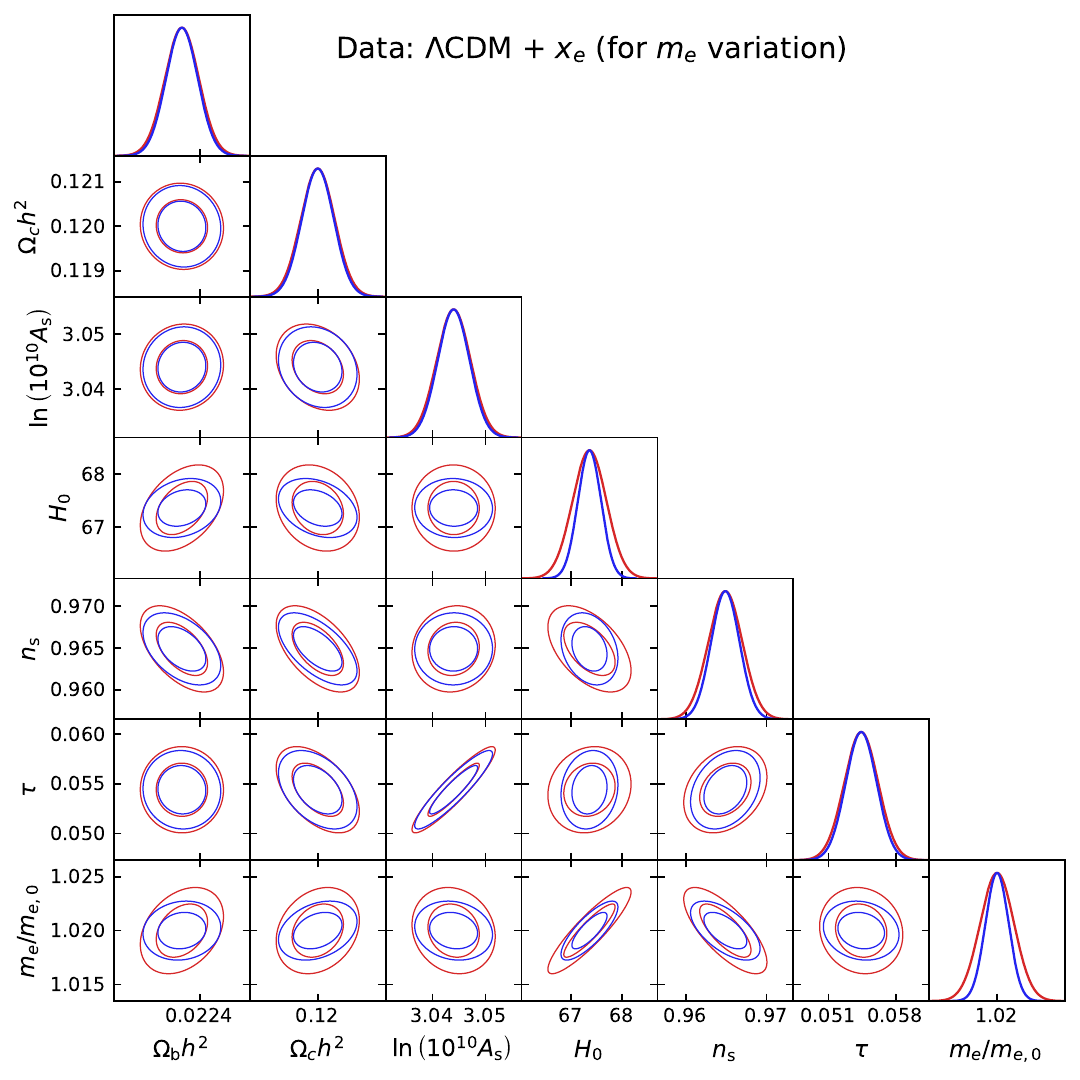}\hfill
    \includegraphics[height=\imgh]{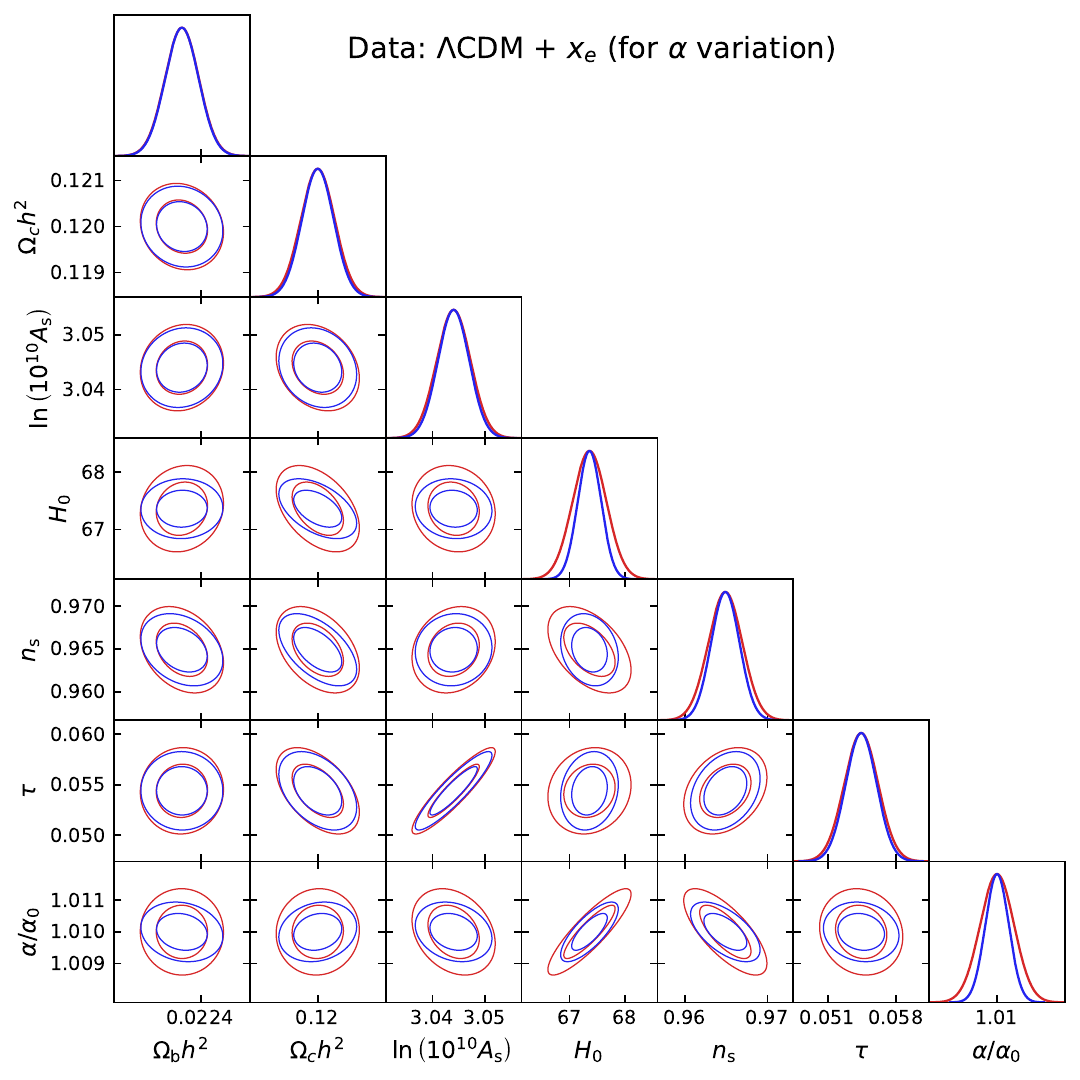}\\
    \includegraphics[height=\imgh]{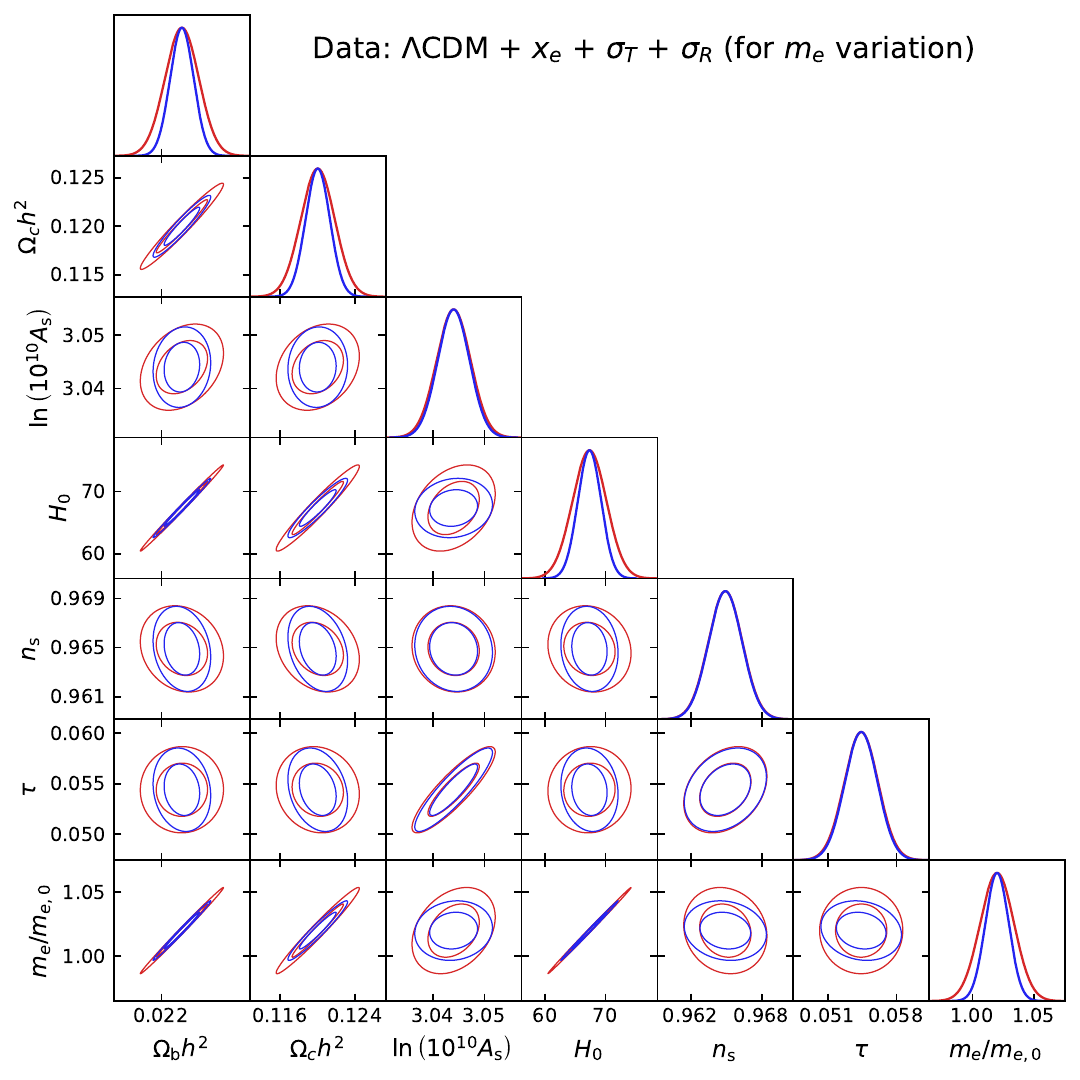}\hfill
    \includegraphics[height=\imgh]{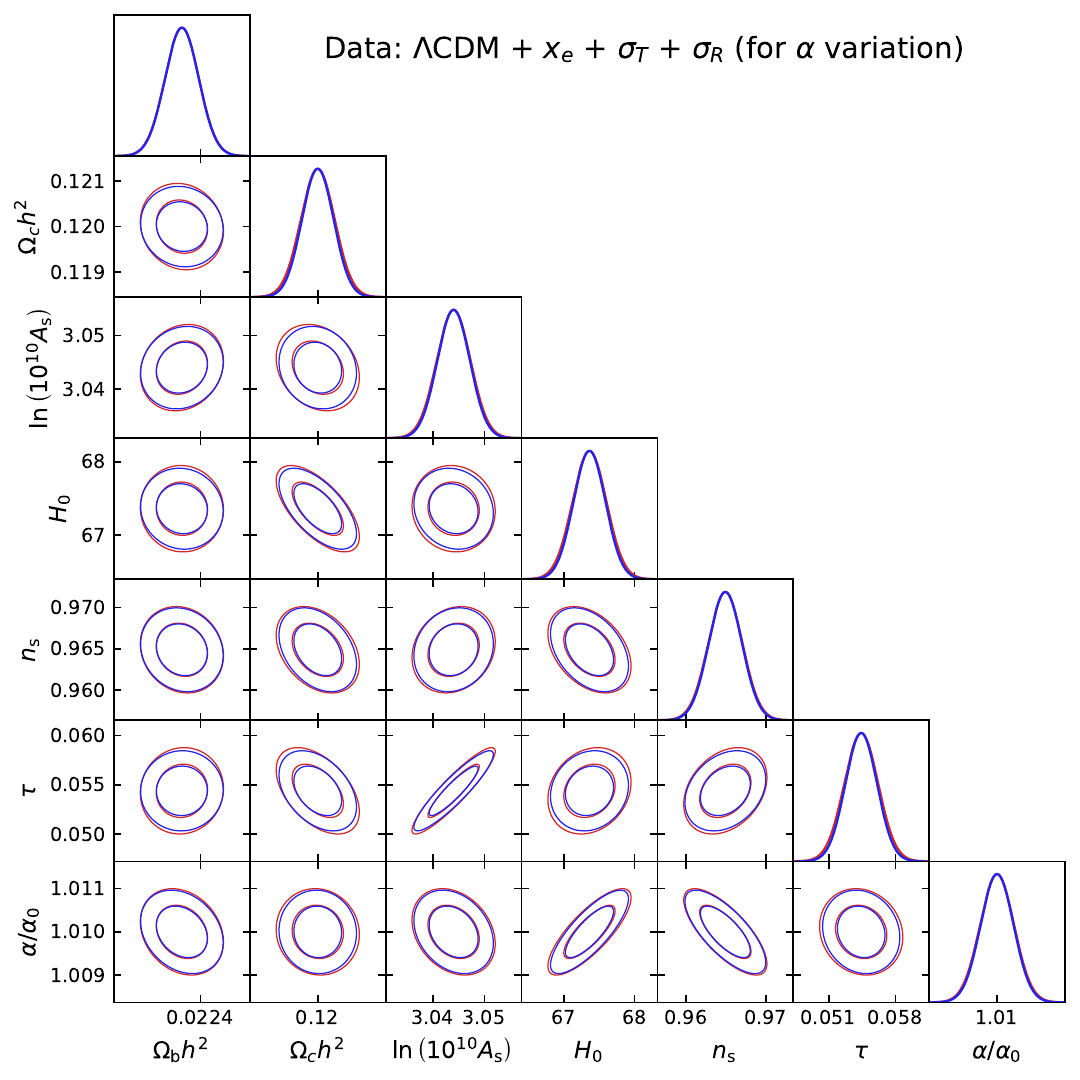}
    \caption{ $1\sigma$ and $2\sigma$ confidence contours obtained for the $\Lambda$CDM model and its extensions when we analyse the data with primary CMB only (red) and primary CMB + RS signal (blue) for PICO.  }
    \label{fig:lcdm_primary_and_rs_contour}
\end{figure}

We first investigate the improvement in the constraints on the six parameters for the standard $\Lambda$CDM model. Including RS tightens the constraints on all parameters, but by different amounts (see Table~\ref{tab:improvement}). The largest gains are for $\tau$ ($7.3\%$), $H_0$ ($6.7\%$) and $\Omega_{\rm c}h^2$ ($6.5\%$), followed by $\ln(10^{10}A_{\rm s})$ ($4.9\%$). The gains for $n_{\rm s}$ ($1.8\%$) and $\Omega_{\rm b}h^2$ ($1.4\%$) are marginal.  The overall improvement is modest because in $\Lambda$CDM, the recombination history is fixed by standard atomic physics, so RS does not open a new degree of freedom. It only provides a frequency-dependent view of the same acoustic structure. Moreover, the RS contribution becomes significant only at high frequencies, where instrumental noise and dust contamination are dominant. 
The improvement differs between parameters because RS adds information mainly through two effects. With RS, the peak of visibility shifts across frequency channels. Each channel therefore probes the sound horizon at a slightly different redshift around last scattering, producing a frequency-dependent shift of the acoustic peaks whose size depends on $H(z)$, and hence on the matter density. %
This constrains $\Omega_{\rm c}h^2$ directly, and $H_0$ through their degeneracy, which is why the precision of both parameters improves by a similar amount ($\simeq6.5\%$). Second, RS boosts the $EE$ power at $10\lesssim\ell\lesssim300$ (see Figure~\ref{fig:diff_power_spectra_lcdm}). This additional polarisation information tightens the constraints on $\tau$ ($7.3\%$) and, through its correlations with $\ln(10^{10}A_{\rm s})$,  the constraint on the primordial amplitude also improves by $4.9\%$. 
In contrast, RS produces no tilt-like change in the total power spectra in the $\Lambda$CDM framework. Furthermore, although RS depends on the baryon density since $n_{\rm H} \propto \Omega_{\rm b}h^2$, in the $\Lambda$CDM framework this information is already encoded in the primary CMB acoustic peak structure. Since the primary CMB already measures $n_{\rm s}$ and $\Omega_{\rm b}h^2$ precisely, from the overall tilt and the relative peak heights, their precision improves only by $\lesssim2\%$ with the addition of RS. Overall, RS shrinks the $\Lambda$CDM posteriors without changing the degeneracy directions, as shown in the contour plots represented on the top panel of Figure~\ref{fig:lcdm_primary_and_rs_contour}.

Next we investigate the two cases of modified recombination history. First we discuss the results for the case in which only the ionisation history $x_{\rm e}(z)$ is modified by varying $m_{\rm e}$ or $\alpha$, but scattering rates are unchanged. The impact of RS is considerably larger for some of the parameters once the recombination history is allowed to vary (see Table~\ref{tab:improvement}). For instance, for varying electron mass, the uncertainty on $m_{\rm e}/m_{\mathrm{e},0}$ improves from $0.0016$ to $0.0011$ and that on $H_0$ from $0.33$ to $0.23\,{\rm km\,s^{-1}Mpc^{-1}}$, i.e.\ by $\simeq31\%$ in both parameters.  Varying $\alpha$ leads to a very similar pattern: $\sigma(\alpha/\alpha_0)$ decreases from $0.0006$ to $0.0004$ and $\sigma(H_0)$ from $0.31$ to $0.21\,{\rm km\,s^{-1}Mpc^{-1}}$ which is a $\simeq31\%$ improvement. These are the parameters RS probes most directly. 
In the primary CMB, a change in $m_{\rm e}$ or $\alpha$ shows up only through the shift of the Thomson visibility. The changes in the Rayleigh visibilities are, however, more pronounced and frequency dependent. The multi-frequency RS signal provides additional, partly independent information on the timing and duration of recombination, beyond what is available from the primary CMB alone. An earlier recombination caused by increased  $m_{\rm e}$ or $\alpha$ values reduces the sound horizon, and with the angular size of the sound horizon, $\theta_*$ fixed by the data, this is compensated almost exactly by a larger $H_0$, giving the near-perfect $m_{\rm e}$--$H_0$ (and $\alpha$--$H_0$) degeneracy as illustrated in the middle-left (and middle-right) panel of Figure~\ref{fig:lcdm_primary_and_rs_contour}. Since RS can sample the recombination timing, this gives improved constraints on $m_{\rm e}$ and $\alpha$ and hence $H_0$ as well through $m_{\rm e}$--$H_0$ (or $\alpha$--$H_0$) correlation. Therefore $H_0$ improves by almost the same amount as $m_{\rm e}/m_{\mathrm{e},0}$ or $\alpha/\alpha_0$. The other parameters improve according to how strongly they are degenerate with the fundamental constants.

The constraint on the spectral index $n_{\rm s}$ improves by $16.4\%$, which is substantially more than in $\Lambda$CDM ($1.8\%$). This improvement is closely related to the changes in the damping tail of CMB power spectra in modified recombination scenarios. A change in the fundamental constant shifts the epoch of recombination and hence modifies the photon diffusion scale, thereby changing the suppression of power at high $\ell$. A change in $n_{\rm s}$ also alters the relative amount of power between large and small scales. From the primary CMB alone, these effects can therefore be partially degenerate, producing a correlation between $n_{\rm s}$ and the fundamental constants. RS helps to break this degeneracy because its contribution to the damping is frequency-dependent. In particular, RS adds extra diffusion damping that grows as $\nu^4$, and therefore its damping effect on the anisotropies varies with frequency, while the primordial tilt is common to all frequency channels. Comparing the damping tail across frequencies therefore helps distinguish changes in the damping caused by the modified recombination history from those associated with $n_{\rm s}$. This additional information reduces the $n_{\rm s}$--$m_{\rm e}$ (or $n_{\rm s}$--$\alpha$) degeneracy, leading to an improvement in the constraint on $n_{\rm s}$. The slightly larger gain for $\alpha$ is consistent with its stronger impact on the recombination history and, consequently, on the damping and visibility structure. In contrast, $\Omega_{\rm c}h^2$, $\ln(10^{10}A_{\rm s})$ and $\tau$ improve by $5$--$9\%$, similar to $\Lambda$CDM. Because the scattering rates are unchanged in this model, the relative peak heights still determine the matter density and the amplitude independently of $m_{\rm e}$ or $\alpha$. These parameters are therefore only weakly correlated with the fundamental constants (quantitatively the correlation coefficients, $|\rho|\lesssim 0.3$), and their gain comes mainly from the same effects as in $\Lambda$CDM. $\Omega_{\rm b}h^2$ improves somewhat more than in $\Lambda$CDM ($6.5\%$ for $m_{\rm e}$). This is because a shift of the recombination redshift changes the baryon loading, $R(z_*) \propto \Omega_{\rm b}h^2 $ at decoupling. Since the RS opacity is linked with  $n_H \propto \Omega_{\rm b}h^2$ this introduces a mild correlation between $\Omega_{\rm b}h^2$ and $m_{\rm e}$, which gives RS more leverage on $\Omega_{\rm b}h^2$ than in the standard $\Lambda$CDM case.
For $\alpha$, $\Omega_{\rm b}h^2$ is essentially uncorrelated with the fundamental constant already in the primary CMB, and its improvement is minimal.

When the variation of $m_{\rm e}$  is consistently propagated to the scattering cross-sections, the constraints from the primary CMB degrade substantially. In particular, the marginalised uncertainties increase to $\sigma(m_{\rm e}/m_{\mathrm{e},0})=0.0138$, $\sigma(H_0)=2.83\,{\rm km\,s^{-1}Mpc^{-1}}$, and $\sigma(\Omega_{\rm b}h^2)=3.1\times10^{-4}$, $\sigma(\Omega_{\rm c}h^2)= 1.81\times10^{-3}$, roughly an order of magnitude weaker than in the $x_{\rm e}$-only modification case. This happens because two effects of a larger $m_{\rm e}$ partly cancel. First, recombination happens earlier, $(1+z_*) \propto m_{\rm e}$, when the Universe is denser, so there are more electrons available to scatter photons. Second, each electron scatters less, because $\sigma_{\rm T}\propto m_{\rm e}^{-2}$. The net Thomson scattering rate at recombination therefore changes only mildly, roughly as $\Gamma_{\rm T} (z_*)=n_{\rm e}\sigma_{\rm T}\propto m_{\rm e}$. This remaining change can be absorbed by a faster expansion, $H(z_*)\propto\sqrt{\omega_{\rm m} (1+z_*)^3}$, obtained by increasing $\Omega_{\rm b}h^2$ and $\Omega_{\rm c}h^2$, with $H_0$ adjusted to keep the angular scale of the acoustic peaks fixed. These adjustments leave the ratio $\Gamma_{\rm T}/H$ unchanged, which decides the shape of the Thomson visibility function \citep{Sekiguchi2021PhRvD,Kylar_L.Greene:2023JCAP}. As a result, the primary CMB can hardly distinguish a change in $m_{\rm e}$ from a combined shift of $H_0$, $\Omega_{\rm b}h^2$ and $\Omega_{\rm c}h^2$, and these four parameters become strongly degenerate, as demonstrated in the bottom-left panel of  Figure~\ref{fig:lcdm_primary_and_rs_contour}.

Rayleigh scattering does not completely break this scaling. Although $\sigma_{\rm R}\propto m_{\rm e}^{-6}$ at fixed observed frequency $\nu$, the photon frequency at the last scattering epoch is $\nu_{\rm phys}=\nu(1+z_*)$, while $1+z_*\propto m_{\rm e}$. Consequently, at a fixed observed frequency,
$\sigma_{\rm R}(\nu,z_*)\propto\nu^4(1+z_*)^4m_{\rm e}^{-6}\propto\nu^4m_{\rm e}^{-2}.$ Since $n_{\rm H}\propto(1+z_*)^3\propto m_{\rm e}^3$, the Rayleigh scattering rate scales approximately as
$ \Gamma_{\rm R}(\nu)
\propto n_{\rm H}\sigma_{\rm R}\propto
\nu^4m_{\rm e}$, which has the same leading dependence on  $m_{\rm e}$ as the Thomson scattering rate, $\Gamma_{\rm T}\propto m_{\rm e}$. Thus, readjustment of other parameters to keep the primary CMB unchanged also preserves RS. However, the multi-frequency RS spectra provide additional, partly independent measurements of the same acoustic and damping structure,  which are not completely redundant with the primary CMB.  These modes provide additional information along the otherwise poorly constrained degeneracy direction, reducing its allowed extent without substantially changing its orientation, as shown in the bottom-left panel of Figure~\ref{fig:lcdm_primary_and_rs_contour}.
As a result, the parameters that are most strongly aligned with this degeneracy improve by nearly the same percentage. In contrast, $\ln(10^{10}A_{\rm s})$, $\tau$, and $n_{\rm s}$ are less strongly aligned with this degeneracy and improve by only $1$--$7\%$.

The behaviour for $\alpha$ is qualitatively different. Although $\alpha$ changes both the atomic binding energies and the scattering cross-sections, the different processes governing recombination do not scale with a common power of $\alpha$. In particular, the rates associated with Lyman-$\alpha$ escape, two-photon decay, and photoionisation have different $\alpha$ dependences from $\sigma_{\rm T}\propto\alpha^2$. Therefore, no analogous rescaling of $\Omega_{\rm b}h^2$, $\Omega_{\rm c}h^2$, and $H_0$ can preserve the Thomson scattering history throughout recombination. The primary CMB consequently already constrains $\alpha/\alpha_0$ tightly, with $\sigma(\alpha/\alpha_0)=0.0004$, leaving little degeneracy for RS to break. The inclusion of RS therefore produces only modest additional improvements, at the level of $\lesssim 8\%$.

\paragraph{\textbf{\textit{For LiteBIRD + Planck:}}} The corresponding results for LiteBIRD + Planck are given in Appendix~\ref{app:B}: Tables~\ref{table:standard_dev_params_LB_Planck} and \ref{tab:improvement_LB_Planck} present the 68\% confidence limits and relative improvement in the error bars of the parameters, respectively. Figure~\ref{fig:pico_errorbar_improve_lb_planck} displays the error-bar improvements as quoted in Table~\ref{tab:improvement_LB_Planck}. The trends are the same as for PICO, but the gains are much smaller for most cases. In $\Lambda$CDM no parameter improves by more than $0.7\%$, and in the $\Lambda$CDM + $x_{\rm e}$ model the largest gains are $1.9\%$ for $H_0$ and $1.4\%$ for $m_{\rm e}/m_{\mathrm{e},0}$ (similarly $1.2\%$ and $0.7\%$ for $H_0$ and $\alpha/\alpha_0$, respectively). This is expected as the RS signal is concentrated at high frequencies, where the noise of the Planck channels limits its detectability, and partially because we conservatively restrict the LiteBIRD multipole range to $\ell_{\rm max}=200$ in this analysis. 
The exception is the $\Lambda$CDM + $x_{\rm e}$ + $\sigma_{\rm T}$ + $\sigma_{\rm R}$ model for varying $m_{\rm e}$, where $\Omega_\mathrm{b}h^2$,
$\Omega_\mathrm{c}h^2$, $H_0$ and $m_{\rm e}/m_{\mathrm{e},0}$ improve by $55$--$58\%$. With the higher noise of LiteBIRD + Planck as compared to PICO, the primary CMB alone constrains the degeneracies among these four parameters discussed above even more weakly, with $\sigma(m_{\rm e}/m_{\mathrm{e},0}) = 0.046$ and $\sigma(H_0) = 9.8$~km\,s$^{-1}$Mpc$^{-1}$. Even the small amount of information that RS adds along this direction of  $m_{\rm e}/m_{\mathrm{e},0}$ therefore reduces these uncertainties substantially, to $0.019$ and $4.1$~km\,s$^{-1}$Mpc$^{-1}$. For $\alpha$, where no such degeneracy exists as also discussed above, the improvement stays below $1\%$.

\begin{table*}[h]
\centering
\small
\setlength{\tabcolsep}{4pt}
\renewcommand{\arraystretch}{1.1}
\begin{tabular}{l l l c c c c c c c}
\hline\hline
Data & Model & RS & $\Omega_\mathrm{b}h^2$ & $\Omega_\mathrm{c}h^2$ & $\ln(10^{10}A_\mathrm{s})$ & $H_0$ & $n_\mathrm{s}$ & $\tau$ & $m_{\rm e}/m_{\mathrm{e},0}$ \\
\hline
\textbf{Primary CMB }
 & $\Lambda$CDM
   & no  & $-0.95\sigma$ & $0.59\sigma$ & $-0.16\sigma$ & $-0.57\sigma$ & $-0.49\sigma$ & $-0.12\sigma$ & -- \\
\textbf{+ RS} & & & $(-0.11\%)$ & $(0.19\%)$ & $(-0.02\%)$ & $(-0.12\%)$ & $(-0.07\%)$ & $(-0.39\%)$ & -- \\
\hline\hline
\textbf{Primary CMB } 
 & $\Lambda$CDM
   & no  & $-11.96\sigma$ & $24.88\sigma$ & $17.29\sigma$ & $-54.46\sigma$ & $-5.49\sigma$ & $5.55\sigma$ & -- \\
 \textbf{+ RS}+$x_{\rm e}$& & & $(-1.40\%)$ & $(7.68\%)$ & $(1.80\%)$ & $(-11.45\%)$ & $(-0.81\%)$ & $(18.03\%)$ & -- \\
 & & yes & $-8.52\sigma$ & $22.80\sigma$ & $20.09\sigma$ & $-54.34\sigma$ & $-4.74\sigma$ & $8.26\sigma$ & -- \\
 & & & $(-0.98\%)$ & $(6.58\%)$ & $(1.98\%)$ & $(-10.66\%)$ & $(-0.69\%)$ & $(24.84\%)$ & -- \\
\cline{2-10}
 & $\Lambda$CDM + $x_{\rm e}$
   & no  & $-0.81\sigma$ & $0.78\sigma$ & $-0.30\sigma$ & $0.35\sigma$ & $-0.89\sigma$ & $-0.22\sigma$ & $0.73\sigma$ \\
 & & & $(-0.10\%)$ & $(0.24\%)$ & $(-0.03\%)$ & $(0.16\%)$ & $(-0.18\%)$ & $(-0.72\%)$ & $(0.10\%)$ \\
\hline\hline
\textbf{Primary CMB } 
 & $\Lambda$CDM
   & no  & $-27.45\sigma$ & $29.17\sigma$ & $20.71\sigma$ & $-65.21\sigma$ & $-27.08\sigma$ & $5.64\sigma$ & -- \\
\textbf{+ RS}+$x_{\rm e}$  & & & $(-3.20\%)$ & $(9.01\%)$ & $(2.15\%)$ & $(-13.72\%)$ & $(-4.00\%)$ & $(18.30\%)$ & -- \\
+ $\sigma_\mathrm{T}$ + $\sigma_\mathrm{R}$ & & yes & $-26.27\sigma$ & $28.64\sigma$ & $24.32\sigma$ & $-67.38\sigma$ & $-26.35\sigma$ & $8.87\sigma$ & -- \\
 & & & $(-3.02\%)$ & $(8.27\%)$ & $(2.40\%)$ & $(-13.22\%)$ & $(-3.82\%)$ & $(26.68\%)$ & -- \\
\cline{2-10}
 & $\Lambda$CDM + $x_{\rm e}$
   & no  & $-17.08\sigma$ & $-5.17\sigma$ & $-0.06\sigma$ & $-13.50\sigma$ & $-1.26\sigma$ & $0.02\sigma$ & $-13.25\sigma$ \\
 & & & $(-2.09\%)$ & $(-1.61\%)$ & $(-0.01\%)$ & $(-6.32\%)$ & $(-0.26\%)$ & $(0.05\%)$ & $(-1.91\%)$ \\
 & & yes & $-17.06\sigma$ & $-6.89\sigma$ & $1.24\sigma$ & $-18.28\sigma$ & $-0.69\sigma$ & $1.42\sigma$ & $-18.63\sigma$ \\
 & & & $(-1.97\%)$ & $(-2.02\%)$ & $(0.12\%)$ & $(-6.08\%)$ & $(-0.12\%)$ & $(4.22\%)$ & $(-1.92\%)$ \\
\cline{2-10}
 & $\Lambda$CDM + $x_{\rm e}$
   & no  & $-0.02\sigma$ & $0.19\sigma$ & $-0.15\sigma$ & $0.03\sigma$ & $-0.45\sigma$ & $-0.13\sigma$ & $0.06\sigma$ \\
 & + $\sigma_\mathrm{T}$ + $\sigma_\mathrm{R}$ & & $(-0.02\%)$ & $(0.27\%)$ & $(-0.02\%)$ & $(0.14\%)$ & $(-0.07\%)$ & $(-0.41\%)$ & $(0.08\%)$ \\
\hline\hline
\end{tabular}
\caption{Biases in units of $\sigma$, with the corresponding percentage shifts in parentheses, in the six $\Lambda$CDM parameters and $m_{\rm e}/m_{\mathrm{e},0}$ for PICO. The data (all including the RS signal) are generated under three scenarios: $\Lambda$CDM, a modified recombination history with $m_{\rm e}/m_{\mathrm{e},0}=1.02$ entering only through $x_{\rm e}$, and the same variation also applied to $\sigma_\mathrm{T}$ and $\sigma_\mathrm{R}$. Each data set is analysed with the models listed, without (``no'') or with (``yes'') modelling RS. When the model matches the data and RS is modelled, the biases vanish by construction and these rows are omitted.}
\label{table:bias_params_}
\end{table*}

\begin{table*}[h]
\centering
\small
\setlength{\tabcolsep}{4pt}
\renewcommand{\arraystretch}{1.1}
\begin{tabular}{l l l c c c c c c c}
\hline\hline
Data & Model & RS & $\Omega_\mathrm{b}h^2$ & $\Omega_\mathrm{c}h^2$ & $\ln(10^{10}A_\mathrm{s})$ & $H_0$ & $n_\mathrm{s}$ & $\tau$ & $\alpha/\alpha_0$ \\
\hline
$\textbf{Primary CMB  }$
 & $\Lambda$CDM
   & no  & $-10.74\sigma$ & $42.10\sigma$ & $29.76\sigma$ & $-77.66\sigma$ & $-9.90\sigma$ & $9.01\sigma$ & -- \\
 \textbf{+ RS +$x_{\rm e}$}& & & $(-1.28\%)$ & $(13.55\%)$ & $(3.11\%)$ & $(-16.96\%)$ & $(-1.47\%)$ & $(29.30\%)$ & -- \\
 & & yes & $-6.47\sigma$ & $39.79\sigma$ & $34.80\sigma$ & $-77.81\sigma$ & $-8.73\sigma$ & $14.01\sigma$ & -- \\
 & & & $(-0.76\%)$ & $(11.94\%)$ & $(3.47\%)$ & $(-15.80\%)$ & $(-1.27\%)$ & $(42.26\%)$ & -- \\
\cline{2-10}
 & $\Lambda$CDM + $x_{\rm e}$
   & no  & $-1.05\sigma$ & $0.66\sigma$ & $-0.44\sigma$ & $0.47\sigma$ & $-1.03\sigma$ & $-0.25\sigma$ & $0.88\sigma$ \\
 & & & $(-0.13\%)$ & $(0.21\%)$ & $(-0.05\%)$ & $(0.21\%)$ & $(-0.22\%)$ & $(-0.80\%)$ & $(0.05\%)$ \\
\hline\hline
\textbf{Primary CMB }
 & $\Lambda$CDM
   & no  & $-2.53\sigma$ & $38.02\sigma$ & $27.43\sigma$ & $-70.48\sigma$ & $1.27\sigma$ & $8.76\sigma$ & -- \\
\textbf{+ RS +$x_{\rm e}$ } & & & $(-0.30\%)$ & $(12.24\%)$ & $(2.87\%)$ & $(-15.39\%)$ & $(0.19\%)$ & $(28.50\%)$ & -- \\
\textbf{+ $\sigma_\mathrm{T}$ + $\sigma_\mathrm{R}$} & & yes & $0.08\sigma$ & $37.27\sigma$ & $31.76\sigma$ & $-72.14\sigma$ & $2.61\sigma$ & $12.89\sigma$ & -- \\
 & & & $(0.01\%)$ & $(11.18\%)$ & $(3.17\%)$ & $(-14.65\%)$ & $(0.38\%)$ & $(38.90\%)$ & -- \\
\cline{2-10}
 & $\Lambda$CDM + $x_{\rm e}$
   & no  & $5.45\sigma$ & $2.83\sigma$ & $-0.83\sigma$ & $6.13\sigma$ & $-0.77\sigma$ & $-0.29\sigma$ & $6.53\sigma$ \\
 & & & $(0.66\%)$ & $(0.90\%)$ & $(-0.09\%)$ & $(2.74\%)$ & $(-0.17\%)$ & $(-0.93\%)$ & $(0.35\%)$ \\
 & & yes & $4.95\sigma$ & $3.55\sigma$ & $1.62\sigma$ & $-1.89\sigma$ & $6.49\sigma$ & $0.55\sigma$ & $-2.12\sigma$ \\
 & & & $(0.59\%)$ & $(1.08\%)$ & $(0.16\%)$ & $(-0.62\%)$ & $(1.20\%)$ & $(1.63\%)$ & $(-0.08\%)$ \\
\cline{2-10}
 & $\Lambda$CDM + $x_{\rm e}$ 
   & no  & $-1.19\sigma$ & $0.61\sigma$ & $-0.40\sigma$ & $0.27\sigma$ & $-1.00\sigma$ & $-0.25\sigma$ & $0.83\sigma$ \\
 &+ $\sigma_\mathrm{T}$ + $\sigma_\mathrm{R}$ & & $(-0.15\%)$ & $(0.20\%)$ & $(-0.04\%)$ & $(0.10\%)$ & $(-0.22\%)$ & $(-0.82\%)$ & $(0.03\%)$ \\
\hline\hline
\end{tabular}
\caption{Biases in units of $\sigma$, with the corresponding percentage shifts in parentheses, for data generated with $\alpha/\alpha_0=1.01$, entering either only through $x_{\rm e}$ or also through $\sigma_\mathrm{T}$ and $\sigma_\mathrm{R}$ for PICO. Each data set is analysed with the models listed, without (``no'') or with (``yes'') modelling RS. When the model matches the data and RS is modelled, the biases vanish by construction and these rows are omitted. The case of $\Lambda$CDM data is identical to Table~\ref{table:bias_params_}.}
\label{table:bias_params_alpha}
\end{table*}
\begin{figure*}
    \centering
    \includegraphics[width=\textwidth]{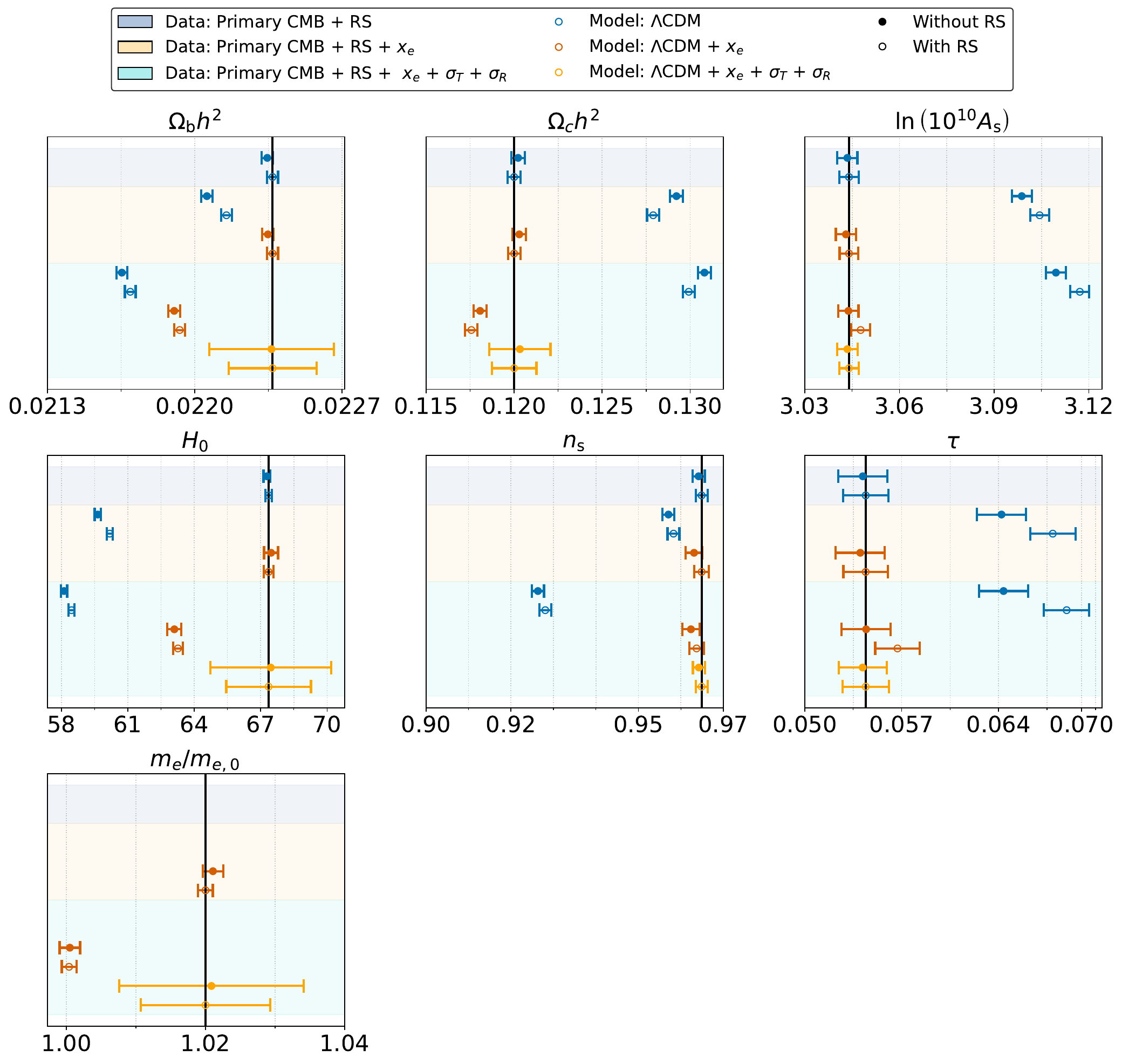}
    \caption{Biased estimated values, $\theta_\mathrm{fid}+\delta\theta$, with their marginalised $1\sigma$ error bars, of the six $\Lambda$CDM parameters and $m_{\rm e}/m_{\mathrm{e},0}$ for PICO, when the simulated
data are analysed with $\Lambda$CDM and its extensions. The shaded horizontal bands correspond to the three simulated data sets, all including the RS signal: $\Lambda$CDM (blue band), a modified recombination history with $m_{\rm e}/m_{\mathrm{e},0}=1.02$ entering only through $x_{\rm e}$ (orange band), and the same variation also applied to $\sigma_\mathrm{T}$ and $\sigma_\mathrm{R}$(cyan band). Within each band, the colour of the points indicates the model used in the analysis:  $\Lambda$CDM (blue),
$\Lambda$CDM + $x_{\rm e}$ (red), and $\Lambda$CDM + $x_{\rm e}$ + $\sigma_\mathrm{T}$ + $\sigma_\mathrm{R}$ (orange). Filled and open markers denote analyses without and with modelling RS, respectively. Each vertical line marks the input value used to generate the data.} 
    \label{fig:prameter_bias_me}
\end{figure*}

\begin{figure*}
    \centering
    \includegraphics[width=\textwidth]{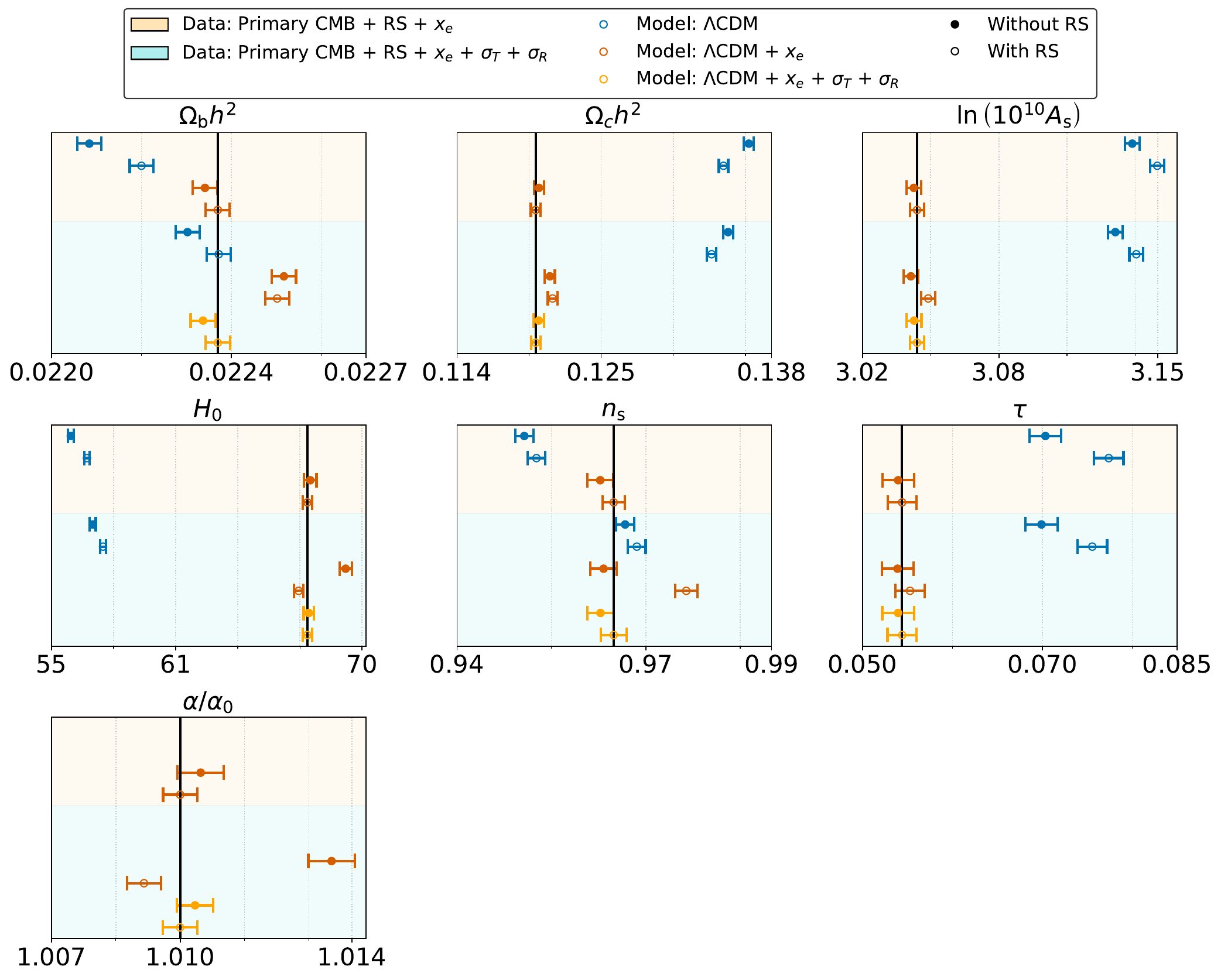}
    \caption{Biased estimated values, $\theta_\mathrm{fid}+\delta\theta$, with their marginalised $1\sigma$ error bars, of the six $\Lambda$CDM parameters and $\alpha/\alpha_0$ for PICO. The data, generated with
$\alpha/\alpha_0=1.01$, include the RS signal, with the variation of $\alpha$ entering only through $x_{\rm e}$ (orange band) or also through $\sigma_\mathrm{T}$ and $\sigma_\mathrm{R}$ (cyan band). Colours, markers and vertical lines are as in Figure~\ref{fig:prameter_bias_me}. The $\Lambda$CDM-data case, identical to Figure~\ref{fig:prameter_bias_me}, is not repeated in this figure.} 
    \label{fig:prameter_bias_alpha}
\end{figure*}

\subsection{Parameter biases}
\label{sec:res_bias}
We now quantify the biases in the inferred parameters using Eq.~\ref{eq:bias}. We generate three sets of mock data, all including the RS signal: (i) standard $\Lambda$CDM; (ii) a modified recombination history in which the variation of the fundamental constant enters only through $x_{\rm e}(z)$; and (iii) the same variation applied also to $\sigma_{\rm T}$ and $\sigma_{\rm R}$. For the electron mass we adopt $m_{\rm e}/m_{\mathrm{e},0}=1.02$ and for the fine-structure constant $\alpha/\alpha_0=1.01$, with all other parameters at their fiducial values. Each data set is analysed with $\Lambda$CDM and with the extended models, with and without modelling the RS contribution. In each case, the bias is projected with the Fisher matrix of the model used in the analysis, and is quoted in units of that model's marginalised $1\sigma$ uncertainty (and in \% in the parentheses). The results are shown in Tables~\ref{table:bias_params_} (for $m_{\rm e}$) and~\ref{table:bias_params_alpha} (for $\alpha$) and illustrated in Figures~\ref{fig:prameter_bias_me} and~\ref{fig:prameter_bias_alpha}, where each point marks the recovered value $\theta_{\rm fid}+\delta\theta$ with the corresponding $1\sigma$ error bar, and the vertical line marks the input value. When the model matches the data and RS is modelled, the bias vanishes by construction. Note that Eq.~\ref{eq:bias} is a linear approximation, valid for $|\delta\theta|\lesssim$ a few $\sigma$; biases of tens of $\sigma$ found below should therefore be read as indicating a severe mis-modelling rather than as precise values.\\

\paragraph{{\textit{For PICO:}}}We begin by discussing the parameter biases expected for the PICO instrument configuration.
\subparagraph{Neglecting the RS in $\Lambda$CDM model:}
If the data follow $\Lambda$CDM but RS is omitted from the model, all biases remain below $1\sigma$. The largest shifts are in $\Omega_{\rm b}h^2$ ($-0.95\sigma$), $\Omega_{\rm c}h^2$ ($+0.59\sigma$) and $H_0$ ($-0.57\sigma$), corresponding to fractional shifts of $\lesssim0.2\%$. Although small, a bias of nearly $1\sigma$ in $\Omega_{\rm b}h^2$ would become relevant when combined with external data such as BBN, showing that RS must be modelled at the sensitivity of next-generation experiments.

\subparagraph{Fitting a modified recombination history with $\Lambda$CDM model:}
If the recombination history in the data is modified only through $x_{\rm e}$ (data set ii) and the analysis assumes $\Lambda$CDM, the inferred parameters are severely biased, whether or not RS is modelled. An increase in $m_{\rm e}$ or $\alpha$ shifts recombination to earlier times and reduces the sound horizon; $\Lambda$CDM can mimic this only by lowering $H_0$ to keep the angular acoustic scale fixed, together with a larger $\Omega_{\rm c}h^2$ and a correlated increase of $\ln(10^{10}A_{\rm s})$ and $\tau$. For $m_{\rm e}/m_{\mathrm{e},0}=1.02$, $H_0$ is shifted by $\simeq-54\sigma$ ($\simeq-11\%$, i.e. $H_0\simeq60\,{\rm km\,s^{-1}Mpc^{-1}}$), $\Omega_{\rm c}h^2$ by $\simeq+23$ to $+25\sigma$ ($\simeq7\%$), $\ln(10^{10}A_{\rm s})$ by $\simeq+17$ to $+20\sigma$ and $\Omega_{\rm b}h^2$ by $\simeq-9$ to $-12\sigma$, while $\tau$ is shifted by $18$--$25\%$. For $\alpha/\alpha_0=1.01$, which shifts the atomic energy levels ($\propto\alpha^2m_{\rm e}$) by the same $2\%$, the shifts are even larger, reflecting the additional, non-uniform $\alpha$-dependence of the recombination rates: $H_0$ by $\simeq-78\sigma$ ($H_0\simeq56$--$57\,{\rm km\,s^{-1}Mpc^{-1}}$), $\Omega_{\rm c}h^2$ by $\simeq+40\sigma$ ($12$--$14\%$) and $\tau$ by $29$--$42\%$. In this case the model mis-specification dominates, and modelling RS within a $\Lambda$CDM fit does not reduce the bias; it only increases its significance, since RS tightens the error bars. Once the fundamental constant is included in the model ($\Lambda$CDM + $x_{\rm e}$), the biases from neglecting RS fall to $\lesssim1\sigma$ for all parameters (at most $-0.89\sigma$ in $n_{\rm s}$ and $0.73\sigma$ in $m_{\rm e}/m_{\mathrm{e},0}$, and $-1.05\sigma$ in $\Omega_{\rm b}h^2$ and $0.88\sigma$ in $\alpha/\alpha_0$), and they vanish when RS is also modelled.

\subparagraph{Mis-modelling the scattering cross-section dependence:}
For data set (iii), in which the variation also rescales $\sigma_{\rm T}$ and $\sigma_{\rm R}$, a $\Lambda$CDM fit again gives very large biases. For $m_{\rm e}$ these reach $\simeq-65\sigma$ in $H_0$ ($\simeq-13.5\%$), $\simeq+29\sigma$ in $\Omega_{\rm c}h^2$, $\simeq-27\sigma$ in both $\Omega_{\rm b}h^2$ and $n_{\rm s}$, and $18$--$27\%$ in $\tau$; for $\alpha$, $H_0$ is shifted by $\simeq-70\sigma$ ($\simeq-15\%$) and $\Omega_{\rm c}h^2$ by $\simeq+37\sigma$, whereas $\Omega_{\rm b}h^2$ and $n_{\rm s}$ are much less affected ($\lesssim3\sigma$).

More importantly, analysing these data with the $\Lambda$CDM + $x_{\rm e}$ model, i.e.\ allowing for a varying constant but neglecting its effect on the cross-sections, leads to qualitatively different outcomes for $m_{\rm e}$ and $\alpha$. For $m_{\rm e}$, large biases remain: $\simeq-17\sigma$ in $\Omega_{\rm b}h^2$, $-13.5\sigma$ ($-18.3\sigma$) in $H_0$ and $-13.3\sigma$ ($-18.6\sigma$) in $m_{\rm e}/m_{\mathrm{e},0}$ without (with) RS. The recovered electron mass is shifted by $\simeq-1.9\%$, to $m_{\rm e}/m_{\mathrm{e},0}\simeq1.000$, so that a true 2\% variation would be inferred as consistent with no variation at all. This is a direct consequence of the scaling degeneracy discussed in Section~\ref{sec:res_constraints}: the reduction of $\sigma_{\rm T}\propto m_{\rm e}^{-2}$ largely compensates the earlier recombination, so that the data resemble a standard recombination history with shifted $\Omega_{\rm b}h^2$ and $H_0$. The fractional shifts are almost identical with and without RS, but their significance increases with RS because the error bars are tighter. For $\alpha$, where no such compensation exists, the recovered value remains close to the input: $\alpha/\alpha_0\simeq1.0135$ ($+6.5\sigma$) without RS and $\simeq1.0092$ ($-2.1\sigma$) with RS, accompanied by biases of $\simeq5$--$6\sigma$ in $\Omega_{\rm b}h^2$ and in $H_0$ (without RS) or $n_{\rm s}$ (with RS). In both cases, only the fully consistent $\Lambda$CDM + $x_{\rm e}$ + $\sigma_{\rm T}$ + $\sigma_{\rm R}$ model recovers the input parameters; neglecting RS in this model produces biases of $\lesssim0.5\sigma$ for $m_{\rm e}$ and $\lesssim1.2\sigma$ for $\alpha$.

\paragraph{\textbf{\textit{For LiteBIRD + Planck:}}}
The corresponding biases for LiteBIRD + Planck are listed in Tables~\ref{table:bias_params_LB_Planck} (for $m_{\rm e}$) and \ref{table:bias_params_alpha_LB_Planck} (for $\alpha$) in Appendix~\ref{app:B}. The corresponding biases are also shown in Figures~\ref{fig:prameter_bias_me_LB_Planck} and ~\ref{fig:prameter_bias_alpha_LB_Planck} respectively. The pattern is the same as for PICO, but the significances are smaller because the uncertainties are larger. Neglecting RS in a $\Lambda$CDM analysis of $\Lambda$CDM data gives biases of at most $0.13\sigma$, so at this sensitivity omitting RS is harmless for the standard model. In contrast, fitting
$\Lambda$CDM to data with a modified recombination history still gives strong biases as expected. 
When the cross-section scaling is neglected (data set for $\Lambda$CDM + $x_{\rm e} + \sigma_{\rm T} + \sigma_{\rm R}$ analysed with $\Lambda$CDM + $x_{\rm e}$), the scaling degeneracy again biases the electron mass: $m_{\rm e}/m_{\mathrm{e},0}$ is shifted by $\simeq -4\sigma$
($\simeq -2.1\%$, i.e. $m_{\rm e}/m_{\mathrm{e},0}\simeq 0.999$), together with $\simeq -4.5\sigma$ in $\Omega_\mathrm{b}h^2$ and $\simeq -4.9\sigma$ in $H_0$. 
For $\alpha$, the corresponding biases remain $\lesssim 2\sigma$. Once the correct model is used, neglecting RS biases all parameters by less than $0.15\sigma$.

\section{Conclusions}\label{sec:conclusion}

In this work, we have studied Rayleigh scattering of CMB photons as a probe of the cosmological
recombination process, with particular focus on recombination scenarios modified by a variation of the
electron mass, $m_{\rm e}$, or the fine-structure constant, $\alpha$. We modelled the RS signal with modified
versions of \textsc{Recfast++} and \textsc{CAMB}, and used a multi-frequency Fisher analysis including
realistic noise and foregrounds to forecast parameter constraints and biases for PICO and LiteBIRD + Planck.

Increasing $m_{\rm e}$ or $\alpha$ increases atomic binding energies, so recombination occurs earlier. Neutral hydrogen and helium therefore appear at higher redshift, which shifts the onset of RS opacity and both the Thomson and Rayleigh visibility functions to earlier times.  
Because the RS contribution scales as $\nu^4$, these
changes imprint a frequency-dependent signature on the $TT, TE$ and especially the $EE$ spectra, which is most
pronounced in the highest-frequency channels.

For PICO, adding RS to the primary CMB improves the $\Lambda$CDM constraints by $\lesssim 7\%$, since
RS does not open any new degree of freedom when the recombination history is fixed. This finding is in good agreement with \citep{Beringue:2021}. The gain is substantially greater once the recombination history is allowed to vary. In the $\Lambda$CDM + $x_{\rm e}$ model, the uncertainties in
$m_{\rm e}/m_{\mathrm{e},0}$, $\alpha/\alpha_0$ and $H_0$ shrink by $\simeq 31\%$, and the error bars of other parameters also improve,
because RS samples the timing of recombination independently at each frequency. When the variation of $m_{\rm e}$
is propagated consistently to the Thomson and Rayleigh cross-sections, the primary CMB alone suffers from an
almost exact scaling degeneracy between $m_{\rm e}$, $\Omega_\mathrm{b}h^2$, $\Omega_\mathrm{c}h^2$ and $H_0$. RS reduces the extent of this
degeneracy and improves all four parameters by $\simeq 30\%$. For $\alpha$, there is no such degeneracy and the
gain is $\lesssim 8\%$. For LiteBIRD + Planck the improvements are below $2\%$, except along the same
$m_{\rm e}$ scaling degeneracy, where they reach $55$--$58\%$ because the primary-CMB constraints are much weaker.

The bias analysis shows that RS must be modelled in future analyses. For PICO, neglecting RS biases the
$\Lambda$CDM parameters by up to $\simeq 1\sigma$, while for LiteBIRD + Planck
the biases remain below $0.15\sigma$. A modified recombination history analysed with $\Lambda$CDM produces
very large biases for both experiments. Most importantly, if the
effect of a varying $m_{\rm e}$ on the scattering cross-sections is neglected, the scaling degeneracy drives the
inferred electron mass back to $m_{\rm e}/m_{\mathrm{e},0}\simeq 1.0$. A genuine $2\%$ variation of $m_{\rm e}$ would then be completely missed because the data is fitted by readjusting the other parameters. A variation of $\alpha$, in contrast, is still recovered close to its input value. Only when the variation is modelled consistently,
including $\sigma_\mathrm{T}$ and $\sigma_\mathrm{R}$, are the input parameters recovered, with residual biases
from neglecting RS of $\lesssim 1\sigma$.

Our forecasts rely on the Fisher formalism, and the bias estimates use a
linear expansion. Biases of tens of $\sigma$ should therefore be read as signs of severe mis-modelling rather
than as precise values, and the constraints along the near-degenerate $m_{\rm e}$ direction would benefit from a full
likelihood analysis. More accurate foreground modelling and calibration parameters, and exploiting the
high-frequency channels of future experiments, will be essential to fully realise the potential of RS as an
independent probe of recombination physics and of proposed modified recombination-based solutions to the Hubble
tension.

\begin{acknowledgments}
DA, RTGS, CHM and JARM acknowledge financial support from the projects PID2023-150398NBI00 and PID2023-151567NB-I00, funded by the Ministry of Science, Innovation and Universities of Spain, AEI/10.13039\\/501100011033 and by FEDER, EU. This work is part of grant CEX2025-001609-S, awarded to the Instituto de Astrofisica de Canarias under the Severo Ochoa Centre of Excellence program and funded by MICIU/AEI/10.13039/501100011033. JC acknowledges the support of the Fundación Occident and the Instituto de Astrofísica de Canarias under the Visiting Researcher Programme 2025-2028 agreed between both institutions. EV acknowledges that the project that gave rise to these results received the support of a fellowship from the a ``Caixa” Foundation (ID B006373). The fellowship code is LCF/BQ/DI24/12070001. JC was also supported by the UKSA grant: LiteBIRD UK ST/Y005945/1.
We thank the Simons Observatory collaboration for making the \fgspectra\ code publicly available, and Antony Lewis for providing the modified version of \camb\ to include the Rayleigh scattering. The paper uses the python packages, numpy \citep{Numpy:2020}, matplotlib \citep{Matplotlib:2007}, scipy \citep{scipy:2020}.

\end{acknowledgments}
\small 

\vspace{3mm}

\bibliographystyle{JHEP}
\bibliography{Lit-2025, refernce}

@ARTICLE{Battaglia2012,
   author = {{Battaglia}, N. and {Bond}, J.~R. and {Pfrommer}, C. and {Sievers}, J.~L.
	},
    title = "{On the Cluster Physics of Sunyaev-Zel'dovich and X-Ray Surveys. II. Deconstructing the Thermal SZ Power Spectrum}",
  journal = {\apj},
archivePrefix = "arXiv",
   eprint = {1109.3711},
     year = 2012,
    month = oct,
   volume = 758,
      eid = {75},
    pages = {75},
      doi = {10.1088/0004-637X/758/2/75},
   adsurl = {http://adsabs.harvard.edu/abs/2012ApJ...758...75B}
}

@ARTICLE{Chluba2016CosmoSpec,
   author = {{Chluba}, J. and {Ali-Ha{\"i}moud}, Y.},
    title = "{COSMOSPEC: fast and detailed computation of the cosmological recombination radiation from hydrogen and helium}",
  journal = {\mnras},
archivePrefix = "arXiv",
   eprint = {1510.03877},
     year = 2016,
    month = mar,
   volume = 456,
    pages = {3494-3508},
      doi = {10.1093/mnras/stv2691},
   adsurl = {http://adsabs.harvard.edu/abs/2016MNRAS.456.3494C}
}

@ARTICLE{Chluba2015PMF,
   author = {{Chluba}, J. and {Paoletti}, D. and {Finelli}, F. and {Rubino-Martin}, J.-A.
	},
    title = "{Effect of primordial magnetic fields on the ionization history}",
  journal = {ArXiv:1503.04827},
archivePrefix = "arXiv",
   eprint = {1503.04827},
     year = 2015,
    month = mar,
   adsurl = {http://adsabs.harvard.edu/abs/2015arXiv150304827C}
}

@ARTICLE{Chluba2010b,
   author = {{Chluba}, J. and {Thomas}, R.~M.},
    title = "{Towards a complete treatment of the cosmological recombination problem}",
  journal = {\mnras},
archivePrefix = "arXiv",
   eprint = {1010.3631},
 primaryClass = "astro-ph.CO",
     year = 2011,
    month = apr,
   volume = 412,
    pages = {748-764},
      doi = {10.1111/j.1365-2966.2010.17940.x},
   adsurl = {http://adsabs.harvard.edu/abs/2011MNRAS.412..748C}
}

@ARTICLE{Scoccola2009,
    author = {{Sc{\'o}ccola}, C.~G. and {Landau}, S.~J. and {Vucetich}, H.
	},
    title = "{WMAP 5-year constraints on time variation of alpha and m\_e}",
  journal = {Memorie della Societ Astronomica Italiana},
archivePrefix = "arXiv",
   eprint = {0910.1083},
 primaryClass = "astro-ph.CO",
     year = 2009,
   volume = 80,
    pages = {814-+},
   adsurl = {http://adsabs.harvard.edu/abs/2009MmSAI..80..814S}
}

@ARTICLE{Kaplinghat1999,
   author = {{Kaplinghat}, M. and {Scherrer}, R.~J. and {Turner}, M.~S.},
    title = "{Constraining variations in the fine-structure constant with the cosmic microwave background}",
  journal = {\prd},
   eprint = {arXiv:astro-ph/9810133},
     year = 1999,
    month = jul,
   volume = 60,
   number = 2,
    pages = {023516-+},
      doi = {10.1103/PhysRevD.60.023516},
   adsurl = {http://adsabs.harvard.edu/abs/1999PhRvD..60b3516K}
}

@ARTICLE{Chluba2010,
   author = {{Chluba}, J. and {Vasil}, G.~M. and {Dursi}, L.~J.},
    title = "{Recombinations to the Rydberg states of hydrogen and their effect during the cosmological recombination epoch}",
  journal = {\mnras},
archivePrefix = "arXiv",
   eprint = {1003.4928},
 primaryClass = "astro-ph.CO",
     year = 2010,
    month = sep,
   volume = 407,
    pages = {599-612},
      doi = {10.1111/j.1365-2966.2010.16940.x},
   adsurl = {http://adsabs.harvard.edu/abs/2010MNRAS.407..599C}
}

@ARTICLE{SeagerRecfast1999,
   author = {{Seager}, S. and {Sasselov}, D.~D. and {Scott}, D.},
    title = "{A New Calculation of the Recombination Epoch}",
  journal = {\apjl},
     year = 1999,
    month = sep,
   volume = 523,
    pages = {L1-L5},
      doi = {10.1086/312250},
   adsurl = {http://cdsads.u-strasbg.fr/cgi-bin/nph-bib_query?bibcode=1999ApJ...523L...1S&db_key=AST}
}

@ARTICLE{Seager2000,
   author = {{Seager}, S. and {Sasselov}, D.~D. and {Scott}, D.},
    title = "{How Exactly Did the Universe Become Neutral?}",
  journal = {\apjs},
   eprint = {astro-ph/9912182},
     year = 2000,
    month = jun,
   volume = 128,
    pages = {407-430},
      doi = {10.1086/313388},
   adsurl = {http://cdsads.u-strasbg.fr/cgi-bin/nph-bib_query?bibcode=2000ApJS..128..407S&db_key=AST}
}

@ARTICLE{Zeldovich68,
   author = {{Zeldovich}, Y.~B. and {Kurt}, V.~G. and {Syunyaev}, R.~A.},
    title = "{Recombination of Hydrogen in the Hot Model of the Universe}",
  journal = {Zhurnal Eksperimental noi i Teoreticheskoi Fiziki},
     year = 1968,
    month = jul,
   volume = 55,
    pages = {278-286},
   adsurl = {http://adsabs.harvard.edu/cgi-bin/nph-bib_query?bibcode=1968ZhETF..55..278Z&db_key=AST}
}

@ARTICLE{CAMB,
   author = {{Lewis}, A. and {Challinor}, A. and {Lasenby}, A.},
    title = "{Efficient Computation of Cosmic Microwave Background Anisotropies in Closed Friedmann-Robertson-Walker Models}",
  journal = {\apj},
  eprinttype = "arxiv",
  eprint = {astro-ph/9911177},
     year = 2000,
    month = aug,
   volume = 538,
    pages = {473-476},
      doi = {10.1086/309179},
   adsurl = {http://adsabs.harvard.edu/abs/2000ApJ...538..473L}
}

@ARTICLE{Battye2001,
   author = {{Battye}, R.~A. and {Crittenden}, R. and {Weller}, J.},
    title = "{Cosmic concordance and the fine structure constant}",
  journal = {\prd},
   eprint = {astro-ph/0008265},
     year = 2001,
    month = feb,
   volume = 63,
   number = 4,
      eid = {043505},
    pages = {043505},
      doi = {10.1103/PhysRevD.63.043505},
   adsurl = {http://adsabs.harvard.edu/abs/2001PhRvD..63d3505B}
}

@article{Hoshiya2023,
author = {{Hoshiya}, Kouki and {Toda}, Yo},
        title = "{Electron mass variation from dark sector interactions and compatibility with cosmological observations}",
      journal = {\prd},
         year = 2023,
        month = feb,
       volume = {107},
       number = {4},
          eid = {043505},
        pages = {043505},
          doi = {10.1103/PhysRevD.107.043505},
archivePrefix = {arXiv},
       eprint = {2202.07714},
 primaryClass = {astro-ph.CO},
       adsurl = {https://ui.adsabs.harvard.edu/abs/2023PhRvD.107d3505H}
}

@ARTICLE{Sekiguchi2021PhRvD,
       author = {{Sekiguchi}, Toyokazu and {Takahashi}, Tomo},
        title = "{Early recombination as a solution to the H$_{0}$ tension}",
      journal = {\prd},
         year = 2021,
        month = apr,
       volume = {103},
       number = {8},
          eid = {083507},
        pages = {083507},
          doi = {10.1103/PhysRevD.103.083507},
archivePrefix = {arXiv},
       eprint = {2007.03381},
 primaryClass = {astro-ph.CO},
       adsurl = {https://ui.adsabs.harvard.edu/abs/2021PhRvD.103h3507S}
}

@ARTICLE{Knox2020,
       author = {{Knox}, L. and {Millea}, M.},
        title = "{Hubble constant hunter's guide}",
      journal = {\prd},
         year = 2020,
        month = feb,
       volume = {101},
       number = {4},
          eid = {043533},
        pages = {043533},
          doi = {10.1103/PhysRevD.101.043533},
archivePrefix = {arXiv},
       eprint = {1908.03663},
 primaryClass = {astro-ph.CO},
       adsurl = {https://ui.adsabs.harvard.edu/abs/2020PhRvD.101d3533K}
}

@ARTICLE{DiValentino2021,
       author = {{Di Valentino}, Eleonora and {Mena}, Olga and {Pan}, Supriya and {Visinelli}, Luca and {Yang}, Weiqiang and {Melchiorri}, Alessandro and {Mota}, David F. and {Riess}, Adam G. and {Silk}, Joseph},
        title = "{In the realm of the Hubble tension-a review of solutions}",
      journal = {Classical and Quantum Gravity},
         year = 2021,
        month = jul,
       volume = {38},
       number = {15},
          eid = {153001},
        pages = {153001},
          doi = {10.1088/1361-6382/ac086d},
archivePrefix = {arXiv},
       eprint = {2103.01183},
 primaryClass = {astro-ph.CO},
       adsurl = {https://ui.adsabs.harvard.edu/abs/2021CQGra..38o3001D}
}

@ARTICLE{Lewis2013Rayleigh,
       author = {{Lewis}, Antony},
        title = "{Rayleigh scattering: blue sky thinking for future CMB observations}",
      journal = {\jcap},
         year = 2013,
        month = aug,
       volume = {2013},
       number = {8},
          eid = {053},
        pages = {053},
          doi = {10.1088/1475-7516/2013/08/053},
archivePrefix = {arXiv},
       eprint = {1307.8148},
 primaryClass = {astro-ph.CO},
       adsurl = {https://ui.adsabs.harvard.edu/abs/2013JCAP...08..053L}
}

@ARTICLE{scipy:2020,
       author = {{Virtanen}, Pauli and {Gommers}, Ralf and {Oliphant}, Travis E. and {Haberland}, Matt and {Reddy}, Tyler and {Cournapeau}, David and {Burovski}, Evgeni and {Peterson}, Pearu and {Weckesser}, Warren and {Bright}, Jonathan and {van der Walt}, St{\'e}fan J. and {Brett}, Matthew and {Wilson}, Joshua and {Millman}, K. Jarrod and {Mayorov}, Nikolay and {Nelson}, Andrew R.~J. and {Jones}, Eric and {Kern}, Robert and {Larson}, Eric and {Carey}, C.~J. and {Polat}, {\.I}lhan and {Feng}, Yu and {Moore}, Eric W. and {VanderPlas}, Jake and {Laxalde}, Denis and {Perktold}, Josef and {Cimrman}, Robert and {Henriksen}, Ian and {Quintero}, E.~A. and {Harris}, Charles R. and {Archibald}, Anne M. and {Ribeiro}, Ant{\^o}nio H. and {Pedregosa}, Fabian and {van Mulbregt}, Paul and {SciPy 1. 0 Contributors}},
        title = "{SciPy 1.0: fundamental algorithms for scientific computing in Python}",
      journal = {Nature Medicine},
         year = 2020,
        month = feb,
       volume = {17},
        pages = {261-272},
          doi = {10.1038/s41592-019-0686-2},
archivePrefix = {arXiv},
       eprint = {1907.10121},
 primaryClass = {cs.MS},
       adsurl = {https://ui.adsabs.harvard.edu/abs/2020NatMe..17..261V}
}

@ARTICLE{Matplotlib:2007,
       author = {{Hunter}, John D.},
        title = "{Matplotlib: A 2D Graphics Environment}",
      journal = {Computing in Science and Engineering},
         year = 2007,
        month = jan,
       volume = {9},
       number = {3},
        pages = {90-95},
          doi = {10.1109/MCSE.2007.55},
       adsurl = {https://ui.adsabs.harvard.edu/abs/2007CSE.....9...90H}
}

@ARTICLE{Numpy:2020,
       author = {{Harris}, Charles R. and {Millman}, K. Jarrod and {van der Walt}, St{\'e}fan J. and {Gommers}, Ralf and {Virtanen}, Pauli and {Cournapeau}, David and {Wieser}, Eric and {Taylor}, Julian and {Berg}, Sebastian and {Smith}, Nathaniel J. and {Kern}, Robert and {Picus}, Matti and {Hoyer}, Stephan and {van Kerkwijk}, Marten H. and {Brett}, Matthew and {Haldane}, Allan and {del R{\'\i}o}, Jaime Fern{\'a}ndez and {Wiebe}, Mark and {Peterson}, Pearu and {G{\'e}rard-Marchant}, Pierre and {Sheppard}, Kevin and {Reddy}, Tyler and {Weckesser}, Warren and {Abbasi}, Hameer and {Gohlke}, Christoph and {Oliphant}, Travis E.},
        title = "{Array programming with NumPy}",
      journal = {\nat},
         year = 2020,
        month = sep,
       volume = {585},
       number = {7825},
        pages = {357-362},
          doi = {10.1038/s41586-020-2649-2},
archivePrefix = {arXiv},
       eprint = {2006.10256},
 primaryClass = {cs.MS},
       adsurl = {https://ui.adsabs.harvard.edu/abs/2020Natur.585..357H}
}

@article{Battaglia_2013,
doi = {10.1088/0004-637X/776/2/83},
url = {https://doi.org/10.1088/0004-637X/776/2/83},
year = {2013},
month = {oct},
publisher = {The American Astronomical Society},
volume = {776},
number = {2},
pages = {83},
author = {Battaglia, N. and Natarajan, A. and Trac, H. and Cen, R. and Loeb, A.},
title = {REIONIZATION ON LARGE SCALES. III. PREDICTIONS FOR LOW-$\ell$ COSMIC MICROWAVE BACKGROUND POLARIZATION AND HIGH-$\ell$ KINETIC SUNYAEV–ZEL'DOVICH OBSERVABLES},
journal = {The Astrophysical Journal},

}

@ARTICLE{Seto:2023,
       author = {{Seto}, Osamu and {Toda}, Yo},
        title = "{Big bang nucleosynthesis constraints on varying electron mass solution to the Hubble tension}",
      journal = {\prd},
         year = 2023,
        month = apr,
       volume = {107},
       number = {8},
          eid = {083512},
        pages = {083512},
          doi = {10.1103/PhysRevD.107.083512},
archivePrefix = {arXiv},
       eprint = {2206.13209},
 primaryClass = {astro-ph.CO},
       adsurl = {https://ui.adsabs.harvard.edu/abs/2023PhRvD.107h3512S}
}

@ARTICLE{PICO:2019,
       author = {{Hanany}, Shaul and {Alvarez}, Marcelo and {Artis}, Emmanuel and {Ashton}, Peter and {Aumont}, Jonathan and {Aurlien}, Ragnhild and {Banerji}, Ranajoy and {Barreiro}, R. Belen and {Bartlett}, James G. and {Basak}, Soumen and {Battaglia}, Nick and {Bock}, Jamie and {Boddy}, Kimberly K. and {Bonato}, Matteo and {Borrill}, Julian and {Bouchet}, Fran{\c{c}}ois and {Boulanger}, Fran{\c{c}}ois and {Burkhart}, Blakesley and {Chluba}, Jens and {Chuss}, David and {Clark}, Susan E. and {Cooperrider}, Joelle and {Crill}, Brendan P. and {De Zotti}, Gianfranco and {Delabrouille}, Jacques and {Di Valentino}, Eleonora and {Didier}, Joy and {Dor{\'e}}, Olivier and {Eriksen}, Hans K. and {Errard}, Josquin and {Essinger-Hileman}, Tom and {Feeney}, Stephen and {Filippini}, Jeffrey and {Fissel}, Laura and {Flauger}, Raphael and {Fuskeland}, Unni and {Gluscevic}, Vera and {Gorski}, Krzysztof M. and {Green}, Dan and {Hensley}, Brandon and {Herranz}, Diego and {Hill}, J. Colin and {Hivon}, Eric and {Hlo{\v{z}}ek}, Ren{\'e}e and {Hubmayr}, Johannes and {Johnson}, Bradley R. and {Jones}, William and {Jones}, Terry and {Knox}, Lloyd and {Kogut}, Al and {L{\'o}pez-Caniego}, Marcos and {Lawrence}, Charles and {Lazarian}, Alex and {Li}, Zack and {Madhavacheril}, Mathew and {Melin}, Jean-Baptiste and {Meyers}, Joel and {Murray}, Calum and {Negrello}, Mattia and {Novak}, Giles and {O'Brient}, Roger and {Paine}, Christopher and {Pearson}, Tim and {Pogosian}, Levon and {Pryke}, Clem and {Puglisi}, Giuseppe and {Remazeilles}, Mathieu and {Rocha}, Graca and {Schmittfull}, Marcel and {Scott}, Douglas and {Shirron}, Peter and {Stephens}, Ian and {Sutin}, Brian and {Tomasi}, Maurizio and {Trangsrud}, Amy and {van Engelen}, Alexander and {Vansyngel}, Flavien and {Wehus}, Ingunn K. and {Wen}, Qi and {Xu}, Siyao and {Young}, Karl and {Zonca}, Andrea},
        title = "{PICO: Probe of Inflation and Cosmic Origins}",
      journal = {arXiv e-prints},
         year = 2019,
        month = feb,
          eid = {arXiv:1902.10541},
        pages = {arXiv:1902.10541},
          doi = {10.48550/arXiv.1902.10541},
archivePrefix = {arXiv},
       eprint = {1902.10541},
 primaryClass = {astro-ph.IM},
       adsurl = {https://ui.adsabs.harvard.edu/abs/2019arXiv190210541H}
}

@article{TODA2025102035,
title = {Constraints on the simultaneous variation of the fine structure constant and the electron mass in light of DESI BAO data},
journal = {Physics of the Dark Universe},
volume = {49},
pages = {102035},
year = {2025},
issn = {2212-6864},
doi = {https://doi.org/10.1016/j.dark.2025.102035},
url = {https://www.sciencedirect.com/science/article/pii/S2212686425002286},
author = {Yo Toda and Osamu Seto}
}

@ARTICLE{SGhsosh:2025,
       author = {{Ghosh}, Subhajit and {Boddy}, Kimberly K. and {Yu}, Tien-Tien},
        title = "{Early Universe Constraints on Variations in Fundamental Constants Induced by Ultralight Scalar Dark Matter}",
      journal = {arXiv e-prints},
         year = 2025,
        month = nov,
          eid = {arXiv:2511.14532},
        pages = {arXiv:2511.14532},
          doi = {10.48550/arXiv.2511.14532},
archivePrefix = {arXiv},
       eprint = {2511.14532},
 primaryClass = {astro-ph.CO},
       adsurl = {https://ui.adsabs.harvard.edu/abs/2025arXiv251114532G}
}

@article{Toda:2024PDU,
title = {Combining pre- and post-recombination new physics to address cosmological tensions: Case study with varying electron mass and sign-switching cosmological constant},
journal = {Physics of the Dark Universe},
volume = {46},
pages = {101676},
year = {2024},
issn = {2212-6864},
doi = {https://doi.org/10.1016/j.dark.2024.101676},
url = {https://www.sciencedirect.com/science/article/pii/S2212686424002589},
author = {Yo Toda and William Giarè and Emre Özülker and Eleonora {Di Valentino} and Sunny Vagnozzi}
}

@article{Hart:2017,
    author = {Hart, Luke and Chluba, Jens},
    title = {New constraints on time-dependent variations of fundamental constants using Planck data},
    journal = {Monthly Notices of the Royal Astronomical Society},
    volume = {474},
    number = {2},
    pages = {1850-1861},
    year = {2017},
    month = {10},
    issn = {0035-8711},
    doi = {10.1093/mnras/stx2783},
    url = {https://doi.org/10.1093/mnras/stx2783},
    eprint = {https://academic.oup.com/mnras/article-pdf/474/2/1850/22636771/stx2783.pdf},
}

@ARTICLE{Schoneberg:2025JCAP,
       author = {{Sch{\"o}neberg}, Nils and {Vacher}, L{\'e}o},
        title = "{The mass effect {\textemdash} variations of the electron mass and their impact on cosmology}",
      journal = {\jcap},
         year = 2025,
        month = mar,
       volume = {2025},
       number = {3},
          eid = {004},
        pages = {004},
          doi = {10.1088/1475-7516/2025/03/004},
archivePrefix = {arXiv},
       eprint = {2407.16845},
 primaryClass = {astro-ph.CO},
       adsurl = {https://ui.adsabs.harvard.edu/abs/2025JCAP...03..004S}
}

@ARTICLE{OColgain:2025MNRAS,
       author = {{{\'O} Colg{\'a}in}, Eoin and {Sheikh-Jabbari}, M.~M.},
        title = "{DESI and SNe: dynamical dark energy, {\ensuremath{\Omega}}$_{m}$ tension or systematics?}",
      journal = {\mnras},
         year = 2025,
        month = sep,
       volume = {542},
       number = {1},
        pages = {L24-L30},
          doi = {10.1093/mnrasl/slaf042},
archivePrefix = {arXiv},
       eprint = {2412.12905},
 primaryClass = {astro-ph.CO},
       adsurl = {https://ui.adsabs.harvard.edu/abs/2025MNRAS.542L..24O}
}

@ARTICLE{Zhao:2017NatAs,
       author = {{Zhao}, Gong-Bo and {Raveri}, Marco and {Pogosian}, Levon and {Wang}, Yuting and {Crittenden}, Robert G. and {Handley}, Will J. and {Percival}, Will J. and {Beutler}, Florian and {Brinkmann}, Jonathan and {Chuang}, Chia-Hsun and {Cuesta}, Antonio J. and {Eisenstein}, Daniel J. and {Kitaura}, Francisco-Shu and {Koyama}, Kazuya and {L'Huillier}, Benjamin and {Nichol}, Robert C. and {Pieri}, Matthew M. and {Rodriguez-Torres}, Sergio and {Ross}, Ashley J. and {Rossi}, Graziano and {S{\'a}nchez}, Ariel G. and {Shafieloo}, Arman and {Tinker}, Jeremy L. and {Tojeiro}, Rita and {Vazquez}, Jose A. and {Zhang}, Hanyu},
        title = "{Dynamical dark energy in light of the latest observations}",
      journal = {Nature Astronomy},
         year = 2017,
        month = aug,
       volume = {1},
        pages = {627-632},
          doi = {10.1038/s41550-017-0216-z},
archivePrefix = {arXiv},
       eprint = {1701.08165},
 primaryClass = {astro-ph.CO},
       adsurl = {https://ui.adsabs.harvard.edu/abs/2017NatAs...1..627Z}
}

@article{Pogosian:2021,
    author = "Pogosian, Levon and Raveri, Marco and Koyama, Kazuya and Martinelli, Matteo and Silvestri, Alessandra and Zhao, Gong-Bo and Li, Jian and Peirone, Simone and Zucca, Alex",
    title = "{Imprints of cosmological tensions in reconstructed gravity}",
    eprint = "2107.12992",
    archivePrefix = "arXiv",
    primaryClass = "astro-ph.CO",
    doi = "10.1038/s41550-022-01808-7",
    journal = "Nature Astron.",
    volume = "6",
    number = "12",
    pages = "1484--1490",
    year = "2022"
}

@article{DiValentino:2017,
    author = "Di Valentino, Eleonora and Melchiorri, Alessandro and Mena, Olga",
    title = "{Can interacting dark energy solve the $H_0$ tension?}",
    eprint = "1704.08342",
    archivePrefix = "arXiv",
    primaryClass = "astro-ph.CO",
    doi = "10.1103/PhysRevD.96.043503",
    journal = "Phys. Rev. D",
    volume = "96",
    number = "4",
    pages = "043503",
    year = "2017"
}

@ARTICLE{Adame:2025JCAP,
       author = {{Adame}, A.~G. and {Aguilar}, J. and {Ahlen}, S. and {Alam}, S. and {Alexander}, D.~M. and {Alvarez}, M. and {Alves}, O. and {Anand}, A. and {Andrade}, U. and {Armengaud}, E. and {Avila}, S. and {Aviles}, A. and {Awan}, H. and {Bahr-Kalus}, B. and {Bailey}, S. and {Baltay}, C. and {Bault}, A. and {Behera}, J. and {BenZvi}, S. and {Bera}, A. and {Beutler}, F. and {Bianchi}, D. and {Blake}, C. and {Blum}, R. and {Brieden}, S. and {Brodzeller}, A. and {Brooks}, D. and {Buckley-Geer}, E. and {Burtin}, E. and {Calderon}, R. and {Canning}, R. and {Carnero Rosell}, A. and {Cereskaite}, R. and {Cervantes-Cota}, J.~L. and {Chabanier}, S. and {Chaussidon}, E. and {Chaves-Montero}, J. and {Chen}, S. and {Chen}, X. and {Claybaugh}, T. and {Cole}, S. and {Cuceu}, A. and {Davis}, T.~M. and {Dawson}, K. and {de la Macorra}, A. and {de Mattia}, A. and {Deiosso}, N. and {Dey}, A. and {Dey}, B. and {Ding}, Z. and {Doel}, P. and {Edelstein}, J. and {Eftekharzadeh}, S. and {Eisenstein}, D.~J. and {Elliott}, A. and {Fagrelius}, P. and {Fanning}, K. and {Ferraro}, S. and {Ereza}, J. and {Findlay}, N. and {Flaugher}, B. and {Font-Ribera}, A. and {Forero-S{\'a}nchez}, D. and {Forero-Romero}, J.~E. and {Frenk}, C.~S. and {Garcia-Quintero}, C. and {Gazta{\~n}aga}, E. and {Gil-Mar{\'\i}n}, H. and {Gontcho a Gontcho}, S. and {Gonzalez-Morales}, A.~X. and {Gonzalez-Perez}, V. and {Gordon}, C. and {Green}, D. and {Gruen}, D. and {Gsponer}, R. and {Gutierrez}, G. and {Guy}, J. and {Hadzhiyska}, B. and {Hahn}, C. and {Hanif}, M.~M.~S. and {Herrera-Alcantar}, H.~K. and {Honscheid}, K. and {Howlett}, C. and {Huterer}, D. and {Ir{\v{s}}i{\v{c}}}, V. and {Ishak}, M. and {Juneau}, S. and {Kara{\c{c}}ayl{\i}}, N.~G. and {Kehoe}, R. and {Kent}, S. and {Kirkby}, D. and {Kremin}, A. and {Krolewski}, A. and {Lai}, Y. and {Lan}, T.-W. and {Landriau}, M. and {Lang}, D. and {Lasker}, J. and {Le Goff}, J.~M. and {Le Guillou}, L. and {Leauthaud}, A. and {Levi}, M.~E. and {Li}, T.~S. and {Linder}, E. and {Lodha}, K. and {Magneville}, C. and {Manera}, M. and {Margala}, D. and {Martini}, P. and {Maus}, M. and {McDonald}, P. and {Medina-Varela}, L. and {Meisner}, A. and {Mena-Fern{\'a}ndez}, J. and {Miquel}, R. and {Moon}, J. and {Moore}, S. and {Moustakas}, J. and {Mueller}, E. and {Mu{\~n}oz-Guti{\'e}rrez}, A. and {Myers}, A.~D. and {Nadathur}, S. and {Napolitano}, L. and {Neveux}, R. and {Newman}, J.~A. and {Nguyen}, N.~M. and {Nie}, J. and {Niz}, G. and {Noriega}, H.~E. and {Padmanabhan}, N. and {Paillas}, E. and {Palanque-Delabrouille}, N. and {Pan}, J. and {Penmetsa}, S. and {Percival}, W.~J. and {Pieri}, M.~M. and {Pinon}, M. and {Poppett}, C. and {Porredon}, A. and {Prada}, F. and {P{\'e}rez-Fern{\'a}ndez}, A. and {P{\'e}rez-R{\`a}fols}, I. and {Rabinowitz}, D. and {Raichoor}, A. and {Ram{\'\i}rez-P{\'e}rez}, C. and {Ramirez-Solano}, S. and {Rashkovetskyi}, M. and {Ravoux}, C. and {Rezaie}, M. and {Rich}, J. and {Rocher}, A. and {Rockosi}, C. and {Roe}, N.~A. and {Rosado-Marin}, A. and {Ross}, A.~J. and {Rossi}, G. and {Ruggeri}, R. and {Ruhlmann-Kleider}, V. and {Samushia}, L. and {Sanchez}, E. and {Saulder}, C. and {Schlafly}, E.~F. and {Schlegel}, D. and {Schubnell}, M. and {Seo}, H. and {Shafieloo}, A. and {Sharples}, R. and {Silber}, J. and {Slosar}, A. and {Smith}, A. and {Sprayberry}, D. and {Tan}, T. and {Tarl{\'e}}, G. and {Taylor}, P. and {Trusov}, S. and {Ure{\~n}a-L{\'o}pez}, L.~A. and {Vaisakh}, R. and {Valcin}, D. and {Valdes}, F. and {Vargas-Maga{\~n}a}, M. and {Verde}, L. and {Walther}, M. and {Wang}, B. and {Wang}, M.~S. and {Weaver}, B.~A. and {Weaverdyck}, N. and {Wechsler}, R.~H. and {Weinberg}, D.~H. and {White}, M. and {Yu}, J. and {Yu}, Y. and {Yuan}, S. and {Y{\`e}che}, C. and {Zaborowski}, E.~A. and {Zarrouk}, P. and {Zhang}, H. and {Zhao}, C. and {Zhao}, R. and {Zhou}, R. and {Zhuang}, T.},
        title = "{DESI 2024 VI: cosmological constraints from the measurements of baryon acoustic oscillations}",
      journal = {\jcap},
         year = 2025,
        month = feb,
       volume = {2025},
       number = {2},
          eid = {021},
        pages = {021},
          doi = {10.1088/1475-7516/2025/02/021},
archivePrefix = {arXiv},
       eprint = {2404.03002},
 primaryClass = {astro-ph.CO},
       adsurl = {https://ui.adsabs.harvard.edu/abs/2025JCAP...02..021A}
}

@ARTICLE{Kamionkowski:2023ARNPS,
       author = {{Kamionkowski}, Marc and {Riess}, Adam G.},
        title = "{The Hubble Tension and Early Dark Energy}",
      journal = {Annual Review of Nuclear and Particle Science},
         year = 2023,
        month = sep,
       volume = {73},
        pages = {153-180},
          doi = {10.1146/annurev-nucl-111422-024107},
archivePrefix = {arXiv},
       eprint = {2211.04492},
 primaryClass = {astro-ph.CO},
       adsurl = {https://ui.adsabs.harvard.edu/abs/2023ARNPS..73..153K}
}

@ARTICLE{wang:2025EPJC,
       author = {{Wang}, Deng and {Mota}, David},
        title = "{Did DESI DR2 truly reveal dynamical dark energy?}",
      journal = {European Physical Journal C},
         year = 2025,
        month = nov,
       volume = {85},
       number = {11},
          eid = {1356},
        pages = {1356},
          doi = {10.1140/epjc/s10052-025-15076-y},
archivePrefix = {arXiv},
       eprint = {2504.15222},
 primaryClass = {astro-ph.CO},
       adsurl = {https://ui.adsabs.harvard.edu/abs/2025EPJC...85.1356W}
}

@article{Ruchika:2023ugh,
    author = "Ruchika and Rathore, Himansh and Roy Choudhury, Shouvik and Rentala, Vikram",
    title = "{A gravitational constant transition within cepheids as supernovae calibrators can solve the Hubble tension}",
    eprint = "2306.05450",
    archivePrefix = "arXiv",
    primaryClass = "astro-ph.CO",
    doi = "10.1088/1475-7516/2024/06/056",
    journal = "JCAP",
    volume = "06",
    pages = "056",
    year = "2024"
}

@article{RoyChoudhury:2020dmd,
    author = "Roy Choudhury, Shouvik and Hannestad, Steen and Tram, Thomas",
    title = "{Updated constraints on massive neutrino self-interactions from cosmology in light of the $H_0$ tension}",
    eprint = "2012.07519",
    archivePrefix = "arXiv",
    primaryClass = "astro-ph.CO",
    doi = "10.1088/1475-7516/2021/03/084",
    journal = "JCAP",
    volume = "03",
    pages = "084",
    year = "2021"
}

@ARTICLE{Chiang:2018,
       author = {{Chiang}, Chi-Ting and {Slosar}, An{\v{z}}e},
        title = "{Inferences of $H_0$ in presence of a non-standard recombination}",
      journal = {arXiv e-prints},
         year = 2018,
        month = nov,
          eid = {arXiv:1811.03624},
        pages = {arXiv:1811.03624},
          doi = {10.48550/arXiv.1811.03624},
archivePrefix = {arXiv},
       eprint = {1811.03624},
 primaryClass = {astro-ph.CO},
       adsurl = {https://ui.adsabs.harvard.edu/abs/2018arXiv181103624C}
}

@ARTICLE{Jedamzik:2020PhRvL,
       author = {{Jedamzik}, Karsten and {Pogosian}, Levon},
        title = "{Relieving the Hubble Tension with Primordial Magnetic Fields}",
      journal = {\prl},
         year = 2020,
        month = oct,
       volume = {125},
       number = {18},
          eid = {181302},
        pages = {181302},
          doi = {10.1103/PhysRevLett.125.181302},
archivePrefix = {arXiv},
       eprint = {2004.09487},
 primaryClass = {astro-ph.CO},
       adsurl = {https://ui.adsabs.harvard.edu/abs/2020PhRvL.125r1302J}
}

@ARTICLE{Mortsell:2022,
       author = {{M{\"o}rtsell}, Edvard and {Goobar}, Ariel and {Johansson}, Joel and {Dhawan}, Suhail},
        title = "{The Hubble Tension Revisited: Additional Local Distance Ladder Uncertainties}",
      journal = {\apj},
         year = 2022,
        month = aug,
       volume = {935},
       number = {1},
          eid = {58},
        pages = {58},
          doi = {10.3847/1538-4357/ac7c19},
archivePrefix = {arXiv},
       eprint = {2106.09400},
 primaryClass = {astro-ph.CO},
       adsurl = {https://ui.adsabs.harvard.edu/abs/2022ApJ...935...58M}
}

@ARTICLE{Mortsell:2022ApJ,
       author = {{M{\"o}rtsell}, Edvard and {Goobar}, Ariel and {Johansson}, Joel and {Dhawan}, Suhail},
        title = "{Sensitivity of the Hubble Constant Determination to Cepheid Calibration}",
      journal = {\apj},
         year = 2022,
        month = jul,
       volume = {933},
       number = {2},
          eid = {212},
        pages = {212},
          doi = {10.3847/1538-4357/ac756e},
archivePrefix = {arXiv},
       eprint = {2105.11461},
 primaryClass = {astro-ph.CO},
       adsurl = {https://ui.adsabs.harvard.edu/abs/2022ApJ...933..212M}
}

@ARTICLE{Pedreira:2024PhRvD,
       author = {{Pedreira}, Igor de O.~C. and {Benetti}, Micol and {Ferreira}, Elisa G.~M. and {Graef}, Leila L. and {Herold}, Laura},
        title = "{Visual tool for assessing tension-resolving models in the H$_{0}$-{\ensuremath{\sigma}}$_{8}$ plane}",
      journal = {\prd},
         year = 2024,
        month = may,
       volume = {109},
       number = {10},
          eid = {103525},
        pages = {103525},
          doi = {10.1103/PhysRevD.109.103525},
archivePrefix = {arXiv},
       eprint = {2311.04977},
 primaryClass = {astro-ph.CO},
       adsurl = {https://ui.adsabs.harvard.edu/abs/2024PhRvD.109j3525P}
}

@article{Karim:2024luk,
doi = {10.1088/1475-7516/2025/02/045},
url = {https://doi.org/10.1088/1475-7516/2025/02/045},
year = {2025},
month = {feb},
publisher = {IOP Publishing},
volume = {2025},
number = {02},
pages = {045},
author = {Karim, Tanveer and Singh, Sukhdeep and Rezaie, Mehdi and Eisenstein, Daniel and Hadzhiyska, Boryana and Speagle, Joshua S. and Aguilar, Jessica Nicole and Ahlen, Steven and Brooks, David and Claybaugh, Todd and de la Macorra, Axel and Ferraro, Simone and Forero-Romero, Jaime E. and Gaztañaga, Enrique and Gontcho, Satya Gontcho A. and Gutierrez, Gaston and Guy, Julien and Honscheid, Klaus and Juneau, Stephanie and Kirkby, David and Krolewski, Alex and Lambert, Andrew and Landriau, Martin and Levi, Michael and Meisner, Aaron and Miquel, Ramon and Moustakas, John and Muñoz-Gutiérrez, Andrea and Myers, Adam and Niz, Gustavo and Palanque-Delabrouille, Nathalie and Percival, Will and Prada, Francisco and Rossi, Graziano and Sanchez, Eusebio and Schlafly, Edward and Schlegel, David and Schubnell, Michael and Sprayberry, David and Tarlé, Gregory and Weaver, Benjamin Alan and Zou, Hu},
title = {Measuring $\sigma8$ using DESI Legacy Imaging Surveys Emission-Line galaxies and Planck CMB lensing, and the impact of dust on parameter inference},
journal = {Journal of Cosmology and Astroparticle Physics}
}

@article{Shaikh_ACT:2023skz,
    author = "Shaikh, S. and others",
    collaboration = "ACT, DES",
    title = "{Cosmology from cross-correlation of ACT-DR4 CMB lensing and DES-Y3 cosmic shear}",
    eprint = "2309.04412",
    archivePrefix = "arXiv",
    primaryClass = "astro-ph.CO",
    reportNumber = "FERMILAB-PUB-23-432-PPD",
    doi = "10.1093/mnras/stad3987",
    journal = "Mon. Not. Roy. Astron. Soc.",
    volume = "528",
    number = "2",
    pages = "2112--2135",
    year = "2024"
}

@article{Qu:2024sfu,
  title = {Atacama Cosmology Telescope DR6 and DESI: Structure growth measurements from the cross-correlation of DESI legacy imaging galaxies and CMB lensing from ACT DR6 and Planck PR4},
  author = {Qu, Frank J. and Hang, Qianjun and Farren, Gerrit and Bolliet, Boris and Aguilar, Jessica Nicole and Ahlen, Steven and Alam, Shadab and Brooks, David and Cai, Yan-Chuan and Calabrese, Erminia and Claybaugh, Todd and de la Macorra, Axel and Devlin, Mark J. and Doel, Peter and Embil-Villagra, Carmen and Ferraro, Simone and Font-Ribera, Andreu and Forero-Romero, Jaime E. and Gazta{\~n}aga, Enrique and Gluscevic, Vera and Gontcho, Satya Gontcho A. and Gutierrez, Gaston and Howlett, Cullan and Kehoe, Robert and Kim, Joshua and Kremin, Anthony and Lambert, Andrew and Landriau, Martin and Le Guillou, Laurent and Levi, Michael and Louis, Thibaut and Meisner, Aaron and Miquel, Ramon and Moustakas, John and Newman, Jeffrey A. and Niz, Gustavo and Peacock, John A. and Percival, Will and Poppett, Claire and Prada, Francisco and P{\'e}rez-R{\`a}fols, Ignasi and Rossi, Graziano and Sanchez, Eusebio and Schlegel, David and Sehgal, Neelima and Shaikh, Shabbir and Sherwin, Blake and Sif{\'o}n, Crist{\'o}bal and Schubnell, Michael and Sprayberry, David and Tarl{\'e}, Gregory and Weaver, Benjamin Alan and Wollack, Edward J. and Zou, Hu},
  journal = {Phys. Rev. D},
  volume = {111},
  issue = {10},
  pages = {103503},
  numpages = {35},
  year = {2025},
  month = {May},
  publisher = {American Physical Society},
  doi = {10.1103/PhysRevD.111.103503},
  url = {https://link.aps.org/doi/10.1103/PhysRevD.111.103503}
}

@ARTICLE{Erminia:2025,
       author = {{Calabrese}, Erminia and {Hill}, J. Colin and {Jense}, Hidde T. and {La Posta}, Adrien and {Abril-Cabezas}, Irene and {Addison}, Graeme E. and {Ade}, Peter A.~R. and {Aiola}, Simone and {Alford}, Tommy and {Alonso}, David and {Amiri}, Mandana and {An}, Rui and {Atkins}, Zachary and {Austermann}, Jason E. and {Barbavara}, Eleonora and {Barbieri}, Nicola and {Battaglia}, Nicholas and {Battistelli}, Elia Stefano and {Beall}, James A. and {Bean}, Rachel and {Beheshti}, Ali and {Beringue}, Benjamin and {Bhandarkar}, Tanay and {Biermann}, Emily and {Bolliet}, Boris and {Bond}, J. Richard and {Capalbo}, Valentina and {Carrero}, Felipe and {Chen}, Shi-Fan and {Chesmore}, Grace and {Cho}, Hsiao-mei and {Choi}, Steve K. and {Clark}, Susan E. and {Cothard}, Nicholas F. and {Coughlin}, Kevin and {Coulton}, William and {Crichton}, Devin and {Crowley}, Kevin T. and {Darwish}, Omar and {Devlin}, Mark J. and {Dicker}, Simon and {Duell}, Cody J. and {Duff}, Shannon M. and {Duivenvoorden}, Adriaan J. and {Dunkley}, Jo and {Dunner}, Rolando and {Embil Villagra}, Carmen and {Fankhanel}, Max and {Farren}, Gerrit S. and {Ferraro}, Simone and {Foster}, Allen and {Freundt}, Rodrigo and {Fuzia}, Brittany and {Gallardo}, Patricio A. and {Garrido}, Xavier and {Gerbino}, Martina and {Giardiello}, Serena and {Gill}, Ajay and {Givans}, Jahmour and {Gluscevic}, Vera and {Goldstein}, Samuel and {Golec}, Joseph E. and {Gong}, Yulin and {Guan}, Yilun and {Halpern}, Mark and {Harrison}, Ian and {Hasselfield}, Matthew and {He}, Adam and {Healy}, Erin and {Henderson}, Shawn and {Hensley}, Brandon and {Herv{\'\i}as-Caimapo}, Carlos and {Hilton}, Gene C. and {Hilton}, Matt and {Hincks}, Adam D. and {Hlo{\v{z}}ek}, Ren{\'e}e and {Ho}, Shuay-Pwu Patty and {Hood}, John and {Hornecker}, Erika and {Huber}, Zachary B. and {Hubmayr}, Johannes and {Huffenberger}, Kevin M. and {Hughes}, John P. and {Ikape}, Margaret and {Irwin}, Kent and {Isopi}, Giovanni and {Joshi}, Neha and {Keller}, Ben and {Kim}, Joshua and {Knowles}, Kenda and {Koopman}, Brian J. and {Kosowsky}, Arthur and {Kramer}, Darby and {Kusiak}, Aleksandra and {Lagu{\"e}}, Alex and {Lakey}, Victoria and {Lattanzi}, Massimiliano and {Lee}, Eunseong and {Li}, Yaqiong and {Li}, Zack and {Limon}, Michele and {Lokken}, Martine and {Louis}, Thibaut and {Lungu}, Marius and {MacCrann}, Niall and {MacInnis}, Amanda and {Madhavacheril}, Mathew S. and {Maldonado}, Diego and {Maldonado}, Felipe and {Mallaby-Kay}, Maya and {Marques}, Gabriela A. and {van Marrewijk}, Joshiwa and {McCarthy}, Fiona and {McMahon}, Jeff and {Mehta}, Yogesh and {Menanteau}, Felipe and {Moodley}, Kavilan and {Morris}, Thomas W. and {Mroczkowski}, Tony and {Naess}, Sigurd and {Namikawa}, Toshiya and {Nati}, Federico and {Nerval}, Simran K. and {Newburgh}, Laura and {Nicola}, Andrina and {Niemack}, Michael D. and {Nolta}, Michael R. and {Orlowski-Scherer}, John and {Pagano}, Luca and {Page}, Lyman A. and {Pandey}, Shivam and {Partridge}, Bruce and {Perez Sarmiento}, Karen and {Prince}, Heather and {Puddu}, Roberto and {Qu}, Frank J. and {Ragavan}, Damien C. and {Ried Guachalla}, Bernardita and {Rogers}, Keir K. and {Rojas}, Felipe and {Sakuma}, Tai and {Schaan}, Emmanuel and {Schmitt}, Benjamin L. and {Sehgal}, Neelima and {Shaikh}, Shabbir and {Sherwin}, Blake D. and {Sierra}, Carlos and {Sievers}, Jon and {Sif{\'o}n}, Crist{\'o}bal and {Simon}, Sara and {Sonka}, Rita and {Spergel}, David N. and {Staggs}, Suzanne T. and {Storer}, Emilie and {Surrao}, Kristen and {Switzer}, Eric R. and {Tampier}, Niklas and {Thiele}, Leander and {Thornton}, Robert and {Trac}, Hy and {Tucker}, Carole and {Ullom}, Joel and {Vale}, Leila R. and {Van Engelen}, Alexander and {Van Lanen}, Jeff and {Vargas}, Cristian and {Vavagiakis}, Eve M. and {Wagoner}, Kasey and {Wang}, Yuhan and {Wenzl}, Lukas and {Wollack}, Edward J. and {Zheng}, Kaiwen and {The Atacama Cosmology Telescope collaboration}},
        title = "{The Atacama Cosmology Telescope: DR6 constraints on extended cosmological models}",
      journal = {\jcap},
         year = 2025,
        month = nov,
       volume = {2025},
       number = {11},
          eid = {063},
        pages = {063},
          doi = {10.1088/1475-7516/2025/11/063},
archivePrefix = {arXiv},
       eprint = {2503.14454},
 primaryClass = {astro-ph.CO},
       adsurl = {https://ui.adsabs.harvard.edu/abs/2025JCAP...11..063C}
}

@ARTICLE{Cosmology_Intertwined_III:2021APh,
       author = {{Di Valentino}, Eleonora and {Anchordoqui}, Luis A. and {Akarsu}, {\"O}zg{\"u}r and {Ali-Haimoud}, Yacine and {Amendola}, Luca and {Arendse}, Nikki and {Asgari}, Marika and {Ballardini}, Mario and {Basilakos}, Spyros and {Battistelli}, Elia and {Benetti}, Micol and {Birrer}, Simon and {Bouchet}, Fran{\c{c}}ois R. and {Bruni}, Marco and {Calabrese}, Erminia and {Camarena}, David and {Capozziello}, Salvatore and {Chen}, Angela and {Chluba}, Jens and {Chudaykin}, Anton and {Colg{\'a}in}, Eoin {\'O}. and {Cyr-Racine}, Francis-Yan and {de Bernardis}, Paolo and {de Cruz P{\'e}rez}, Javier and {Delabrouille}, Jacques and {Dunkley}, Jo and {Escamilla-Rivera}, Celia and {Fert{\'e}}, Agn{\`e}s and {Finelli}, Fabio and {Freedman}, Wendy and {Frusciante}, Noemi and {Giusarma}, Elena and {G{\'o}mez-Valent}, Adri{\`a} and {Handley}, Will and {Harrison}, Ian and {Hart}, Luke and {Heavens}, Alan and {Hildebrandt}, Hendrik and {Holz}, Daniel and {Huterer}, Dragan and {Ivanov}, Mikhail M. and {Joudaki}, Shahab and {Kamionkowski}, Marc and {Karwal}, Tanvi and {Knox}, Lloyd and {Kumar}, Suresh and {Lamagna}, Luca and {Lesgourgues}, Julien and {Lucca}, Matteo and {Marra}, Valerio and {Masi}, Silvia and {Matarrese}, Sabino and {Mazumdar}, Arindam and {Melchiorri}, Alessandro and {Mena}, Olga and {Mersini-Houghton}, Laura and {Miranda}, Vivian and {Moreno-Pulido}, Cristian and {Mota}, David F. and {Muir}, Jessica and {Mukherjee}, Ankan and {Niedermann}, Florian and {Notari}, Alessio and {Nunes}, Rafael C. and {Pace}, Francesco and {Paliathanasis}, Andronikos and {Palmese}, Antonella and {Pan}, Supriya and {Paoletti}, Daniela and {Pettorino}, Valeria and {Piacentini}, Francesco and {Poulin}, Vivian and {Raveri}, Marco and {Riess}, Adam G. and {Salzano}, Vincenzo and {Saridakis}, Emmanuel N. and {Sen}, Anjan A. and {Shafieloo}, Arman and {Shajib}, Anowar J. and {Silk}, Joseph and {Silvestri}, Alessandra and {Sloth}, Martin S. and {Smith}, Tristan L. and {Sol{\`a} Peracaula}, Joan and {van de Bruck}, Carsten and {Verde}, Licia and {Visinelli}, Luca and {Wandelt}, Benjamin D. and {Wang}, Deng and {Wang}, Jian-Min and {Yadav}, Anil K. and {Yang}, Weiqiang},
        title = "{Cosmology Intertwined III: f{\ensuremath{\sigma}}$_{8}$ and S$_{8}$}",
      journal = {Astroparticle Physics},
         year = 2021,
        month = sep,
       volume = {131},
          eid = {102604},
        pages = {102604},
          doi = {10.1016/j.astropartphys.2021.102604},
archivePrefix = {arXiv},
       eprint = {2008.11285},
 primaryClass = {astro-ph.CO},
       adsurl = {https://ui.adsabs.harvard.edu/abs/2021APh...13102604D}
}

@article{Riess:2022ApJ,
doi = {10.3847/2041-8213/ac5c5b},
url = {https://doi.org/10.3847/2041-8213/ac5c5b},
year = {2022},
month = {jul},
publisher = {The American Astronomical Society},
volume = {934},
number = {1},
pages = {L7},
author = {Riess, Adam G. and Yuan, Wenlong and Macri, Lucas M. and Scolnic, Dan and Brout, Dillon and Casertano, Stefano and Jones, David O. and Murakami, Yukei and Anand, Gagandeep S. and Breuval, Louise and Brink, Thomas G. and Filippenko, Alexei V. and Hoffmann, Samantha and Jha, Saurabh W. and D’arcy Kenworthy, W. and Mackenty, John and Stahl, Benjamin E. and Zheng, WeiKang},
title = {A Comprehensive Measurement of the Local Value of the Hubble Constant with 1 km s$^{-1}$ Mpc$^{-1}$  Uncertainty from the Hubble Space Telescope and the SH0ES Team},
journal = {The Astrophysical Journal Letters}
}

@ARTICLE{planck-VI:2020,
       author = {{Planck Collaboration} and {Aghanim}, N. and {Akrami}, Y. and {Ashdown}, M. and {Aumont}, J. and {Baccigalupi}, C. and {Ballardini}, M. and {Banday}, A.~J. and {Barreiro}, R.~B. and {Bartolo}, N. and {Basak}, S. and {Battye}, R. and {Benabed}, K. and {Bernard}, J. -P. and {Bersanelli}, M. and {Bielewicz}, P. and {Bock}, J.~J. and {Bond}, J.~R. and {Borrill}, J. and {Bouchet}, F.~R. and {Boulanger}, F. and {Bucher}, M. and {Burigana}, C. and {Butler}, R.~C. and {Calabrese}, E. and {Cardoso}, J. -F. and {Carron}, J. and {Challinor}, A. and {Chiang}, H.~C. and {Chluba}, J. and {Colombo}, L.~P.~L. and {Combet}, C. and {Contreras}, D. and {Crill}, B.~P. and {Cuttaia}, F. and {de Bernardis}, P. and {de Zotti}, G. and {Delabrouille}, J. and {Delouis}, J. -M. and {Di Valentino}, E. and {Diego}, J.~M. and {Dor{\'e}}, O. and {Douspis}, M. and {Ducout}, A. and {Dupac}, X. and {Dusini}, S. and {Efstathiou}, G. and {Elsner}, F. and {En{\ss}lin}, T.~A. and {Eriksen}, H.~K. and {Fantaye}, Y. and {Farhang}, M. and {Fergusson}, J. and {Fernandez-Cobos}, R. and {Finelli}, F. and {Forastieri}, F. and {Frailis}, M. and {Fraisse}, A.~A. and {Franceschi}, E. and {Frolov}, A. and {Galeotta}, S. and {Galli}, S. and {Ganga}, K. and {G{\'e}nova-Santos}, R.~T. and {Gerbino}, M. and {Ghosh}, T. and {Gonz{\'a}lez-Nuevo}, J. and {G{\'o}rski}, K.~M. and {Gratton}, S. and {Gruppuso}, A. and {Gudmundsson}, J.~E. and {Hamann}, J. and {Handley}, W. and {Hansen}, F.~K. and {Herranz}, D. and {Hildebrandt}, S.~R. and {Hivon}, E. and {Huang}, Z. and {Jaffe}, A.~H. and {Jones}, W.~C. and {Karakci}, A. and {Keih{\"a}nen}, E. and {Keskitalo}, R. and {Kiiveri}, K. and {Kim}, J. and {Kisner}, T.~S. and {Knox}, L. and {Krachmalnicoff}, N. and {Kunz}, M. and {Kurki-Suonio}, H. and {Lagache}, G. and {Lamarre}, J. -M. and {Lasenby}, A. and {Lattanzi}, M. and {Lawrence}, C.~R. and {Le Jeune}, M. and {Lemos}, P. and {Lesgourgues}, J. and {Levrier}, F. and {Lewis}, A. and {Liguori}, M. and {Lilje}, P.~B. and {Lilley}, M. and {Lindholm}, V. and {L{\'o}pez-Caniego}, M. and {Lubin}, P.~M. and {Ma}, Y. -Z. and {Mac{\'\i}as-P{\'e}rez}, J.~F. and {Maggio}, G. and {Maino}, D. and {Mandolesi}, N. and {Mangilli}, A. and {Marcos-Caballero}, A. and {Maris}, M. and {Martin}, P.~G. and {Martinelli}, M. and {Mart{\'\i}nez-Gonz{\'a}lez}, E. and {Matarrese}, S. and {Mauri}, N. and {McEwen}, J.~D. and {Meinhold}, P.~R. and {Melchiorri}, A. and {Mennella}, A. and {Migliaccio}, M. and {Millea}, M. and {Mitra}, S. and {Miville-Desch{\^e}nes}, M. -A. and {Molinari}, D. and {Montier}, L. and {Morgante}, G. and {Moss}, A. and {Natoli}, P. and {N{\o}rgaard-Nielsen}, H.~U. and {Pagano}, L. and {Paoletti}, D. and {Partridge}, B. and {Patanchon}, G. and {Peiris}, H.~V. and {Perrotta}, F. and {Pettorino}, V. and {Piacentini}, F. and {Polastri}, L. and {Polenta}, G. and {Puget}, J. -L. and {Rachen}, J.~P. and {Reinecke}, M. and {Remazeilles}, M. and {Renzi}, A. and {Rocha}, G. and {Rosset}, C. and {Roudier}, G. and {Rubi{\~n}o-Mart{\'\i}n}, J.~A. and {Ruiz-Granados}, B. and {Salvati}, L. and {Sandri}, M. and {Savelainen}, M. and {Scott}, D. and {Shellard}, E.~P.~S. and {Sirignano}, C. and {Sirri}, G. and {Spencer}, L.~D. and {Sunyaev}, R. and {Suur-Uski}, A. -S. and {Tauber}, J.~A. and {Tavagnacco}, D. and {Tenti}, M. and {Toffolatti}, L. and {Tomasi}, M. and {Trombetti}, T. and {Valenziano}, L. and {Valiviita}, J. and {Van Tent}, B. and {Vibert}, L. and {Vielva}, P. and {Villa}, F. and {Vittorio}, N. and {Wandelt}, B.~D. and {Wehus}, I.~K. and {White}, M. and {White}, S.~D.~M. and {Zacchei}, A. and {Zonca}, A.},
        title = "{Planck 2018 results. VI. Cosmological parameters}",
      journal = {\aap},
         year = 2020,
        month = sep,
       volume = {641},
          eid = {A6},
        pages = {A6},
          doi = {10.1051/0004-6361/201833910},
archivePrefix = {arXiv},
       eprint = {1807.06209},
 primaryClass = {astro-ph.CO},
       adsurl = {https://ui.adsabs.harvard.edu/abs/2020A&A...641A...6P}
}

@article{Bennett:2013,
doi = {10.1088/0067-0049/208/2/20},
url = {https://dx.doi.org/10.1088/0067-0049/208/2/20},
year = {2013},
month = {sep},
publisher = {The American Astronomical Society},
volume = {208},
number = {2},
pages = {20},
author = {C. L. Bennett and D. Larson and J. L. Weiland and N. Jarosik and G. Hinshaw and N. Odegard and K. M. Smith and R. S. Hill and B. Gold and M. Halpern and E. Komatsu and M. R. Nolta and L. Page and D. N. Spergel and E. Wollack and J. Dunkley and A. Kogut and M. Limon and S. S. Meyer and G. S. Tucker and E. L. Wright},
title = {NINE-YEAR WILKINSON MICROWAVE ANISOTROPY PROBE (WMAP) OBSERVATIONS: FINAL MAPS AND RESULTS},
journal = {The Astrophysical Journal Supplement Series}
}

@ARTICLE{Planck-I:2020,
       author = {{Planck Collaboration} and {Aghanim}, N. and {Akrami}, Y. and {Arroja}, F. and {Ashdown}, M. and {Aumont}, J. and {Baccigalupi}, C. and {Ballardini}, M. and {Banday}, A.~J. and {Barreiro}, R.~B. and {Bartolo}, N. and {Basak}, S. and {Battye}, R. and {Benabed}, K. and {Bernard}, J. -P. and {Bersanelli}, M. and {Bielewicz}, P. and {Bock}, J.~J. and {Bond}, J.~R. and {Borrill}, J. and {Bouchet}, F.~R. and {Boulanger}, F. and {Bucher}, M. and {Burigana}, C. and {Butler}, R.~C. and {Calabrese}, E. and {Cardoso}, J. -F. and {Carron}, J. and {Casaponsa}, B. and {Challinor}, A. and {Chiang}, H.~C. and {Colombo}, L.~P.~L. and {Combet}, C. and {Contreras}, D. and {Crill}, B.~P. and {Cuttaia}, F. and {de Bernardis}, P. and {de Zotti}, G. and {Delabrouille}, J. and {Delouis}, J. -M. and {D{\'e}sert}, F. -X. and {Di Valentino}, E. and {Dickinson}, C. and {Diego}, J.~M. and {Donzelli}, S. and {Dor{\'e}}, O. and {Douspis}, M. and {Ducout}, A. and {Dupac}, X. and {Efstathiou}, G. and {Elsner}, F. and {En{\ss}lin}, T.~A. and {Eriksen}, H.~K. and {Falgarone}, E. and {Fantaye}, Y. and {Fergusson}, J. and {Fernandez-Cobos}, R. and {Finelli}, F. and {Forastieri}, F. and {Frailis}, M. and {Franceschi}, E. and {Frolov}, A. and {Galeotta}, S. and {Galli}, S. and {Ganga}, K. and {G{\'e}nova-Santos}, R.~T. and {Gerbino}, M. and {Ghosh}, T. and {Gonz{\'a}lez-Nuevo}, J. and {G{\'o}rski}, K.~M. and {Gratton}, S. and {Gruppuso}, A. and {Gudmundsson}, J.~E. and {Hamann}, J. and {Handley}, W. and {Hansen}, F.~K. and {Helou}, G. and {Herranz}, D. and {Hildebrandt}, S.~R. and {Hivon}, E. and {Huang}, Z. and {Jaffe}, A.~H. and {Jones}, W.~C. and {Karakci}, A. and {Keih{\"a}nen}, E. and {Keskitalo}, R. and {Kiiveri}, K. and {Kim}, J. and {Kisner}, T.~S. and {Knox}, L. and {Krachmalnicoff}, N. and {Kunz}, M. and {Kurki-Suonio}, H. and {Lagache}, G. and {Lamarre}, J. -M. and {Langer}, M. and {Lasenby}, A. and {Lattanzi}, M. and {Lawrence}, C.~R. and {Le Jeune}, M. and {Leahy}, J.~P. and {Lesgourgues}, J. and {Levrier}, F. and {Lewis}, A. and {Liguori}, M. and {Lilje}, P.~B. and {Lilley}, M. and {Lindholm}, V. and {L{\'o}pez-Caniego}, M. and {Lubin}, P.~M. and {Ma}, Y. -Z. and {Mac{\'\i}as-P{\'e}rez}, J.~F. and {Maggio}, G. and {Maino}, D. and {Mandolesi}, N. and {Mangilli}, A. and {Marcos-Caballero}, A. and {Maris}, M. and {Martin}, P.~G. and {Martinelli}, M. and {Mart{\'\i}nez-Gonz{\'a}lez}, E. and {Matarrese}, S. and {Mauri}, N. and {McEwen}, J.~D. and {Meerburg}, P.~D. and {Meinhold}, P.~R. and {Melchiorri}, A. and {Mennella}, A. and {Migliaccio}, M. and {Millea}, M. and {Mitra}, S. and {Miville-Desch{\^e}nes}, M. -A. and {Molinari}, D. and {Moneti}, A. and {Montier}, L. and {Morgante}, G. and {Moss}, A. and {Mottet}, S. and {M{\"u}nchmeyer}, M. and {Natoli}, P. and {N{\o}rgaard-Nielsen}, H.~U. and {Oxborrow}, C.~A. and {Pagano}, L. and {Paoletti}, D. and {Partridge}, B. and {Patanchon}, G. and {Pearson}, T.~J. and {Peel}, M. and {Peiris}, H.~V. and {Perrotta}, F. and {Pettorino}, V. and {Piacentini}, F. and {Polastri}, L. and {Polenta}, G. and {Puget}, J. -L. and {Rachen}, J.~P. and {Reinecke}, M. and {Remazeilles}, M. and {Renault}, C. and {Renzi}, A. and {Rocha}, G. and {Rosset}, C. and {Roudier}, G. and {Rubi{\~n}o-Mart{\'\i}n}, J.~A. and {Ruiz-Granados}, B. and {Salvati}, L. and {Sandri}, M. and {Savelainen}, M. and {Scott}, D. and {Shellard}, E.~P.~S. and {Shiraishi}, M. and {Sirignano}, C. and {Sirri}, G. and {Spencer}, L.~D. and {Sunyaev}, R. and {Suur-Uski}, A. -S. and {Tauber}, J.~A. and {Tavagnacco}, D. and {Tenti}, M. and {Terenzi}, L. and {Toffolatti}, L. and {Tomasi}, M. and {Trombetti}, T. and {Valiviita}, J. and {Van Tent}, B. and {Vibert}, L. and {Vielva}, P. and {Villa}, F. and {Vittorio}, N. and {Wandelt}, B.~D. and {Wehus}, I.~K. and {White}, M. and {White}, S.~D.~M. and {Zacchei}, A. and {Zonca}, A.},
        title = "{Planck 2018 results. I. Overview and the cosmological legacy of Planck}",
      journal = {\aap},
         year = 2020,
        month = sep,
       volume = {641},
          eid = {A1},
        pages = {A1},
          doi = {10.1051/0004-6361/201833880},
archivePrefix = {arXiv},
       eprint = {1807.06205},
 primaryClass = {astro-ph.CO},
       adsurl = {https://ui.adsabs.harvard.edu/abs/2020A&A...641A...1P}
}

@article{Takahara:1991,
    author = {Takahara, Fumio and Sasaki, Shin},
    title = {Effect of Rayleigh Scattering on Cosmic Microwave Background Anisotropies},
    journal = {Progress of Theoretical Physics},
    volume = {86},
    number = {5},
    pages = {1021-1030},
    year = {1991},
    month = {11},
    issn = {0033-068X},
    doi = {10.1143/ptp/86.5.1021},
    url = {https://doi.org/10.1143/ptp/86.5.1021},
    eprint = {https://academic.oup.com/ptp/article-pdf/86/5/1021/5430029/86-5-1021.pdf},
}

@article{Yu:2001,
doi = {10.1086/322482},
url = {https://doi.org/10.1086/322482},
year = {2001},
month = {sep},
publisher = {},
volume = {558},
number = {1},
pages = {23},
author = {Yu, Qingjuan and Spergel, David N. and Ostriker, Jeremiah P.},
title = {Rayleigh Scattering and Microwave Background
Fluctuations},
journal = {The Astrophysical Journal}
}

@article{Tarafdar:1967,
    author = {Tarafdar, S. P. and Vardya, M. S.},
    title = {The Rayleigh Scattering Cross-Sections of He, C, N and O},
    journal = {Monthly Notices of the Royal Astronomical Society},
    volume = {145},
    number = {2},
    pages = {171-180},
    year = {1969},
    month = {07},
    issn = {0035-8711},
    doi = {10.1093/mnras/145.2.171},
    url = {https://doi.org/10.1093/mnras/145.2.171},
    eprint = {https://academic.oup.com/mnras/article-pdf/145/2/171/9360740/mnras145-0171.pdf},
}

@article{Rayleigh01081881,
author = {Lord Rayleigh},
title = {X. On the electromagnetic theory of light },
journal = {The London, Edinburgh, and Dublin Philosophical Magazine and Journal of Science},
volume = {12},
number = {73},
pages = {81--101},
year = {1881},
publisher = {Taylor \& Francis},
doi = {10.1080/14786448108627074},


URL = { 
        https://doi.org/10.1080/14786448108627074
},
eprint = { 
    
        https://doi.org/10.1080/14786448108627074
}}

@ARTICLE{Carron.J:2013,
       author = {{Carron}, J.},
        title = "{On the assumption of Gaussianity for cosmological two-point statistics and parameter dependent covariance matrices}",
      journal = {\aap},
         year = 2013,
        month = mar,
       volume = {551},
          eid = {A88},
        pages = {A88},
          doi = {10.1051/0004-6361/201220538},
archivePrefix = {arXiv},
       eprint = {1204.4724},
 primaryClass = {astro-ph.CO},
       adsurl = {https://ui.adsabs.harvard.edu/abs/2013A&A...551A..88C}
}

@ARTICLE{Tegmark:1997ApJ,
       author = {{Tegmark}, Max and {Taylor}, Andy N. and {Heavens}, Alan F.},
        title = "{Karhunen-Lo{\`e}ve Eigenvalue Problems in Cosmology: How Should We Tackle Large Data Sets?}",
      journal = {\apj},
         year = 1997,
        month = may,
       volume = {480},
       number = {1},
        pages = {22-35},
          doi = {10.1086/303939},
archivePrefix = {arXiv},
       eprint = {astro-ph/9603021},
 primaryClass = {astro-ph},
       adsurl = {https://ui.adsabs.harvard.edu/abs/1997ApJ...480...22T}
}

@article{LiteBIRD:2022cnt,
    author = "Allys, E. and others",
    collaboration = "LiteBIRD",
    title = "{Probing Cosmic Inflation with the LiteBIRD Cosmic Microwave Background Polarization Survey}",
    eprint = "2202.02773",
    archivePrefix = "arXiv",
    primaryClass = "astro-ph.IM",
    doi = "10.1093/ptep/ptac150",
    journal = "PTEP",
    volume = "2023",
    number = "4",
    pages = "042F01",
    year = "2023"
}

@article{Lee:2005,
    author = {Lee, Hee-Won},
    title = {Exact low-energy expansion of the Rayleigh scattering cross-section by atomic hydrogen},
    journal = {Monthly Notices of the Royal Astronomical Society},
    volume = {358},
    number = {4},
    pages = {1472-1476},
    year = {2005},
    month = {04},
    issn = {0035-8711},
    doi = {10.1111/j.1365-2966.2005.08859.x},
    url = {https://doi.org/10.1111/j.1365-2966.2005.08859.x},
    eprint = {https://academic.oup.com/mnras/article-pdf/358/4/1472/18660986/358-4-1472.pdf},
}

@INPROCEEDINGS{SPTpol:2009,
   author = {{McMahon}, J.~J. and {Aird}, K.~A. and {Benson}, B.~A. and {Bleem}, L.~E. and
	{Britton}, J. and {Carlstrom}, J.~E. and {Chang}, C.~L. and
	{Cho}, H.~S. and {de Haan}, T. and {Crawford}, T.~M. and {Crites}, A.~T. and
	},
    title = "{SPTpol: an instrument for CMB polarization}",
booktitle = {American Institute of Physics Conference Series},
     year = 2009,
   series = {American Institute of Physics Conference Series},
   volume = 1185,
   editor = "{B.~Young, B.~Cabrera, \& A.~Miller}",
    month = dec,
    pages = {511-514},
      doi = {10.1063/1.3292391},
   adsurl = {http://adsabs.harvard.edu/abs/2009AIPC.1185..511M}
}

@INPROCEEDINGS{ACTPol:2010,
   author = {{Niemack}, M.~D. and {Ade}, P.~A.~R. and {Aguirre}, J. and {Barrientos}, F. and
	{Beall}, J.~A. and {Bond}, J.~R. and {Britton}, J. and {Cho}, H.~M. and
	{Das}, S. and {Devlin}, M.~J. and {Dicker}, S. and {Dunkley}, J.},
    title = "{ACTPol: a polarization-sensitive receiver for the Atacama Cosmology Telescope}",
booktitle = {Society of Photo-Optical Instrumentation Engineers (SPIE) Conference Series},
     year = 2010,
   series = {Presented at the Society of Photo-Optical Instrumentation Engineers (SPIE) Conference},
   volume = 7741,
archivePrefix = "arXiv",
   eprint = {1006.5049},
 primaryClass = "astro-ph.IM",
    month = jul,
      doi = {10.1117/12.857464},
   adsurl = {http://adsabs.harvard.edu/abs/2010SPIE.7741E..51N}
}

@ARTICLE{Hart:2020PCA,
       author = {{Hart}, Luke and {Chluba}, Jens},
        title = "{Improved model-independent constraints on the recombination era and development of a direct projection method}",
      journal = {\mnras},
         year = 2020,
        month = jul,
       volume = {495},
       number = {4},
        pages = {4210-4226},
          doi = {10.1093/mnras/staa1426},
archivePrefix = {arXiv},
       eprint = {1912.04682},
 primaryClass = {astro-ph.CO},
       adsurl = {https://ui.adsabs.harvard.edu/abs/2020MNRAS.495.4210H}
}

@ARTICLE{Kylar_L.Greene:2023JCAP,
       author = {{Greene}, Kylar L. and {Cyr-Racine}, Francis-Yan},
        title = "{Thomson scattering: one rate to rule them all}",
      journal = {\jcap},
         year = 2023,
        month = oct,
       volume = {2023},
       number = {10},
          eid = {065},
        pages = {065},
          doi = {10.1088/1475-7516/2023/10/065},
archivePrefix = {arXiv},
       eprint = {2306.06165},
 primaryClass = {astro-ph.CO},
       adsurl = {https://ui.adsabs.harvard.edu/abs/2023JCAP...10..065G}
}

@ARTICLE{Planck-IV:2020,
       author = {{Planck Collaboration} and {Akrami}, Y. and {Ashdown}, M. and {Aumont}, J. and {Baccigalupi}, C. and {Ballardini}, M. and {Banday}, A.~J. and {Barreiro}, R.~B. and {Bartolo}, N. and {Basak}, S. and {Benabed}, K. and {Bersanelli}, M. and {Bielewicz}, P. and {Bond}, J.~R. and {Borrill}, J. and {Bouchet}, F.~R. and {Boulanger}, F. and {Bucher}, M. and {Burigana}, C. and {Calabrese}, E. and {Cardoso}, J.-F. and {Carron}, J. and {Casaponsa}, B. and {Challinor}, A. and {Colombo}, L.~P.~L. and {Combet}, C. and {Crill}, B.~P. and {Cuttaia}, F. and {de Bernardis}, P. and {de Rosa}, A. and {de Zotti}, G. and {Delabrouille}, J. and {Delouis}, J.-M. and {Di Valentino}, E. and {Dickinson}, C. and {Diego}, J.~M. and {Donzelli}, S. and {Dor{\'e}}, O. and {Ducout}, A. and {Dupac}, X. and {Efstathiou}, G. and {Elsner}, F. and {En{\ss}lin}, T.~A. and {Eriksen}, H.~K. and {Falgarone}, E. and {Fernandez-Cobos}, R. and {Finelli}, F. and {Forastieri}, F. and {Frailis}, M. and {Fraisse}, A.~A. and {Franceschi}, E. and {Frolov}, A. and {Galeotta}, S. and {Galli}, S. and {Ganga}, K. and {G{\'e}nova-Santos}, R.~T. and {Gerbino}, M. and {Ghosh}, T. and {Gonz{\'a}lez-Nuevo}, J. and {G{\'o}rski}, K.~M. and {Gratton}, S. and {Gruppuso}, A. and {Gudmundsson}, J.~E. and {Handley}, W. and {Hansen}, F.~K. and {Helou}, G. and {Herranz}, D. and {Hildebrandt}, S.~R. and {Huang}, Z. and {Jaffe}, A.~H. and {Karakci}, A. and {Keih{\"a}nen}, E. and {Keskitalo}, R. and {Kiiveri}, K. and {Kim}, J. and {Kisner}, T.~S. and {Krachmalnicoff}, N. and {Kunz}, M. and {Kurki-Suonio}, H. and {Lagache}, G. and {Lamarre}, J.-M. and {Lasenby}, A. and {Lattanzi}, M. and {Lawrence}, C.~R. and {Le Jeune}, M. and {Levrier}, F. and {Liguori}, M. and {Lilje}, P.~B. and {Lindholm}, V. and {L{\'o}pez-Caniego}, M. and {Lubin}, P.~M. and {Ma}, Y.-Z. and {Mac{\'\i}as-P{\'e}rez}, J.~F. and {Maggio}, G. and {Maino}, D. and {Mandolesi}, N. and {Mangilli}, A. and {Marcos-Caballero}, A. and {Maris}, M. and {Martin}, P.~G. and {Mart{\'\i}nez-Gonz{\'a}lez}, E. and {Matarrese}, S. and {Mauri}, N. and {McEwen}, J.~D. and {Meinhold}, P.~R. and {Melchiorri}, A. and {Mennella}, A. and {Migliaccio}, M. and {Miville-Desch{\^e}nes}, M.-A. and {Molinari}, D. and {Moneti}, A. and {Montier}, L. and {Morgante}, G. and {Natoli}, P. and {Oppizzi}, F. and {Pagano}, L. and {Paoletti}, D. and {Partridge}, B. and {Peel}, M. and {Pettorino}, V. and {Piacentini}, F. and {Polenta}, G. and {Puget}, J.-L. and {Rachen}, J.~P. and {Reinecke}, M. and {Remazeilles}, M. and {Renzi}, A. and {Rocha}, G. and {Roudier}, G. and {Rubi{\~n}o-Mart{\'\i}n}, J.~A. and {Ruiz-Granados}, B. and {Salvati}, L. and {Sandri}, M. and {Savelainen}, M. and {Scott}, D. and {Seljebotn}, D.~S. and {Sirignano}, C. and {Spencer}, L.~D. and {Suur-Uski}, A.-S. and {Tauber}, J.~A. and {Tavagnacco}, D. and {Tenti}, M. and {Thommesen}, H. and {Toffolatti}, L. and {Tomasi}, M. and {Trombetti}, T. and {Valiviita}, J. and {Van Tent}, B. and {Vielva}, P. and {Villa}, F. and {Vittorio}, N. and {Wandelt}, B.~D. and {Wehus}, I.~K. and {Zacchei}, A. and {Zonca}, A.},
        title = "{Planck 2018 results. IV. Diffuse component separation}",
      journal = {\aap},
         year = 2020,
        month = sep,
       volume = {641},
          eid = {A4},
        pages = {A4},
          doi = {10.1051/0004-6361/201833881},
archivePrefix = {arXiv},
       eprint = {1807.06208},
 primaryClass = {astro-ph.CO},
       adsurl = {https://ui.adsabs.harvard.edu/abs/2020A&A...641A...4P}
}

@ARTICLE{Planck-xi:2020,
       author = {{Planck Collaboration} and {Akrami}, Y. and {Ashdown}, M. and {Aumont}, J. and {Baccigalupi}, C. and {Ballardini}, M. and {Banday}, A.~J. and {Barreiro}, R.~B. and {Bartolo}, N. and {Basak}, S. and {Benabed}, K. and {Bernard}, J.-P. and {Bersanelli}, M. and {Bielewicz}, P. and {Bond}, J.~R. and {Borrill}, J. and {Bouchet}, F.~R. and {Boulanger}, F. and {Bracco}, A. and {Bucher}, M. and {Burigana}, C. and {Calabrese}, E. and {Cardoso}, J.-F. and {Carron}, J. and {Chiang}, H.~C. and {Combet}, C. and {Crill}, B.~P. and {de Bernardis}, P. and {de Zotti}, G. and {Delabrouille}, J. and {Delouis}, J.-M. and {Di Valentino}, E. and {Dickinson}, C. and {Diego}, J.~M. and {Ducout}, A. and {Dupac}, X. and {Efstathiou}, G. and {Elsner}, F. and {En{\ss}lin}, T.~A. and {Falgarone}, E. and {Fantaye}, Y. and {Ferri{\`e}re}, K. and {Finelli}, F. and {Forastieri}, F. and {Frailis}, M. and {Fraisse}, A.~A. and {Franceschi}, E. and {Frolov}, A. and {Galeotta}, S. and {Galli}, S. and {Ganga}, K. and {G{\'e}nova-Santos}, R.~T. and {Ghosh}, T. and {Gonz{\'a}lez-Nuevo}, J. and {G{\'o}rski}, K.~M. and {Gruppuso}, A. and {Gudmundsson}, J.~E. and {Guillet}, V. and {Handley}, W. and {Hansen}, F.~K. and {Herranz}, D. and {Huang}, Z. and {Jaffe}, A.~H. and {Jones}, W.~C. and {Keih{\"a}nen}, E. and {Keskitalo}, R. and {Kiiveri}, K. and {Kim}, J. and {Krachmalnicoff}, N. and {Kunz}, M. and {Kurki-Suonio}, H. and {Lamarre}, J.-M. and {Lasenby}, A. and {Le Jeune}, M. and {Levrier}, F. and {Liguori}, M. and {Lilje}, P.~B. and {Lindholm}, V. and {L{\'o}pez-Caniego}, M. and {Lubin}, P.~M. and {Ma}, Y.-Z. and {Mac{\'\i}as-P{\'e}rez}, J.~F. and {Maggio}, G. and {Maino}, D. and {Mandolesi}, N. and {Mangilli}, A. and {Martin}, P.~G. and {Mart{\'\i}nez-Gonz{\'a}lez}, E. and {Matarrese}, S. and {McEwen}, J.~D. and {Meinhold}, P.~R. and {Melchiorri}, A. and {Migliaccio}, M. and {Miville-Desch{\^e}nes}, M.-A. and {Molinari}, D. and {Moneti}, A. and {Montier}, L. and {Morgante}, G. and {Natoli}, P. and {Pagano}, L. and {Paoletti}, D. and {Pettorino}, V. and {Piacentini}, F. and {Polenta}, G. and {Puget}, J.-L. and {Rachen}, J.~P. and {Reinecke}, M. and {Remazeilles}, M. and {Renzi}, A. and {Rocha}, G. and {Rosset}, C. and {Roudier}, G. and {Rubi{\~n}o-Mart{\'\i}n}, J.~A. and {Ruiz-Granados}, B. and {Salvati}, L. and {Sandri}, M. and {Savelainen}, M. and {Scott}, D. and {Soler}, J.~D. and {Spencer}, L.~D. and {Tauber}, J.~A. and {Tavagnacco}, D. and {Toffolatti}, L. and {Tomasi}, M. and {Trombetti}, T. and {Valiviita}, J. and {Vansyngel}, F. and {Van Tent}, B. and {Vielva}, P. and {Villa}, F. and {Vittorio}, N. and {Wehus}, I.~K. and {Zacchei}, A. and {Zonca}, A.},
        title = "{Planck 2018 results. XI. Polarized dust foregrounds}",
      journal = {\aap},
         year = 2020,
        month = sep,
       volume = {641},
          eid = {A11},
        pages = {A11},
          doi = {10.1051/0004-6361/201832618},
archivePrefix = {arXiv},
       eprint = {1801.04945},
 primaryClass = {astro-ph.GA},
       adsurl = {https://ui.adsabs.harvard.edu/abs/2020A&A...641A..11P}
}

@ARTICLE{Dunkley:2011ApJ,
       author = {{Dunkley}, J. and {Hlozek}, R. and {Sievers}, J. and {Acquaviva}, V. and {Ade}, P.~A.~R. and {Aguirre}, P. and {Amiri}, M. and {Appel}, J.~W. and {Barrientos}, L.~F. and {Battistelli}, E.~S. and {Bond}, J.~R. and {Brown}, B. and {Burger}, B. and {Chervenak}, J. and {Das}, S. and {Devlin}, M.~J. and {Dicker}, S.~R. and {Bertrand Doriese}, W. and {D{\"u}nner}, R. and {Essinger-Hileman}, T. and {Fisher}, R.~P. and {Fowler}, J.~W. and {Hajian}, A. and {Halpern}, M. and {Hasselfield}, M. and {Hern{\'a}ndez-Monteagudo}, C. and {Hilton}, G.~C. and {Hilton}, M. and {Hincks}, A.~D. and {Huffenberger}, K.~M. and {Hughes}, D.~H. and {Hughes}, J.~P. and {Infante}, L. and {Irwin}, K.~D. and {Juin}, J.~B. and {Kaul}, M. and {Klein}, J. and {Kosowsky}, A. and {Lau}, J.~M. and {Limon}, M. and {Lin}, Y.-T. and {Lupton}, R.~H. and {Marriage}, T.~A. and {Marsden}, D. and {Mauskopf}, P. and {Menanteau}, F. and {Moodley}, K. and {Moseley}, H. and {Netterfield}, C.~B. and {Niemack}, M.~D. and {Nolta}, M.~R. and {Page}, L.~A. and {Parker}, L. and {Partridge}, B. and {Reid}, B. and {Sehgal}, N. and {Sherwin}, B. and {Spergel}, D.~N. and {Staggs}, S.~T. and {Swetz}, D.~S. and {Switzer}, E.~R. and {Thornton}, R. and {Trac}, H. and {Tucker}, C. and {Warne}, R. and {Wollack}, E. and {Zhao}, Y.},
        title = "{The Atacama Cosmology Telescope: Cosmological Parameters from the 2008 Power Spectrum}",
      journal = {\apj},
         year = 2011,
        month = sep,
       volume = {739},
       number = {1},
          eid = {52},
        pages = {52},
          doi = {10.1088/0004-637X/739/1/52},
archivePrefix = {arXiv},
       eprint = {1009.0866},
 primaryClass = {astro-ph.CO},
       adsurl = {https://ui.adsabs.harvard.edu/abs/2011ApJ...739...52D}
}

@ARTICLE{Addison:2012MNRAS,
       author = {{Addison}, G.~E. and {Dunkley}, J. and {Spergel}, D.~N.},
        title = "{Modelling the correlation between the thermal Sunyaev Zel'dovich effect and the cosmic infrared background}",
      journal = {\mnras},
         year = 2012,
        month = dec,
       volume = {427},
       number = {2},
        pages = {1741-1754},
          doi = {10.1111/j.1365-2966.2012.21664.x},
archivePrefix = {arXiv},
       eprint = {1204.5927},
 primaryClass = {astro-ph.CO},
       adsurl = {https://ui.adsabs.harvard.edu/abs/2012MNRAS.427.1741A}
}

@ARTICLE{Choi:2020JCAP,
       author = {{Choi}, Steve K. and {Hasselfield}, Matthew and {Ho}, Shuay-Pwu Patty and {Koopman}, Brian and {Lungu}, Marius and {Abitbol}, Maximilian H. and {Addison}, Graeme E. and {Ade}, Peter A.~R. and {Aiola}, Simone and {Alonso}, David and {Amiri}, Mandana and {Amodeo}, Stefania and {Angile}, Elio and {Austermann}, Jason E. and {Baildon}, Taylor and {Battaglia}, Nick and {Beall}, James A. and {Bean}, Rachel and {Becker}, Daniel T. and {Bond}, J. Richard and {Bruno}, Sarah Marie and {Calabrese}, Erminia and {Calafut}, Victoria and {Campusano}, Luis E. and {Carrero}, Felipe and {Chesmore}, Grace E. and {Cho}, Hsiao-mei and {Clark}, Susan E. and {Cothard}, Nicholas F. and {Crichton}, Devin and {Crowley}, Kevin T. and {Darwish}, Omar and {Datta}, Rahul and {Denison}, Edward V. and {Devlin}, Mark J. and {Duell}, Cody J. and {Duff}, Shannon M. and {Duivenvoorden}, Adriaan J. and {Dunkley}, Jo and {D{\"u}nner}, Rolando and {Essinger-Hileman}, Thomas and {Fankhanel}, Max and {Ferraro}, Simone and {Fox}, Anna E. and {Fuzia}, Brittany and {Gallardo}, Patricio A. and {Gluscevic}, Vera and {Golec}, Joseph E. and {Grace}, Emily and {Gralla}, Megan and {Guan}, Yilun and {Hall}, Kirsten and {Halpern}, Mark and {Han}, Dongwon and {Hargrave}, Peter and {Henderson}, Shawn and {Hensley}, Brandon and {Hill}, J. Colin and {Hilton}, Gene C. and {Hilton}, Matt and {Hincks}, Adam D. and {Hlo{\v{z}}ek}, Ren{\'e}e and {Hubmayr}, Johannes and {Huffenberger}, Kevin M. and {Hughes}, John P. and {Infante}, Leopoldo and {Irwin}, Kent and {Jackson}, Rebecca and {Klein}, Jeff and {Knowles}, Kenda and {Kosowsky}, Arthur and {Lakey}, Vincent and {Li}, Dale and {Li}, Yaqiong and {Li}, Zack and {Lokken}, Martine and {Louis}, Thibaut and {MacInnis}, Amanda and {Madhavacheril}, Mathew and {Maldonado}, Felipe and {Mallaby-Kay}, Maya and {Marsden}, Danica and {Maurin}, Lo{\"\i}c and {McMahon}, Jeff and {Menanteau}, Felipe and {Moodley}, Kavilan and {Morton}, Tim and {Naess}, Sigurd and {Namikawa}, Toshiya and {Nati}, Federico and {Newburgh}, Laura and {Nibarger}, John P. and {Nicola}, Andrina and {Niemack}, Michael D. and {Nolta}, Michael R. and {Orlowski-Sherer}, John and {Page}, Lyman A. and {Pappas}, Christine G. and {Partridge}, Bruce and {Phakathi}, Phumlani and {Prince}, Heather and {Puddu}, Roberto and {Qu}, Frank J. and {Rivera}, Jesus and {Robertson}, Naomi and {Rojas}, Felipe and {Salatino}, Maria and {Schaan}, Emmanuel and {Schillaci}, Alessandro and {Schmitt}, Benjamin L. and {Sehgal}, Neelima and {Sherwin}, Blake D. and {Sierra}, Carlos and {Sievers}, Jon and {Sifon}, Cristobal and {Sikhosana}, Precious and {Simon}, Sara and {Spergel}, David N. and {Staggs}, Suzanne T. and {Stevens}, Jason and {Storer}, Emilie and {Sunder}, Dhaneshwar D. and {Switzer}, Eric R. and {Thorne}, Ben and {Thornton}, Robert and {Trac}, Hy and {Treu}, Jesse and {Tucker}, Carole and {Vale}, Leila R. and {Van Engelen}, Alexander and {Van Lanen}, Jeff and {Vavagiakis}, Eve M. and {Wagoner}, Kasey and {Wang}, Yuhan and {Ward}, Jonathan T. and {Wollack}, Edward J. and {Xu}, Zhilei and {Zago}, Fernando and {Zhu}, Ningfeng},
        title = "{The Atacama Cosmology Telescope: a measurement of the Cosmic Microwave Background power spectra at 98 and 150 GHz}",
      journal = {\jcap},
         year = 2020,
        month = dec,
       volume = {2020},
       number = {12},
          eid = {045},
        pages = {045},
          doi = {10.1088/1475-7516/2020/12/045},
archivePrefix = {arXiv},
       eprint = {2007.07289},
 primaryClass = {astro-ph.CO},
       adsurl = {https://ui.adsabs.harvard.edu/abs/2020JCAP...12..045C}
}

@ARTICLE{Beringue:2025JCAP,
       author = {{Beringue}, Benjamin and {Surrao}, Kristen M. and {Hill}, J. Colin and {Atkins}, Zachary and {Battaglia}, Nicholas and {Bolliet}, Boris and {Calabrese}, Erminia and {Choi}, Steve K. and {Clark}, Susan E. and {Duivenvoorden}, Adriaan J. and {Dunkley}, Jo and {Giardiello}, Serena and {Goldstein}, Samuel and {Hensley}, Brandon S. and {Hlo{\v{z}}ek}, Ren{\'e}e and {Jense}, Hidde T. and {Kramer}, Darby and {La Posta}, Adrien and {Louis}, Thibaut and {Mehta}, Yogesh and {Moodley}, Kavilan and {Naess}, Sigurd and {Partridge}, Bruce and {Qu}, Frank J. and {Ried Guachalla}, Bernardita and {Sehgal}, Neelima and {Sif{\'o}n}, Crist{\'o}bal and {Staggs}, Suzanne T. and {Trac}, Hy and {Van Engelen}, Alexander and {Wollack}, Edward J.},
        title = "{The Atacama Cosmology Telescope: DR6 power spectrum foreground model and validation}",
      journal = {\jcap},
         year = 2025,
        month = oct,
       volume = {2025},
       number = {10},
          eid = {082},
        pages = {082},
          doi = {10.1088/1475-7516/2025/10/082},
archivePrefix = {arXiv},
       eprint = {2506.06274},
 primaryClass = {astro-ph.CO},
       adsurl = {https://ui.adsabs.harvard.edu/abs/2025JCAP...10..082B}
}

@ARTICLE{Planck-XXX:2014,
       author = {{Planck Collaboration} and {Ade}, P.~A.~R. and {Aghanim}, N. and {Armitage-Caplan}, C. and {Arnaud}, M. and {Ashdown}, M. and {Atrio-Barandela}, F. and {Aumont}, J. and {Baccigalupi}, C. and {Banday}, A.~J. and {Barreiro}, R.~B. and {Bartlett}, J.~G. and {Battaner}, E. and {Benabed}, K. and {Beno{\^\i}t}, A. and {Benoit-L{\'e}vy}, A. and {Bernard}, J.-P. and {Bersanelli}, M. and {Bethermin}, M. and {Bielewicz}, P. and {Blagrave}, K. and {Bobin}, J. and {Bock}, J.~J. and {Bonaldi}, A. and {Bond}, J.~R. and {Borrill}, J. and {Bouchet}, F.~R. and {Boulanger}, F. and {Bridges}, M. and {Bucher}, M. and {Burigana}, C. and {Butler}, R.~C. and {Cardoso}, J.-F. and {Catalano}, A. and {Challinor}, A. and {Chamballu}, A. and {Chen}, X. and {Chiang}, H.~C. and {Chiang}, L.-Y. and {Christensen}, P.~R. and {Church}, S. and {Clements}, D.~L. and {Colombi}, S. and {Colombo}, L.~P.~L. and {Couchot}, F. and {Coulais}, A. and {Crill}, B.~P. and {Curto}, A. and {Cuttaia}, F. and {Danese}, L. and {Davies}, R.~D. and {Davis}, R.~J. and {de Bernardis}, P. and {de Rosa}, A. and {de Zotti}, G. and {Delabrouille}, J. and {Delouis}, J.-M. and {D{\'e}sert}, F.-X. and {Dickinson}, C. and {Diego}, J.~M. and {Dole}, H. and {Donzelli}, S. and {Dor{\'e}}, O. and {Douspis}, M. and {Dupac}, X. and {Efstathiou}, G. and {En{\ss}lin}, T.~A. and {Eriksen}, H.~K. and {Finelli}, F. and {Forni}, O. and {Frailis}, M. and {Franceschi}, E. and {Galeotta}, S. and {Ganga}, K. and {Ghosh}, T. and {Giard}, M. and {Giraud-H{\'e}raud}, Y. and {Gonz{\'a}lez-Nuevo}, J. and {G{\'o}rski}, K.~M. and {Gratton}, S. and {Gregorio}, A. and {Gruppuso}, A. and {Hansen}, F.~K. and {Hanson}, D. and {Harrison}, D. and {Helou}, G. and {Henrot-Versill{\'e}}, S. and {Hern{\'a}ndez-Monteagudo}, C. and {Herranz}, D. and {Hildebrandt}, S.~R. and {Hivon}, E. and {Hobson}, M. and {Holmes}, W.~A. and {Hornstrup}, A. and {Hovest}, W. and {Huffenberger}, K.~M. and {Jaffe}, A.~H. and {Jaffe}, T.~R. and {Jones}, W.~C. and {Juvela}, M. and {Kalberla}, P. and {Keih{\"a}nen}, E. and {Kerp}, J. and {Keskitalo}, R. and {Kisner}, T.~S. and {Kneissl}, R. and {Knoche}, J. and {Knox}, L. and {Kunz}, M. and {Kurki-Suonio}, H. and {Lacasa}, F. and {Lagache}, G. and {L{\"a}hteenm{\"a}ki}, A. and {Lamarre}, J.-M. and {Langer}, M. and {Lasenby}, A. and {Laureijs}, R.~J. and {Lawrence}, C.~R. and {Leonardi}, R. and {Le{\'o}n-Tavares}, J. and {Lesgourgues}, J. and {Liguori}, M. and {Lilje}, P.~B. and {Linden-V{\o}rnle}, M. and {L{\'o}pez-Caniego}, M. and {Lubin}, P.~M. and {Mac{\'\i}as-P{\'e}rez}, J.~F. and {Maffei}, B. and {Maino}, D. and {Mandolesi}, N. and {Maris}, M. and {Marshall}, D.~J. and {Martin}, P.~G. and {Mart{\'\i}nez-Gonz{\'a}lez}, E. and {Masi}, S. and {Massardi}, M. and {Matarrese}, S. and {Matthai}, F. and {Mazzotta}, P. and {Melchiorri}, A. and {Mendes}, L. and {Mennella}, A. and {Migliaccio}, M. and {Mitra}, S. and {Miville-Desch{\^e}nes}, M.-A. and {Moneti}, A. and {Montier}, L. and {Morgante}, G. and {Mortlock}, D. and {Munshi}, D. and {Murphy}, J.~A. and {Naselsky}, P. and {Nati}, F. and {Natoli}, P. and {Netterfield}, C.~B. and {N{\o}rgaard-Nielsen}, H.~U. and {Noviello}, F. and {Novikov}, D. and {Novikov}, I. and {Osborne}, S. and {Oxborrow}, C.~A. and {Paci}, F. and {Pagano}, L. and {Pajot}, F. and {Paladini}, R. and {Paoletti}, D. and {Partridge}, B. and {Pasian}, F. and {Patanchon}, G. and {Perdereau}, O. and {Perotto}, L. and {Perrotta}, F. and {Piacentini}, F. and {Piat}, M. and {Pierpaoli}, E. and {Pietrobon}, D. and {Plaszczynski}, S. and {Pointecouteau}, E. and {Polenta}, G. and {Ponthieu}, N. and {Popa}, L. and {Poutanen}, T. and {Pratt}, G.~W. and {Pr{\'e}zeau}, G. and {Prunet}, S. and {Puget}, J.-L. and {Rachen}, J.~P. and {Reach}, W.~T. and {Rebolo}, R. and {Reinecke}, M. and {Remazeilles}, M. and {Renault}, C. and {Ricciardi}, S. and {Riller}, T. and {Ristorcelli}, I. and {Rocha}, G. and {Rosset}, C. and {Roudier}, G. and {Rowan-Robinson}, M. and {Rubi{\~n}o-Mart{\'\i}n}, J.~A.},
        title = "{Planck 2013 results. XXX. Cosmic infrared background measurements and implications for star formation}",
      journal = {\aap},
         year = 2014,
        month = nov,
       volume = {571},
          eid = {A30},
        pages = {A30},
          doi = {10.1051/0004-6361/201322093},
archivePrefix = {arXiv},
       eprint = {1309.0382},
 primaryClass = {astro-ph.CO},
       adsurl = {https://ui.adsabs.harvard.edu/abs/2014A&A...571A..30P}
}

@ARTICLE{Louis:2025JCAP,
       author = {{Louis}, Thibaut and {La Posta}, Adrien and {Atkins}, Zachary and {Jense}, Hidde T. and {Abril-Cabezas}, Irene and {Addison}, Graeme E. and {Ade}, Peter A.~R. and {Aiola}, Simone and {Alford}, Tommy and {Alonso}, David and {Amiri}, Mandana and {An}, Rui and {Austermann}, Jason E. and {Barbavara}, Eleonora and {Battaglia}, Nicholas and {Battistelli}, Elia Stefano and {Beall}, James A. and {Bean}, Rachel and {Beheshti}, Ali and {Beringue}, Benjamin and {Bhandarkar}, Tanay and {Biermann}, Emily and {Bolliet}, Boris and {Bond}, J. Richard and {Calabrese}, Erminia and {Capalbo}, Valentina and {Carrero}, Felipe and {Chen}, Shi-Fan and {Chesmore}, Grace and {Cho}, Hsiao-mei and {Choi}, Steve K. and {Clark}, Susan E. and {Cothard}, Nicholas F. and {Coughlin}, Kevin and {Coulton}, William and {Crichton}, Devin and {Crowley}, Kevin T. and {Darwish}, Omar and {Devlin}, Mark J. and {Dicker}, Simon and {Duell}, Cody J. and {Duff}, Shannon M. and {Duivenvoorden}, Adriaan J. and {Dunkley}, Jo and {Dunner}, Rolando and {Embil Villagra}, Carmen and {Fankhanel}, Max and {Farren}, Gerrit S. and {Ferraro}, Simone and {Foster}, Allen and {Freundt}, Rodrigo and {Fuzia}, Brittany and {Gallardo}, Patricio A. and {Garrido}, Xavier and {Gerbino}, Martina and {Giardiello}, Serena and {Gill}, Ajay and {Givans}, Jahmour and {Gluscevic}, Vera and {Goldstein}, Samuel and {Golec}, Joseph E. and {Gong}, Yulin and {Guan}, Yilun and {Halpern}, Mark and {Harrison}, Ian and {Hasselfield}, Matthew and {Healy}, Erin and {Henderson}, Shawn and {Hensley}, Brandon and {Herv{\'\i}as-Caimapo}, Carlos and {Hill}, J. Colin and {Hilton}, Gene C. and {Hilton}, Matt and {Hincks}, Adam D. and {Hlo{\v{z}}ek}, Ren{\'e}e and {Ho}, Shuay-Pwu Patty and {Hood}, John and {Hornecker}, Erika and {Huber}, Zachary B. and {Hubmayr}, Johannes and {Huffenberger}, Kevin M. and {Hughes}, John P. and {Ikape}, Margaret and {Irwin}, Kent and {Isopi}, Giovanni and {Joshi}, Neha and {Keller}, Ben and {Kim}, Joshua and {Knowles}, Kenda and {Koopman}, Brian J. and {Kosowsky}, Arthur and {Kramer}, Darby and {Kusiak}, Aleksandra and {Lagu{\"e}}, Alex and {Lakey}, Victoria and {Lee}, Eunseong and {Li}, Yaqiong and {Li}, Zack and {Limon}, Michele and {Lokken}, Martine and {Lungu}, Marius and {MacCrann}, Niall and {MacInnis}, Amanda and {Madhavacheril}, Mathew S. and {Maldonado}, Diego and {Maldonado}, Felipe and {Mallaby-Kay}, Maya and {Marques}, Gabriela A. and {van Marrewijk}, Joshiwa and {McCarthy}, Fiona and {McMahon}, Jeff and {Mehta}, Yogesh and {Menanteau}, Felipe and {Moodley}, Kavilan and {Morris}, Thomas W. and {Mroczkowski}, Tony and {Naess}, Sigurd and {Namikawa}, Toshiya and {Nati}, Federico and {Nerval}, Simran K. and {Newburgh}, Laura and {Nicola}, Andrina and {Niemack}, Michael D. and {Nolta}, Michael R. and {Orlowski-Scherer}, John and {Pagano}, Luca and {Page}, Lyman A. and {Pandey}, Shivam and {Partridge}, Bruce and {Perez Sarmiento}, Karen and {Prince}, Heather and {Puddu}, Roberto and {Qu}, Frank J. and {Ragavan}, Damien C. and {Ried Guachalla}, Bernardita and {Rogers}, Keir K. and {Rojas}, Felipe and {Sakuma}, Tai and {Schaan}, Emmanuel and {Schmitt}, Benjamin L. and {Sehgal}, Neelima and {Shaikh}, Shabbir and {Sherwin}, Blake D. and {Sierra}, Carlos and {Sievers}, Jon and {Sif{\'o}n}, Crist{\'o}bal and {Simon}, Sara and {Sonka}, Rita and {Spergel}, David N. and {Staggs}, Suzanne T. and {Storer}, Emilie and {Surrao}, Kristen and {Switzer}, Eric R. and {Tampier}, Niklas and {Thornton}, Robert and {Trac}, Hy and {Tucker}, Carole and {Ullom}, Joel and {Vale}, Leila R. and {Van Engelen}, Alexander and {Van Lanen}, Jeff and {Vargas}, Cristian and {Vavagiakis}, Eve M. and {Wagoner}, Kasey and {Wang}, Yuhan and {Wenzl}, Lukas and {Wollack}, Edward J. and {Zheng}, Kaiwen and {The Atacama Cosmology Telescope collaboration}},
        title = "{The Atacama Cosmology Telescope: DR6 power spectra, likelihoods and {\ensuremath{\Lambda}}CDM parameters}",
      journal = {\jcap},
         year = 2025,
        month = nov,
       volume = {2025},
       number = {11},
          eid = {062},
        pages = {062},
          doi = {10.1088/1475-7516/2025/11/062},
archivePrefix = {arXiv},
       eprint = {2503.14452},
 primaryClass = {astro-ph.CO},
       adsurl = {https://ui.adsabs.harvard.edu/abs/2025JCAP...11..062L}
}

@ARTICLE{Hart:2020Hubble,
       author = {{Hart}, Luke and {Chluba}, Jens},
        title = "{Updated fundamental constant constraints from Planck 2018 data and possible relations to the Hubble tension}",
      journal = {\mnras},
         year = 2020,
        month = apr,
       volume = {493},
       number = {3},
        pages = {3255-3263},
          doi = {10.1093/mnras/staa412},
archivePrefix = {arXiv},
       eprint = {1912.03986},
 primaryClass = {astro-ph.CO},
       adsurl = {https://ui.adsabs.harvard.edu/abs/2020MNRAS.493.3255H}
}

@article{Schoneberg:2021qvd,
    author = {Sch{\"o}neberg, Nils and Franco Abell{\'a}n, Guillermo and P{\'e}rez S{\'a}nchez, Andrea and Witte, Samuel J. and Poulin, Vivian and Lesgourgues, Julien},
    title = "{The H0 Olympics: A fair ranking of proposed models}",
    eprint = "2107.10291",
    archivePrefix = "arXiv",
    primaryClass = "astro-ph.CO",
    doi = "10.1016/j.physrep.2022.07.001",
    journal = "Phys. Rept.",
    volume = "984",
    pages = "1--55",
    year = "2022"
}

@ARTICLE{Beringue:2021,
       author = {{Beringue}, Benjamin and {Meerburg}, P. Daniel and {Meyers}, Joel and {Battaglia}, Nicholas},
        title = "{Cosmology with Rayleigh scattering of the cosmic microwave background}",
      journal = {\jcap},
         year = 2021,
        month = jan,
       volume = {2021},
       number = {1},
          eid = {060},
        pages = {060},
          doi = {10.1088/1475-7516/2021/01/060},
archivePrefix = {arXiv},
       eprint = {2008.11688},
 primaryClass = {astro-ph.CO},
       adsurl = {https://ui.adsabs.harvard.edu/abs/2021JCAP...01..060B}
}

\appendix
\section{Derivation of Fisher matrix and parameter bias due to incorrect fiducial model}\label{sec:derivations}
The log-likelihood of a Gaussian random vector, $\textbf{a}$, having a covariance matrix, $\textbf{C}$ 
is given by
\begin{equation}
    \ln \mathcal{L} = -\frac{1}{2}\big [ \ln \,\det \textbf{C} + \textbf{a}^{\dag}\textbf{C}^{-1} \textbf{a}\big] + constant,
\end{equation}
where $\mathbf{C}$ is a block-diagonal matrix, with the matrix $\mathbf{C}_\ell$ of size $2\ell + 1$ as its elements for each multipole $\ell$.
Taking the derivative with respect to  parameter $\theta_{i}$ and using the log-determinant matrix identities $\partial_{i} \ln \,\det \textbf{C} = \mathrm{Tr} (\textbf{C}^{-1}\partial_{i} \textbf{C})$ and  $\partial_{i} \textbf{C}^{-1} = -\textbf{C}^{-1} \,\partial_{i}\textbf{C}\,\textbf{C}^{-1}$, we obtain

\begin{equation}\label{eq:first_derivative}
    \frac{\partial \ln\mathcal{L} }{\partial\theta_{i}} = -\frac{1}{2}\big[\mathrm{Tr}\, (\textbf{C}^{-1}\partial_{i}\textbf{C}) - \textbf{a}^{\dag}\textbf{C}^{-1}\partial_{i}\textbf{C} \textbf{C}^{-1}\textbf{a} \big].
\end{equation}
Taking the second derivative with respect to the parameter $\theta_{j}$ and using the matrix identity, $\big <\textbf{a}^{\dag}A\textbf{a}\big> = \mathrm{Tr} (A\textbf{C})$ where $\big<{\textbf{a}^{\dag} \textbf{a}}\big> =\textbf{C}$, we derive the Fisher matrix as

\begin{equation}
  \mathrm{F}_{ij} = - \big <\frac{\partial^2 \ln\mathcal{L}}{\partial\theta_{i}\partial\theta_{j}}\big> = \frac{1}{2}\mathrm{Tr}\big(\textbf{C}^{-1} \partial_{i}\textbf{C}\textbf{C}^{-1}\partial_{j}\textbf{C}\big).
\end{equation}
For CMB data, since the covariance matrix $\textbf{C}$ is block-diagonal for each $\ell$ of size $(2\ell + 1)$, the Fisher matrix for a sky-fraction of $f_{\rm sky}$ is given by
\begin{equation}
    \mathrm{F}_{ij} = \sum_{\ell} \frac{(2\ell+1)f_{\rm sky}}{2}\mathrm{Tr} \big (\textbf{C}_{\ell}^{-1} \frac{\partial \textbf{C}_{\ell}}{\partial\theta_{i}} \textbf{C}_{\ell}^{-1}  \frac{\partial \textbf{C}_{\ell}}{\partial\theta_{j}} \big).
\end{equation}
Now we show the derivation of the bias in the parameters due to the mismatch between model $\textbf{C}_{\ell}$ and true data $\textbf{C}^{\rm true}_{\ell}$. Due to the systematic difference $\Delta \textbf{C}_{\ell} = \textbf{C}^{\rm true}_{\ell} - \textbf{C}_{\ell}$ between the true power spectrum and our model's expectation, the maximum of the likelihood will be shifted to $\hat{\theta}$ from the true parameter values $\theta_{\rm fid}$ by a bias $\delta \theta$. Therefore the estimator $\hat{\theta}$ satisfies the condition,
\begin{equation}\label{eq:score}
    \frac{\partial \ln \mathcal{L}}{\partial \theta_{i}}\Bigg|_{\hat{\theta}} = 0.  
\end{equation}
Expanding Eq.~\ref{eq:score} around the true parameter values $\theta_{\rm fid}$ we get,
\begin{equation}\label{eq:score2}
    \frac{\partial \ln \mathcal{L}}{\partial \theta_{i}}\Bigg|_{\hat{\theta}} \simeq \frac{\partial \ln \mathcal{L}}{\partial \theta_{i}}\Bigg|_{\theta_{\rm fid}} + \delta\theta_{j}\frac{\partial^2 \ln \mathcal{L}}{\partial \theta_{i} \partial \theta_{j}}\Bigg|_{\theta_{\rm fid}} = 0 
\end{equation}
Taking the expectation value of Eq.~\ref{eq:score2} and using the definition of the Fisher matrix, $\mathrm{F}_{ij} = -\Biggl \langle \frac{\partial^2 \ln \mathcal{L}}{\partial\theta_{i}\partial\theta_j} \biggr  \rangle\Bigg|_{\theta_{\rm fid}}$, we obtain
\begin{equation}\label{eq:fish_bias}
    \Biggl \langle \frac{\partial \ln \mathcal{L}}{\partial \theta_{i}}\biggr\rangle\Bigg|_{\theta_{\rm fid}} - \sum_{j} \mathrm{F}_{ij}\delta \theta_{j} = 0.
\end{equation}
Therefore, the bias vector is
\begin{equation}\label{eq:bias_exp}
    \delta\theta_{i} = \sum_{j} \mathrm{F}^{-1}_{ij} \Biggl \langle \frac{\partial \ln \mathcal{L}}{\partial \theta_{j}}\biggr\rangle\Bigg|_{\theta_{\rm fid}}.
\end{equation}
Using the expression in Eq.~\ref{eq:first_derivative} and considering $<\textbf{a}^{\dag}\textbf{a}> = \textbf{C}^{\rm true}$ and the matrix identity, $<\textbf{a}^{\dag}A\textbf{a}> = \mathrm{Tr} (A\textbf{C}^{\rm true})$, Eq~\ref{eq:bias_exp} can be expressed as,
\begin{equation}
\Biggl \langle \frac{\partial \ln \mathcal{L}}{\partial \theta_{i}}\biggr\rangle = -\frac{1}{2}\Big [ \mathrm{Tr}\, (\textbf{C}^{-1}\partial_{i}\textbf{C}) - \mathrm{Tr} (\textbf{C}^{-1}\partial_{i}\textbf{C} \textbf{C}^{-1}\textbf{C}^{\rm true}) \Big].
\end{equation}
Using $\textbf{C}^{\rm true} = \textbf{C} + \Delta \textbf{C}$ and considering that CMB power spectra are diagonal in $\ell$ and $m$ space, from Eq.~\ref{eq:bias_exp}, we get the final expression for the parameter bias vector,

\begin{equation}
    \delta \theta_{i} = \sum_{j}  (\mathrm{F}^{-1})_{ij}\sum_{\ell}\frac{(2\ell + 1)}{2}f_{\rm sky}\mathrm{Tr}[\textbf{C}_{\ell}^{-1}\frac{\partial \textbf{C}_{\ell}}{\partial \theta_{j}}\textbf{C}_{\ell}^{-1} \Delta \textbf{C}_{\ell}],
\end{equation}
where $\Delta \textbf{C}_{\ell} = \textbf{C}^{\rm true}_{\ell} - \textbf{C}_{\ell} (\theta_{\rm fid})$.

\section{Results with LiteBIRD + Planck}\label{app:B}

\begin{table*}[h]
\centering
\renewcommand{\arraystretch}{1.15}
\begin{tabular}{l c c c c c c c c}
\hline\hline
\multicolumn{1}{c}{Model} & $\Omega_\mathrm{b}h^2$ & $\Omega_\mathrm{c}h^2$ & $\ln(10^{10}A_\mathrm{s})$ &
$H_0$ & $n_\mathrm{s}$ & $\tau$ & $m_{\rm e}/m_{\mathrm{e},0}$ & $\alpha/\alpha_0$ \\
 & [$\times10^{-5}$] & [$\times10^{-4}$] & [$\times10^{-3}$] & & [$\times10^{-3}$] & [$\times10^{-3}$] &
[$\times10^{-3}$] & [$\times10^{-4}$] \\
\hline\hline
\textbf{$\Lambda$CDM} & & & & & & & & \\
\quad Without RS & 9.3 & 8.2 & 3.9 & 0.36 & 2.4 & 1.9 & -- & -- \\
\quad With RS    & 9.2 & 8.1 & 3.9 & 0.36 & 2.4 & 1.9 & -- & -- \\
\hline\hline
\textbf{$\Lambda$CDM + $x_{\rm e}$} & & & & & & & & \\
\quad Without RS & 9.8 & 9.9 & 4.1 & 0.89 & 3.8 & 1.9 & 5.0 & -- \\
\quad With RS    & 9.7 & 9.8 & 4.1 & 0.88 & 3.8 & 1.9 & 4.9 & -- \\
\hline
\quad Without RS & 9.3 & 9.4 & 4.4 & 0.79 & 4.4 & 1.9 & -- & 17 \\
\quad With RS    & 9.2 & 9.4 & 4.4 & 0.78 & 4.3 & 1.9 & -- & 17 \\
\hline\hline
\textbf{$\Lambda$CDM + $x_{\rm e}$ + $\sigma_\mathrm{T}$ + $\sigma_\mathrm{R}$} & & & & & & & & \\
\quad Without RS & 103 & 57  & 4.3 & 9.8  & 2.6 & 1.9 & 46  & -- \\
\quad With RS    & 44  & 26  & 4.0 & 4.1  & 2.4 & 1.9 & 19  & -- \\
\hline
\quad Without RS & 9.5 & 8.9 & 4.4 & 0.59 & 4.2 & 1.9 & -- & 13 \\
\quad With RS    & 9.4 & 8.8 & 4.4 & 0.58 & 4.2 & 1.9 & -- & 13 \\
\hline\hline
\end{tabular}
\caption{68\% confidence intervals for the six $\Lambda$CDM parameters and the additional parameter $m_{\rm e}/m_{\mathrm{e},0}$ (and $\alpha/\alpha_0$) obtained for the $\Lambda$CDM model and its possible extensions, comparing analyses with and without including the RS signal along with the Primary CMB for LiteBIRD + Planck.}\label{table:standard_dev_params_LB_Planck}
\end{table*}

\begin{table*}[h]
\centering
\renewcommand{\arraystretch}{1.1}
\begin{tabular}{l c c c c c c c c}
\hline\hline
Model & $\Omega_\mathrm{b}h^2$ & $\Omega_\mathrm{c}h^2$ & $\ln(10^{10}A_{\rm s})$ & $H_0$ & $n_\mathrm{s}$ & $\tau$ & $m_{\rm e}/m_{\mathrm{e},0}$ & $\alpha/\alpha_{0}$\\
\hline
$\Lambda$CDM & 0.4 & 0.6 & 0.0 & 0.7 & 0.2 & 0.2 & -- & -- \\
$\Lambda$CDM + $x_{\rm e}$ (for $m_{\rm e}$) & 0.9 & 0.2 & 0.1 & 1.9 & 0.6 & 0.1 & 1.4 & -- \\
$\Lambda$CDM + $x_{\rm e}$ (for $\alpha$) & 0.5 & 0.2 & 0.2 & 1.2 & 0.4 & 0.2 & -- & 0.7 \\
$\Lambda$CDM + $x_{\rm e}$ + $\sigma_{\rm T}$ + $\sigma_{\rm R}$ (for $m_{\rm e}$) & 57.4 & 54.5 & 6.7 & 57.8 & 6.6 & 1.5 & 57.6 & -- \\
$\Lambda$CDM + $x_{\rm e}$ + $\sigma_{\rm T}$ + $\sigma_{\rm R}$ (for $\alpha$) & 0.3 & 0.4 & 0.0 & 0.7 & 0.1 & 0.2 & -- & 0.2 \\
\hline\hline
\end{tabular}
\caption{Fractional improvement (in \%), $1-\sigma_\theta^{\rm with\,RS}/\sigma_\theta^{\rm without\,RS}$, in the marginalised $1\sigma$ uncertainties quoted in  Table~\ref{table:standard_dev_params_LB_Planck} when the RS signal is included for LiteBIRD + Planck.}
\label{tab:improvement_LB_Planck}
\end{table*}

\begin{figure}
    \includegraphics[width=.5\linewidth]{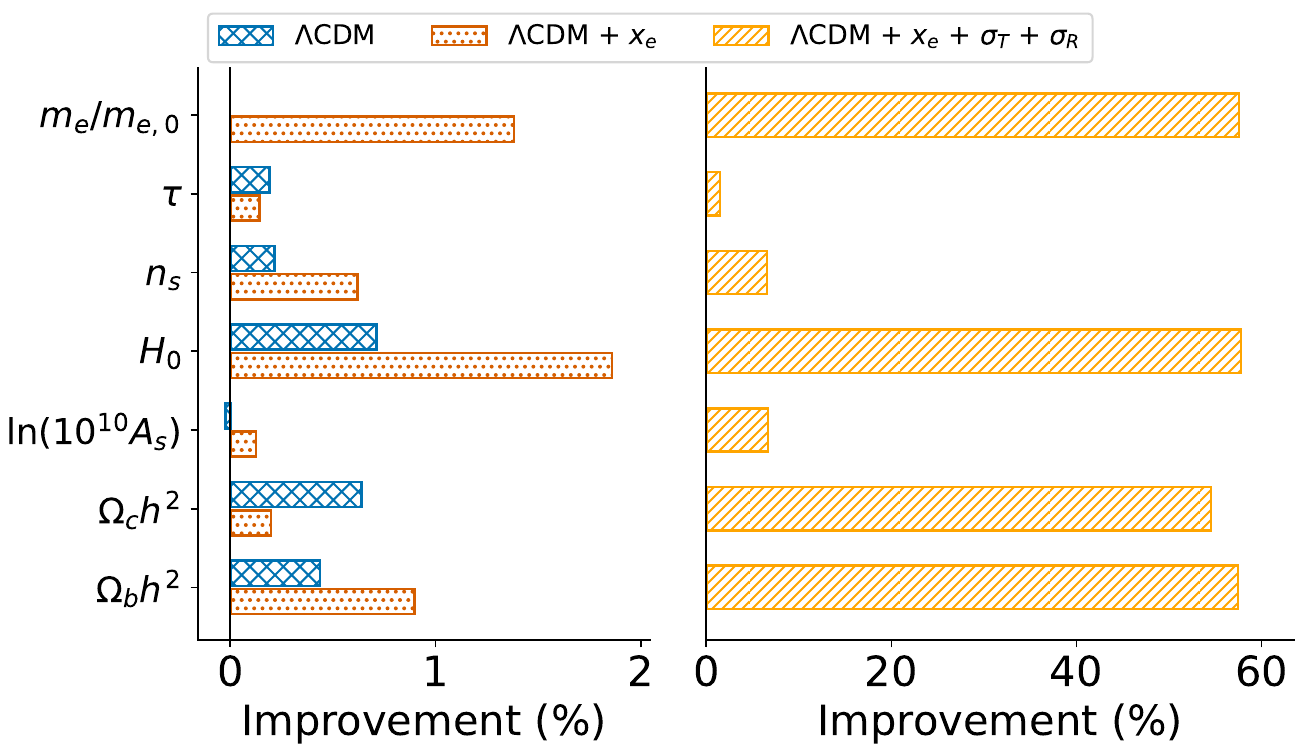}
    \includegraphics[width=.5\linewidth]{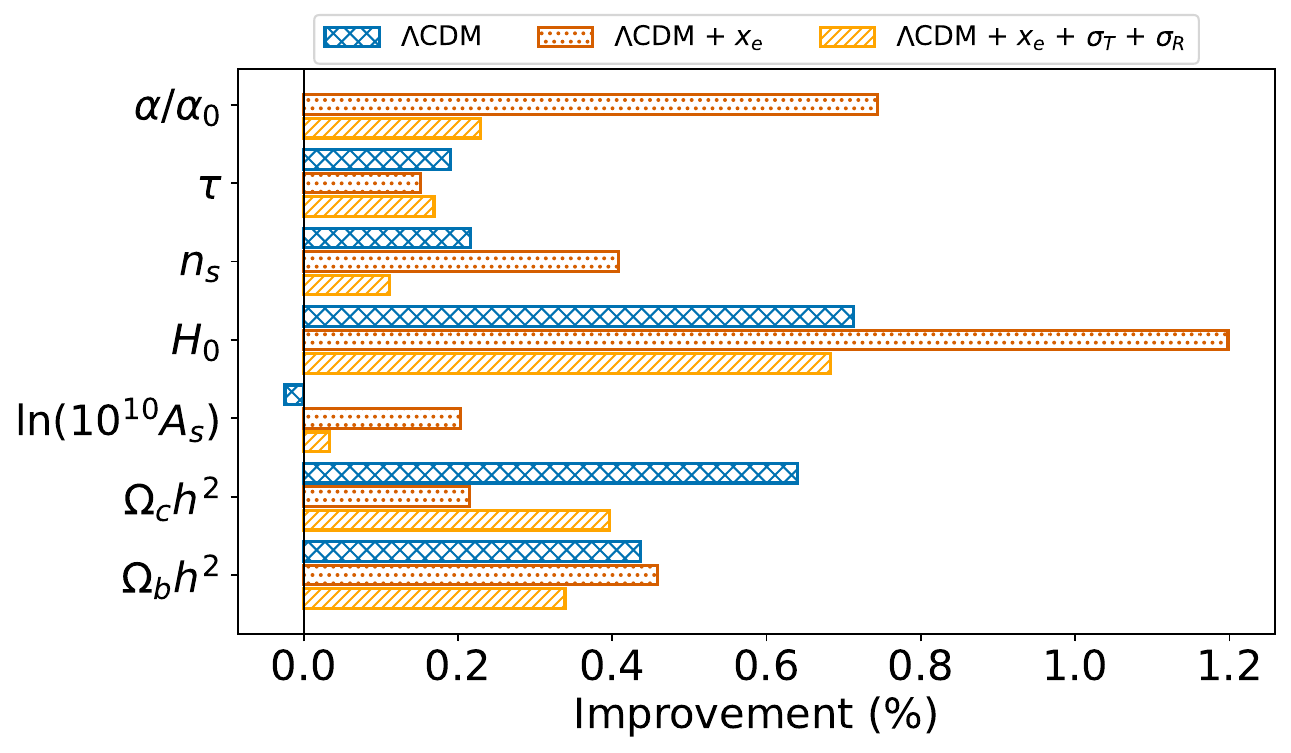}
    
    \caption{Improvement in the error bars $1-\sigma_\theta^{\rm with\,RS}/\sigma_\theta^{\rm without\,RS}$ of the cosmological parameters and the electron mass (left) or the fine-structure constant (right) when we include the RS signal with the primary CMB signal for LiteBIRD + Planck.  The blue, red and orange bars correspond to the $\Lambda$CDM and the two modified recombination cases, respectively. For clarity, in the left-hand panel, the improvements for the $\Lambda$CDM + $x_{\rm e} + \sigma_{\rm T} + \sigma_{\rm R}$ model are separately plotted since the values are much larger than in the other two cases.}
    \label{fig:pico_errorbar_improve_lb_planck}
\end{figure}

\begin{figure*}
    \centering
    \includegraphics[width=\textwidth]{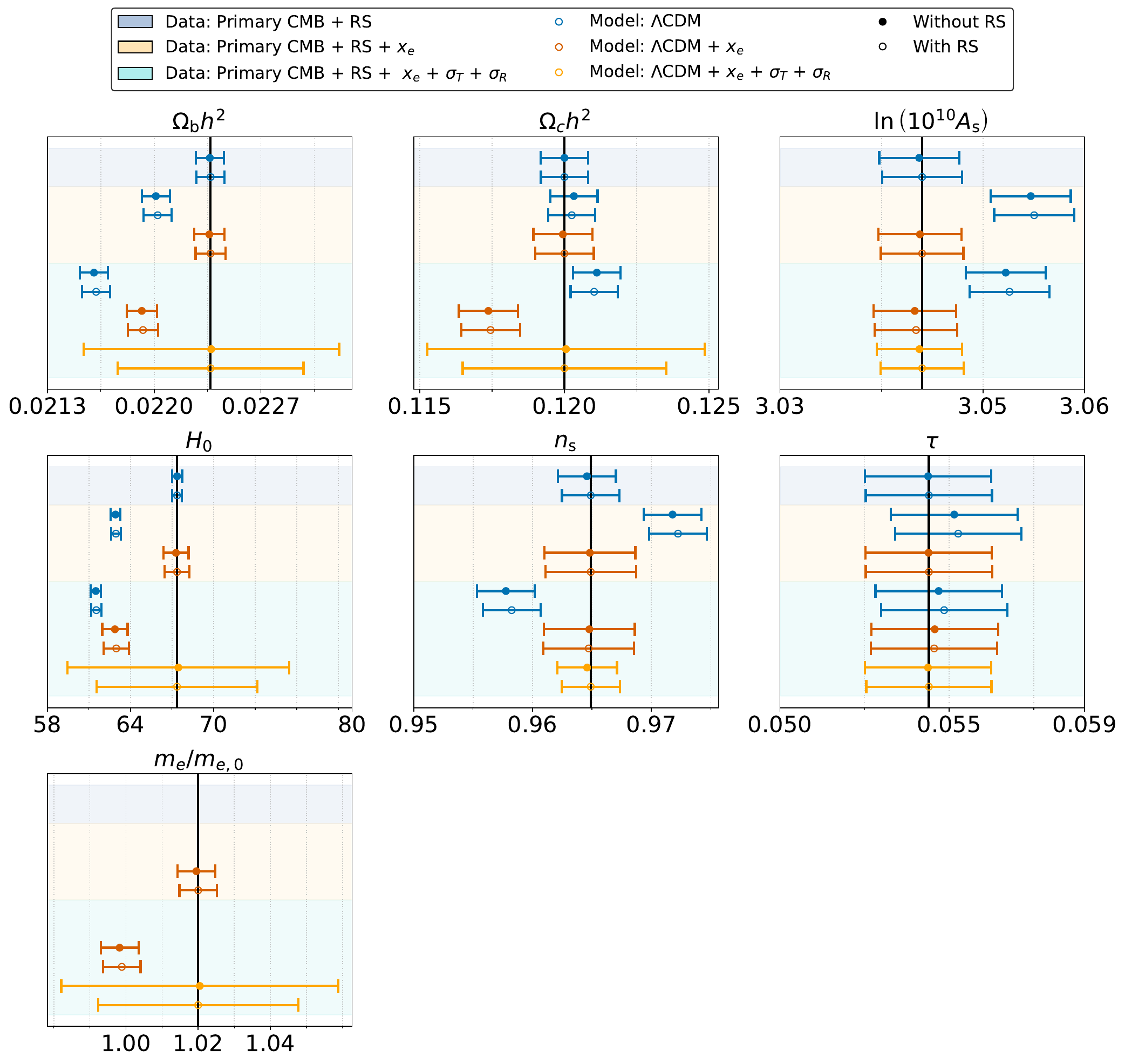}
    \caption{Recovered biased values, $\theta_\mathrm{fid}+\delta\theta$ of the six $\Lambda$CDM parameters and $m_{\rm e}/m_{\mathrm{e},0}$ with their marginalised $1\sigma$ error bars  for LiteBIRD + Planck. The data sets, models, colours and markers are as in Figure~\ref{fig:prameter_bias_me}.} 
    \label{fig:prameter_bias_me_LB_Planck}
\end{figure*}

\begin{figure*}
    \centering
    \includegraphics[width=\textwidth]{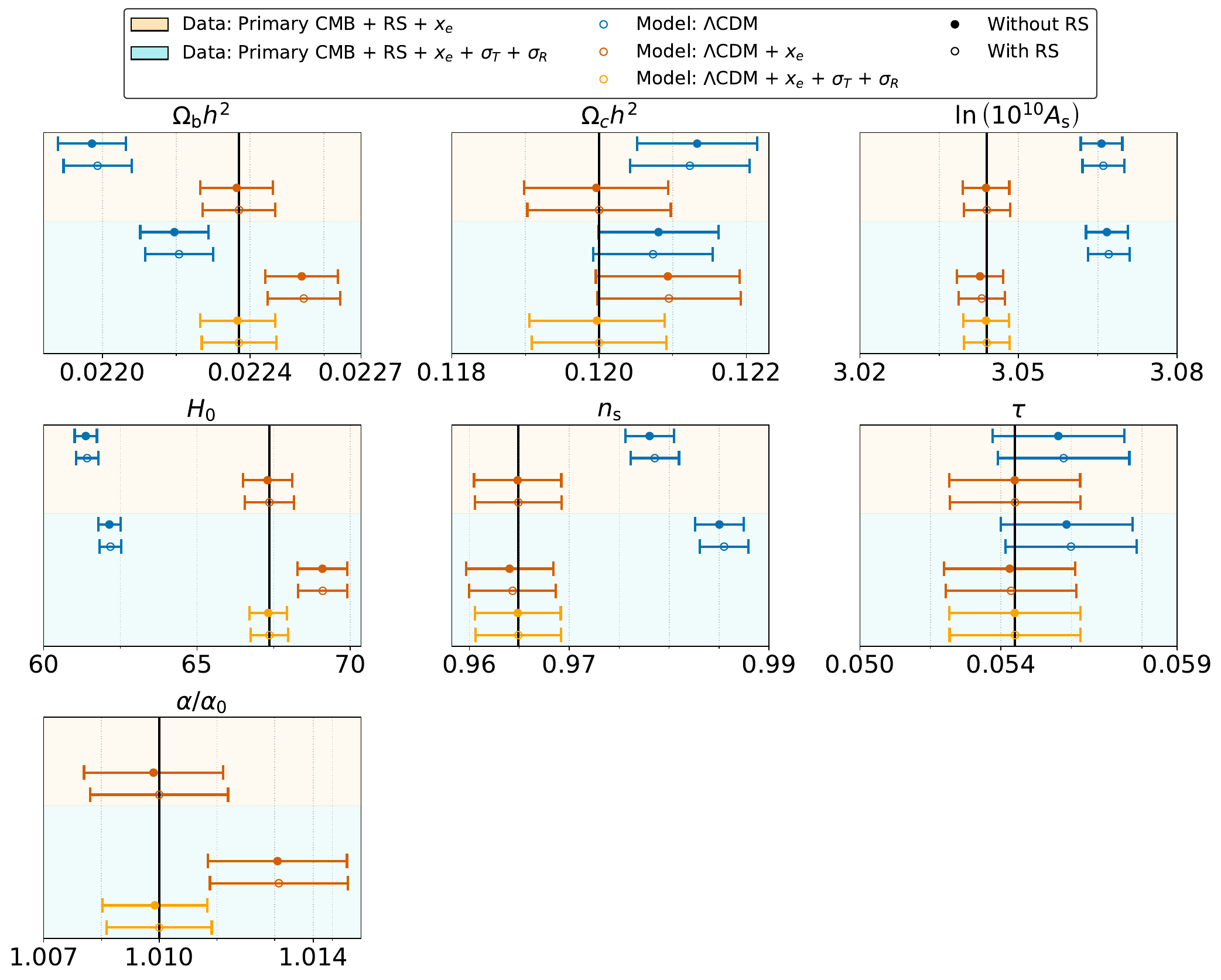}
    \caption{Recovered biased values, $\theta_\mathrm{fid}+\delta\theta$ of the six $\Lambda$CDM parameters and $\alpha/\alpha_{0}$ with their marginalised $1\sigma$ error bars,  for LiteBIRD + Planck. The data sets, models, colours and markers are as in Figure~\ref{fig:prameter_bias_alpha}.}
    \label{fig:prameter_bias_alpha_LB_Planck}
\end{figure*}
\begin{table*}[h]
\centering
\small
\setlength{\tabcolsep}{4pt}
\renewcommand{\arraystretch}{1.1}
\begin{tabular}{l l l c c c c c c c}
\hline\hline
Data & Model & RS & $\Omega_\mathrm{b}h^2$ & $\Omega_\mathrm{c}h^2$ & $\ln(10^{10}A_\mathrm{s})$ & $H_0$ & $n_\mathrm{s}$ & $\tau$ & $m_{\rm e}/m_{\mathrm{e},0}$ \\
\hline
\textbf{Primary CMB } & $\Lambda$CDM & no & $-0.05\sigma$ & $0.00\sigma$ & $-0.07\sigma$ & $0.00\sigma$ & $-0.13\sigma$ & $-0.01\sigma$ & -- \\
\textbf{+ RS} &  &  & $(-0.02\%)$ & $(0.00\%)$ & $(-0.01\%)$ & $(0.00\%)$ & $(-0.03\%)$ & $(-0.04\%)$ & -- \\
\hline\hline
\textbf{Primary CMB } & $\Lambda$CDM & no & $-3.88\sigma$ & $0.40\sigma$ & $2.72\sigma$ & $-12.23\sigma$ & $2.83\sigma$ & $0.40\sigma$ & -- \\
\textbf{+ RS}+$x_{\rm e}$ &  &  & $(-1.61\%)$ & $(0.27\%)$ & $(0.35\%)$ & $(-6.60\%)$ & $(0.71\%)$ & $(1.38\%)$ & -- \\
 &  & yes & $-3.76\sigma$ & $0.31\sigma$ & $2.80\sigma$ & $-12.23\sigma$ & $3.02\sigma$ & $0.46\sigma$ & -- \\
 &  &  & $(-1.55\%)$ & $(0.21\%)$ & $(0.36\%)$ & $(-6.55\%)$ & $(0.76\%)$ & $(1.59\%)$ & -- \\
\cline{2-10}
 & $\Lambda$CDM + $x_{\rm e}$ & no & $-0.08\sigma$ & $-0.05\sigma$ & $-0.05\sigma$ & $-0.08\sigma$ & $-0.02\sigma$ & $0.00\sigma$ & $-0.09\sigma$ \\
 &  &  & $(-0.04\%)$ & $(-0.05\%)$ & $(-0.01\%)$ & $(-0.11\%)$ & $(-0.01\%)$ & $(-0.02\%)$ & $(-0.05\%)$ \\
\hline\hline
\textbf{Primary CMB } & $\Lambda$CDM & no & $-8.27\sigma$ & $1.37\sigma$ & $2.09\sigma$ & $-16.16\sigma$ & $-2.94\sigma$ & $0.15\sigma$ & -- \\
\textbf{+ RS}+$x_{\rm e}$ &  &  & $(-3.42\%)$ & $(0.93\%)$ & $(0.27\%)$ & $(-8.71\%)$ & $(-0.74\%)$ & $(0.53\%)$ & -- \\
+ $\sigma_\mathrm{T}$ + $\sigma_\mathrm{R}$ &  & yes & $-8.15\sigma$ & $1.26\sigma$ & $2.18\sigma$ & $-16.16\sigma$ & $-2.74\sigma$ & $0.24\sigma$ & -- \\
 &  &  & $(-3.36\%)$ & $(0.86\%)$ & $(0.28\%)$ & $(-8.65\%)$ & $(-0.69\%)$ & $(0.83\%)$ & -- \\
\cline{2-10}
 & $\Lambda$CDM + $x_{\rm e}$ & no & $-4.57\sigma$ & $-2.58\sigma$ & $-0.18\sigma$ & $-4.91\sigma$ & $-0.03\sigma$ & $0.09\sigma$ & $-4.17\sigma$ \\
 &  &  & $(-2.02\%)$ & $(-2.19\%)$ & $(-0.02\%)$ & $(-6.66\%)$ & $(-0.01\%)$ & $(0.32\%)$ & $(-2.13\%)$ \\
 &  & yes & $-4.50\sigma$ & $-2.52\sigma$ & $-0.15\sigma$ & $-4.84\sigma$ & $-0.05\sigma$ & $0.08\sigma$ & $-4.07\sigma$ \\
 &  &  & $(-1.98\%)$ & $(-2.13\%)$ & $(-0.02\%)$ & $(-6.52\%)$ & $(-0.02\%)$ & $(0.28\%)$ & $(-2.07\%)$ \\
\cline{2-10}
 & $\Lambda$CDM + $x_{\rm e}$ & no & $0.01\sigma$ & $0.01\sigma$ & $-0.06\sigma$ & $0.01\sigma$ & $-0.12\sigma$ & $-0.01\sigma$ & $0.01\sigma$ \\
 & + $\sigma_\mathrm{T}$ + $\sigma_\mathrm{R}$ &  & $(0.02\%)$ & $(0.05\%)$ & $(-0.01\%)$ & $(0.14\%)$ & $(-0.03\%)$ & $(-0.05\%)$ & $(0.04\%)$ \\
\hline\hline
\end{tabular}
\caption{Same as Table~\ref{table:bias_params_}, but for LiteBIRD + Planck: biases in units of
$\sigma$, with the corresponding percentage shifts in parentheses, in the six $\Lambda$CDM parameters and
$m_{\rm e}/m_{\mathrm{e},0}$, for data generated with $m_{\rm e}/m_{\mathrm{e},0}=1.02$ and analysed with different models.}
\label{table:bias_params_LB_Planck}
\end{table*}
\begin{table*}[h]
\centering
\small
\setlength{\tabcolsep}{4pt}
\renewcommand{\arraystretch}{1.1}
\begin{tabular}{l l l c c c c c c c}
\hline\hline
Data & Model & RS & $\Omega_\mathrm{b}h^2$ & $\Omega_\mathrm{c}h^2$ & $\ln(10^{10}A_\mathrm{s})$ & $H_0$ & $n_\mathrm{s}$ & $\tau$ & $\alpha/\alpha_0$ \\
\hline
\textbf{Primary CMB } & $\Lambda$CDM & no & $-4.30\sigma$ & $1.63\sigma$ & $5.51\sigma$ & $-16.49\sigma$ & $5.40\sigma$ & $0.66\sigma$ & -- \\
\textbf{+ RS +$x_{\rm e}$} &  &  & $(-1.78\%)$ & $(1.11\%)$ & $(0.71\%)$ & $(-8.89\%)$ & $(1.36\%)$ & $(2.26\%)$ & -- \\
 &  & yes & $-4.16\sigma$ & $1.52\sigma$ & $5.60\sigma$ & $-16.48\sigma$ & $5.62\sigma$ & $0.74\sigma$ & -- \\
 &  &  & $(-1.71\%)$ & $(1.03\%)$ & $(0.72\%)$ & $(-8.82\%)$ & $(1.41\%)$ & $(2.53\%)$ & -- \\
\cline{2-10}
 & $\Lambda$CDM + $x_{\rm e}$ & no & $-0.06\sigma$ & $-0.04\sigma$ & $-0.04\sigma$ & $-0.07\sigma$ & $-0.02\sigma$ & $-0.01\sigma$ & $-0.08\sigma$ \\
 &  &  & $(-0.03\%)$ & $(-0.03\%)$ & $(-0.01\%)$ & $(-0.09\%)$ & $(-0.01\%)$ & $(-0.02\%)$ & $(-0.01\%)$ \\
\hline\hline
\textbf{Primary CMB } & $\Lambda$CDM & no & $-1.89\sigma$ & $0.99\sigma$ & $5.77\sigma$ & $-14.36\sigma$ & $8.27\sigma$ & $0.78\sigma$ & -- \\
\textbf{+ RS +$x_{\rm e}$ } &  &  & $(-0.78\%)$ & $(0.67\%)$ & $(0.75\%)$ & $(-7.75\%)$ & $(2.09\%)$ & $(2.69\%)$ & -- \\
\textbf{+ $\sigma_\mathrm{T}$ + $\sigma_\mathrm{R}$} &  & yes & $-1.76\sigma$ & $0.90\sigma$ & $5.85\sigma$ & $-14.38\sigma$ & $8.49\sigma$ & $0.85\sigma$ & -- \\
 &  &  & $(-0.73\%)$ & $(0.61\%)$ & $(0.76\%)$ & $(-7.70\%)$ & $(2.13\%)$ & $(2.92\%)$ & -- \\
\cline{2-10}
 & $\Lambda$CDM + $x_{\rm e}$ & no & $1.73\sigma$ & $0.96\sigma$ & $-0.29\sigma$ & $2.14\sigma$ & $-0.20\sigma$ & $-0.08\sigma$ & $1.70\sigma$ \\
 &  &  & $(0.76\%)$ & $(0.78\%)$ & $(-0.04\%)$ & $(2.57\%)$ & $(-0.09\%)$ & $(-0.28\%)$ & $(0.30\%)$ \\
 &  & yes & $1.78\sigma$ & $0.97\sigma$ & $-0.22\sigma$ & $2.17\sigma$ & $-0.13\sigma$ & $-0.06\sigma$ & $1.74\sigma$ \\
 &  &  & $(0.78\%)$ & $(0.79\%)$ & $(-0.03\%)$ & $(2.58\%)$ & $(-0.06\%)$ & $(-0.21\%)$ & $(0.31\%)$ \\
\cline{2-10}
 & $\Lambda$CDM + $x_{\rm e}$ & no & $-0.04\sigma$ & $-0.03\sigma$ & $-0.04\sigma$ & $-0.06\sigma$ & $-0.01\sigma$ & $-0.01\sigma$ & $-0.08\sigma$ \\
 & + $\sigma_\mathrm{T}$ + $\sigma_\mathrm{R}$ &  & $(-0.02\%)$ & $(-0.02\%)$ & $(-0.01\%)$ & $(-0.06\%)$ & $(-0.01\%)$ & $(-0.02\%)$ & $(-0.01\%)$ \\
\hline\hline
\end{tabular}
\caption{Same as Table~\ref{table:bias_params_alpha}, but for LiteBIRD + Planck: biases in units of
$\sigma$, with the corresponding percentage shifts in parentheses, in the six $\Lambda$CDM parameters and
$\alpha/\alpha_0$, for data generated with $\alpha/\alpha_0=1.01$ and analysed with different models. }
\label{table:bias_params_alpha_LB_Planck}
\end{table*}

\end{document}